# Large-scale machine learning-based phenotyping significantly improves genomic discovery for optic nerve head morphology

Babak Alipanahi[1,†,*], Farhad Hormozdiari[2,†], Babak Behsaz[2,†], Justin Cosentino[1,†], Zachary R. McCaw[1,†], Emanuel Schorsch[1], D Sculley[2], Elizabeth H. Dorfman[1], Sonia Phene[1], Naama Hammel[1], Andrew Carroll[1], Anthony P. Khawaja[3,4,‡], Cory Y. McLean[2,‡,*]

[1]Google, Health, Palo Alto, CA
[2]Google, Health, Cambridge, MA
[3]NIHR Biomedical Research Centre at Moorfields Eye Hospital & UCL Institute of Ophthalmology, London, UK
[4]Department of Public Health & Primary Care, University of Cambridge, Cambridge, UK
[†]These authors contributed equally to this work.
[‡]These authors contributed equally to this work.
[*]To whom correspondence should be addressed: cym@google.com or babaka@google.com.

## Abstract

Genome-wide association studies (GWAS) require accurate cohort phenotyping, but expert labeling can be costly, time-intensive, and variable. Here we develop a machine learning (ML) model to predict glaucomatous optic nerve head features from color fundus photographs. We used the model to predict vertical cup-to-disc ratio (VCDR), a diagnostic parameter and cardinal endophenotype for glaucoma, in 65,680 Europeans in the UK Biobank (UKB). A GWAS of ML-based VCDR identified 299 independent genome-wide significant (GWS; $P \leq 5\times10^{-8}$) hits in 156 loci. The ML-based GWAS replicated 62 of 65 GWS loci from a recent VCDR GWAS in the UKB for which two ophthalmologists manually labeled images for 67,040 Europeans. The ML-based GWAS also identified 92 novel loci, significantly expanding our understanding of the genetic etiologies of glaucoma and VCDR. Pathway analyses support the biological significance of the novel hits to VCDR, with select loci near genes involved in neuronal and synaptic biology or known to cause severe Mendelian ophthalmic disease. Finally, the ML-based GWAS results significantly improve polygenic prediction of VCDR and primary open-angle glaucoma in the independent EPIC-Norfolk cohort.

## Introduction

Genome-wide association studies (GWAS) require accurate phenotyping of large cohorts, but expert phenotyping can be costly and time-intensive. On the other hand, self-reported phenotyping, while cost-effective and often insightful (Tung et al. 2011), can be inaccurate for nuanced phenotypes such as osteoarthritis (Deveza et al. 2017) or infeasible to obtain for complex quantitative phenotypes. Population-scale biobanks, such as the UK Biobank (UKB) (Sudlow et al. 2015) and Biobank Japan (Nagai et al. 2017) that contain genomics, biomedical data, and health records for hundreds of thousands of individuals provide opportunities to study complex disorders and traits (DeBoever et al. 2020). GWAS of individual blood- and urine-based biomarkers, which can be assayed accurately with high throughput, have shed light on disease etiology (Wheeler et al. 2017; Sinnott-Armstrong et al. 2019).

Advances in deep learning have enabled the extraction of medically relevant features from high-dimensional data, such as using cardiac magnetic resonance imaging to infer cardiac and aortic dimensions (Bai et al. 2020), color fundus photographs to detect glaucoma risk (Phene et al. 2019), and optical coherence tomography images to predict age-related macular degeneration progression (Yim et al. 2020). Using medically relevant features extracted from biobank data by machine learning (ML) models as GWAS phenotypes provides an opportunity to identify genetic signals influencing these traits.

Here we propose training an ML model to automatically phenotype a large cohort for genomic discovery. The proposed paradigm has two phases: in the "model training" phase, a database of expert-labeled samples (for which genomics data are not required) is used to train and validate a phenotype prediction model (Fig. 1a). In the "model application" phase, the model is applied to biobank data to predict phenotypes of interest, which are then analyzed for genomic associations (Fig. 1b). This paradigm has several advantages: first, model application is scalable and efficient. Second, a single model can predict multiple phenotypes simultaneously. Third, the model can be applied retrospectively to existing data, resulting in new phenotypes or more accurate predictions for the existing phenotypes. Fourth, multiple lines of evidence can be integrated to predict a single phenotype, which would be prohibitively expensive if performed manually.

As a proof of concept, we investigate predicting glaucoma-related features from fundus images and performing genomic discovery on the predicted features. Glaucoma is an optic neuropathy that results from progressive retinal ganglion cell degeneration (Jonas et al. 2017) and is the leading cause of irreversible blindness globally (Pascolini and Mariotti 2012), affecting more than 80 million people worldwide (Tham et al. 2014). Moreover, glaucoma is one of the most heritable common human diseases, with heritability estimates of 70% (K. Wang et al. 2017), and there is evidence for effective genomic risk prediction (Khawaja et al. 2018; Craig et al. 2020).

The hallmark diagnostic feature of glaucoma is optic disc cupping (Jonas et al. 2017). The vertical cup-to-disc ratio (VCDR; Fig. 1c), a quantitative indicator for optic nerve head morphology and a frequently reported quantitative measure of cupping, is an important endophenotype of glaucoma (Foster et al. 2002; Gordon et al. 2002; Czudowska et al. 2010; Springelkamp et al. 2017). With the advent of very large biobank studies and routine retinal imaging in community optometric practices, there is huge potential for furthering our understanding of glaucoma through population-level analysis of VCDR; however, human grading of optic disc images to ascertain VCDR is costly and infeasible at large scale.

Here we developed an ML model using 81,830 non-UKB, ophthalmologist-labeled fundus images to predict image gradability, VCDR, and referable glaucoma risk. We used the model to predict VCDR in 65,680 UKB participants of European ancestry from 175,337 fundus images. We then performed a GWAS on the ML-based VCDR phenotype (hereafter, “ML-based GWAS”) and compared the results to prior VCDR GWAS, including a recent VCDR GWAS using phenotypes derived from expert-labeled UKB fundus images (Craig et al. 2020). We show that ML-based phenotypes are accurate and substantially more efficient to obtain than expert-phenotyped VCDR measurements, identify novel genetic associations with plausible links to known VCDR biology, and produce more accurate polygenic risk scores for predicting VCDR in an independent population.

# Results

## Overview of the ML-based phenotyping method

We used 81,830 fundus images graded by a panel of experts that passed our labeling guideline assessment (Supplementary Note) to train a phenotype prediction model that jointly predicts image gradability, VCDR, and referable glaucoma risk (Fig. 1d). We split these images into *train*, *tune*, and *test* sets; training images were graded by 1–2 eye care providers with varied expertise, while images in the two latter sets were each graded by three glaucoma specialist experts. We benchmarked model performance on all data splits (Fig. 1e–g; Supplementary Table 1). On the test set of 1,076 test images, the model achieved a Pearson's correlation of $R$=0.91 between predicted and graded VCDR (95% confidence interval [CI]=0.90–0.92) and root mean square error (RMSE) of 0.079 (95% CI=0.074–0.085). Additionally, we validated model generalizability on 2,115 UKB fundus images each graded by 2–3 experts (hereafter, “UKB test set”), which achieved similar predictive performance to the test set (Fig. 1h; $R$=0.89, 95% CI=0.88–0.90; RMSE=0.092, 95% CI=0.088-0.096; Supplementary Table 1).

## ML-based GWAS replicates a manual phenotyping VCDR GWAS and discovers 92 additional novel loci

We applied the VCDR prediction model to the entire set of 175,337 UKB fundus images. After removing 21,400 images predicted to be ungradable (e.g. under- or over-exposed, or out of focus), aggregating predicted VCDR values across left and right eyes and the first and second visits for each individual, subsetting the cohort to individuals of European ancestry, and performing cohort quality control, a cohort of 65,680 individuals with VCDR phenotype remained for further analysis (Supplementary Note and Supplementary Fig. 1). To control for confounding factors (e.g., population structure) and increase power, we added age at the time of visit, sex, average image gradability, number of fundus images used in VCDR calculation, normalized refractive error, genotyping array type, and the top 15 genetic principal components as covariates.

We performed the ML-based GWAS using BOLT-LMM (Supplementary Note). While genomic inflation $\lambda_{GC}$ was 1.20 (Supplementary Fig. 2), the stratified linkage disequilibrium score regression-based (S-LDSC) intercept (Bulik-Sullivan et al. 2015) was 1.06 (s.e.m=0.02), indicating that most test statistic inflation can be attributed to polygenicity rather than population structure. The SNP-based heritability in the ML-based GWAS was 0.43 (s.e.m=0.03), a majority of the 56% heritability estimated for VCDR by twin and family-based studies (Asefa et al. 2019). The ML-based GWAS identified 299 independent genome-wide significant (GWS) hits ($R^2 \leq 0.1$, $P \leq 5 \times 10^{-8}$) at 156 independent GWS loci after merging hits within 250kb together (Fig. 2a, Supplementary Tables 2 and 3). Based on Sum of Single Effects Regression (G. Wang et al. 2020), the number of causal variants within the 156 independent GWS loci was conservatively estimated at 813 (Supplementary Note; Supplementary Tables 4 and 5).

To understand the influence of training dataset size on model performance and GWAS results, we retrained the ML model with as little as 10% of the full training set. Performance curves indicate that using fewer than 8,000 training images achieved a Pearson's correlation $R$=0.83 (95% CI=0.81–0.84) on the UKB test set, identified 131 GWS loci, and replicated 123 of the 156 loci identified in the full model (Supplementary Figs. 3 and 4). An analysis of the implications of phenotyping accuracy on genomic discovery suggested that the difference in power for the model trained with 10% of the training data and the model trained with all data would maximally reach 15% (Supplementary Fig. 5).

Next, we compared the ML-based GWAS results with those from the two largest existing VCDR GWAS. First, we compared with the VCDR meta-analysis from the International Glaucoma Genetics Consortium (IGGC) in 23,899 Europeans (Springelkamp et al. 2017), for which all summary statistics are publicly available (see URLs). The ML-based GWAS replicated all 22 GWS loci, exhibited strong genetic correlation (0.95, s.e.m=0.03, $P = 2.1 \times 10^{-167}$) with the IGGC GWAS (Fig. 2b; Table 1), and effect size regression analysis showed a slope significantly different from zero (slope=0.983, s.e.m.=0.041, $P = 1 \times 10^{-61}$) and indistinguishable from one ($P$=0.67; Supplementary Fig. S7; Supplementary Note). Second, we compared with a GWAS on 67,040 manually-phenotyped UKB fundus images (Craig et al. 2020), for which only the independent genome-wide significant SNPs are publicly available. The ML-based GWAS replicated 62 out of 65 GWS loci with very similar estimated effect sizes (Fig. 2b,c; Table 1) and more significant $P$-values (Supplementary Fig. 6). The three loci not replicated at the GWS level in the ML-based GWAS were all Bonferroni-replicated (adjusting for 65 tests), with $P$-values ranging from $5.5 \times 10^{-8}$ to $6.6 \times 10^{-5}$. Third, we compared our results with a meta-analysis of the Craig *et al.* and IGGC VCDR GWAS (Craig et al. 2020). The ML-based GWAS replicated 82 of the 90 loci at GWS level, with the remaining eight loci Bonferroni-replicated with $P$-values ranging from $1.4 \times 10^{-7}$ to $6.6 \times 10^{-5}$ (Table 1).

Finally, we performed a meta-analysis of our ML-based GWAS with the IGGC VCDR GWAS, which resulted in 189 GWS loci (Supplementary Note; Table 1 and Supplementary Tables 6 and 7). This ML-based meta-analysis replicated 63 out of 65 of Craig *et al.*'s discovery GWAS and 85 out of 90 Craig *et al.*'s meta-analysis at GWS level (Table 1). Taken together, these comparisons demonstrate that the ML-based GWAS accurately identifies known VCDR associations and additionally identifies over 90 novel loci (Fig. 2b), substantially increasing our understanding of the genetic underpinnings of this complex trait.

To assess the biological plausibility of the novel loci identified in the ML-based GWAS, we compared gene set enrichment analyses of the 156 ML-based loci to those of the 65 Craig *et al* loci using FUMA (Watanabe et al. 2017). Nine eye-related gene sets were significantly enriched in both sets of loci. The enrichment odds ratios (ML-based enrichment over Craig *et al*. enrichment) of all nine gene sets were greater than one, suggesting improved identification of functionally relevant pathways in the ML-based loci (Supplementary Fig. 8). To assess effects of distal *cis*-regulatory interactions, we also performed enrichment analyses for the two sets of loci using GREAT (McLean et al. 2010). Consistent with the FUMA results, the ML-based loci were more significantly enriched than the Craig *et al*. loci across all tested ontologies (Supplementary Fig. 9). The ML-based loci were significantly enriched for 22 gene sets, the majority of which are developmental and seven of which are eye-related (Supplementary Table 8). In contrast, the Craig *et al*. loci were significantly enriched for only three gene sets; two of these are eye-related sets that were also enriched in the ML-based results (Supplementary Table 8).

## Biological significance of select novel VCDR-associated loci

Several of the VCDR-associated loci discovered in this study are known to be associated with intraocular pressure (IOP), including rs1361108 near *CENPW* (Gao et al. 2018), rs2570981 in *SNCAIP* (Gao et al. 2018), rs6999835 near *PKIA* (Khawaja et al. 2018), and rs351364 in *WNT2B* (Khawaja et al. 2018). This suggests that a proportion of the genetic variation in VCDR is mediated via IOP and pathophysiological processes affecting the anterior segment of the eye, consistent with IOP being a strong risk factor for glaucoma (Chan et al. 2017). Indeed, we observed that 13% (14 of 107) of the GWS loci from the latest IOP meta-analysis (Khawaja et al. 2018) were GWS in the ML-based VCDR GWAS. In addition, the overall genetic correlation between our ML-based VCDR GWAS and the IOP GWAS meta-analysis is 0.19 (s.e.m.=0.02, $P=5.5\times10^{-15}$), indicating that VCDR is partially explained by IOP. Moreover, a Mendelian randomization (MR) analysis followed by Egger regression (Egger et al. 1997) suggests that IOP has a strong directional association with ML-based VCDR (Intercept=0.001, SE=0.002, $P$=0.7; Slope=0.072, SE=0.020, $P=4\times10^{-4}$), whereas the reverse analysis provided no evidence for a directional association between ML-based VCDR and IOP (Supplementary Note; Supplementary Fig. 10).

VCDR is an objective quantification of the proportion of neuronal tissue at the head of the optic nerve (Fig. 1c). Interestingly, several VCDR-associated loci discovered in this study encompass genes involved in neuronal and synaptic biology, and thus may influence VCDR via direct effects on the retina and optic nerve rather than via IOP. *NCKIPSD* (rs7633840) is involved in the formation and maintenance of dendritic spines, and modulates synaptic activity in neurons (Lee et al. 2006). *CPLX4* (rs77759734) is required for the maintenance of synaptic ultrastructure in the adult retina (Reim et al. 2009). *MARK2* (rs199826712) has roles in neuronal cell polarity and the regulation of neuronal migration (Sapir et al. 2008). These loci complement additional neuronal loci also discovered by Craig *et al.*; some notable examples include *MYO16* (rs10162202), *TRIM71* (rs56131903), and *FLRT2* (rs1289426). An increase in VCDR may be due not only to loss of retinal ganglion cell neurons, but also loss of neural supporting tissue, such as glial cells. One of our novel VCDR-associated loci is an indel on chromosome 8 (8:131606303_CTGTT_C), near *ASAP1*; this locus has been associated with glioma (Melin et al. 2017), suggesting glial cells as potential mediators of the VCDR association.

Several genes at the novel VCDR-associated loci harbor mutations that cause severe Mendelian ophthalmic disease. Here, for the first time, we report common variants at these genes that are associated with VCDR variation at a population level. Three of our novel loci are at *ADAMTSL3* (rs59199978), *PITX2* (rs2661764) and *FOXC1* (rs2745572), all of which are associated with syndromic ocular anterior segment dysgenesis, which in turn causes raised IOP and secondary glaucoma. *ADAMTSL3* is an important paralog of *ADAMTSL1* — which itself is also associated with VCDR in our GWAS. A mutation in *ADAMTSL1* has been reported to cause inherited anterior segment dysgenesis and secondary congenital glaucoma (Hendee et al. 2017). Mutations in *PITX2* and *FOXC1* cause Axenfeld-Rieger syndrome (Seifi and Walter 2018). Common variants at these loci may mark more subtle effects on ocular anterior segment development, resulting in subclinical changes in IOP and VCDR that are apparent on a population level. While *FOXC1* variants have been previously associated with glaucoma (Bailey et al. 2016), this is the first time they have been associated with population-variation in VCDR. Mutations in *PRSS56*, a gene at one of our novel VCDR-associated loci, cause microphthalmia in humans (Gal et al. 2011). Another two of our VCDR-associated loci are at *EYA1* and *EYA2* (eyes absent homologs 1 and 2), genes that are important for eye development in Drosophila. *EYA1* has been implicated in ocular anterior segment anomalies and cataract (Azuma et al. 2000). We also replicate some of the loci identified by Craig *et al.* such as *ELP4*, which has been associated with aniridia (Wawrocka and Krawczynski 2018), a condition characterized by the absence of an iris and that can predispose patients to glaucoma (D'Elia et al. 2007; Wawrocka and Krawczynski 2018).

## ML-based GWAS improves VCDR polygenic risk scores

We developed two polygenic risk scores (PRS) from the ML-based VCDR GWAS using pruning and thresholding (P+T) (Chatterjee, Shi, and García-Closas 2016) and Elastic Net (Zou and Hastie 2005) methods. These PRS were evaluated in two test sets: A holdout set of 2,076 subjects from UKB with VCDR measured by 2–3 experts, and a set of 5,868 subjects from the European Prospective Investigation into Cancer Norfolk (EPIC-Norfolk) cohort with VCDR measured by scanning laser ophthalmoscopy (HRT) (Hayat et al. 2014). Since the EPIC-Norfolk imputation was done using the HRC v1 (Haplotype Reference Consortium) panel, which excludes indels (McCarthy et al. 2016), we subset the ML-based GWAS summary statistics to HRC v1.

For the P+T model, subsetting to HRC v1 results in 282 hits, down from 299 original hits. Using the effect sizes from the ML-based GWAS, this model achieves a Pearson's correlation $R$=0.37 (95% CI=0.33–40) in the UKB adjudicated cohort. The P+T model from Craig *et al.* GWAS does not include 18 out of 76 SNPs (absent in HRC v1) and achieves a Pearson's correlation $R$=0.29 (95% CI=0.25–0.33). The performance metrics of the ML-based Craig *et al.* P+T models when not subset to HRC v1 are shown in Supplementary Fig. 11. Performance in the EPIC-Norfolk set was slightly lower but the P+T model still explained 9.6% of the total variance (Fig. 3a). In both sets, the ML-based P+T model outperformed the Craig *et al.* P+T model (UKB: Δ$R$=0.079, $P$ < 0.031, $n$=2,076; EPIC: Δ$R$=0.082, $P < 5.9\times10^{-4}$, $n$=5,868, permutation test).

We then used the ML-based VCDR values from UKB to train Elastic Net models; after removing all images used in building the adjudicated test set, the training set contained 62,969 samples. In contrast to the P+T model in which GWAS marginal effect sizes are used as PRS weights, Elastic Net jointly learns all weights in a supervised manner. To make up for the 18 missing Craig *et al.* SNPs, we identified linkage disequilibrium-based proxies for all of the missing hits in HRC v1 and included them in training the Elastic Net model. The ML-based Elastic Net model numerically improved upon the P+T model in both UKB ($R$=0.38, 95% CI=0.34–0.41) and EPIC ($R$=0.33, 95% CI=0.30–0.35) sets (Fig. 3b). The Elastic Net model explains 14.2% and 10.6% of total VCDR variation in the UKB and EPIC-Norfolk sets, respectively. The Craig *et al.* Elastic Net model has a more pronounced improvement — probably due to the addition of proxy SNPs — but the ML-based model still significantly outperforms it (UKB: Δ$R$=0.064, $P < 9.6\times10^{-3}$, $n$=2,076; EPIC: Δ$R$=0.053, $P < 6.8\times10^{-4}$, $n$=5,868, permutation test).

## Relationship of primary open-angle glaucoma and VCDR

To study the relationship between primary open-angle glaucoma (POAG) and VCDR, we defined POAG status in UKB using a combination of self-report and hospital episode ICD 9/10 codes (Supplementary Note). ML-based VCDR has moderate predictive power for POAG, with an area under the ROC curve (AUC) of 0.76 ($n$=65,193, 95% CI=0.74–0.78, POAG prevalence=1.9%) and area under the precision-recall curve (AUPRC) of 0.14 (95% CI=0.12–0.16). After binning individuals by ML-based VCDR, we computed odds ratios (ORs) in each bin versus the bottom bin (Fig. 4a). The most extreme bin (VCDR>0.7, $n$=385), which corresponds to a diagnostic criterion for glaucoma (Foster et al. 2002), has an OR of 74.3 (95% CI=57.0–94.3) versus the bottom bin (VCDR<0.3, $n$=30,752).

We then performed mediation analysis (MA) to study the association of VCDR with glaucoma. Similar to MR, MA evaluates the association between an intermediary or mediating phenotype (here, VCDR) and an outcome phenotype (here, glaucoma). However, whereas in MR the SNP set is selected based on association with the mediator, due to the limited availability of glaucoma summary statistics from the study by (Gharahkhani et al. 2020), the SNP set for MA was selected based on association with the outcome. Since, contrary to MR's exclusion restriction, the included SNPs may have affected glaucoma through a pathway other than VCDR (e.g. IOP), the per-SNP estimates of association were meta-analyzed using Egger regression (Egger et al. 1997), which is robust to this assumption (Bowden et al. 2017). The Egger slope of 5.7 (SE=1.8, $P=3\times10^{-3}$) differs significantly from zero, providing evidence that VCDR, as ascertained by our ML-based models, is strongly associated with the odds of glaucoma (Supplementary Fig. 12). We note that the Egger intercept of 0.04 also differs significantly from zero ($P=7\times10^{-7}$), indicating the presence of directional pleiotropy; that is, variants included in the analysis, on average, were associated with an increase in the odds of POAG through a pathway other than VCDR.

As shown above, VCDR is an informative endophenotype for glaucoma and we hypothesize that its PRS should also be predictive of POAG. Indeed, 32 out of 118 loci previously associated with POAG (Gharahkhani et al. 2020) were significantly associated with ML-based VCDR in this study. We applied the ML-based Elastic Net model to the UKB individuals of European ancestry that do not have fundus images (*n*=98,151) to estimate their genetic VCDR. As expected, this genetic model performs noticeably worse than the model using a direct measurement of the VCDR phenotype (AUC=0.56, 95% CI=0.55–0.57; AUPRC=0.07, 95% CI=0.066–0.073, *n*=98,151, POAG prevalence=5.5%). Nonetheless, when we binned samples by VCDR elastic net PRS, participants in the highest bin (PRS $Z>2.5$, *n*=567) had a considerably higher POAG prevalence (OR=3.4, 95% CI=2.6–4.3; Fig 4b) than those in the lowest bin (PRS $Z<-0.1$, *n*=46,136).

In addition to VCDR, the ML model was trained to predict referable glaucoma risk (Phene et al. 2019); this model output can be interpreted as the probability a specialist would refer an individual for detailed glaucoma evaluation. Because the model output is a continuous value, we can evaluate the contribution of features other than VCDR to referable glaucoma risk by regressing out the VCDR signal. We computed glaucoma risk liability as the logit transform of the ML-based glaucoma probability, which is highly correlated with ML-based VCDR (Fig. 4c, Pearson's $R$=0.91, *n*=65,680, $P<1\times10^{-300}$). While a large VCDR is the cardinal feature of a glaucomatous optic nerve, there are other features which suggest glaucoma that are difficult to quantify (e.g. bayoneting or baring of blood vessels, and hemorrhages). To examine the genetic associations with glaucomatous optic disc features other than VCDR, we carried out a GWAS of ML-based glaucoma risk conditioned on ML-predicted VCDR, using BOLT-LMM. The observed SNP heritability was 0.062 (s.e.m=0.013) with genomic inflation of 1.04 and S-LDSC-based intercept of 1.01 (s.e.m=$9.8\times10^{-3}$; Supplementary Fig. 13) and the GWAS identified eight GWS loci (Supplementary Tables 9 and 10). Interestingly, two of these loci, *OCA2-HERC2* (Fig. 4d; rs12913832, $P=2.2\times10^{-66}$) and *TYR* (rs1126809, $P=5.8\times10^{-13}$), have been previously associated with macular inner retinal thickness (retinal nerve fiber layer and ganglion cell inner plexiform layer) as derived from UKB optical coherence tomography images (Currant et al. 2020). These inner retinal parameters have diagnostic utility for glaucoma that is considered complementary to VCDR, and may be particularly efficacious at detecting early glaucoma (Khawaja et al. 2020). Moreover, it is not currently possible to ascertain the thickness of the inner retina from fundus images, which are two-dimensional. Together, this suggests that ML-based phenotyping has the potential to identify glaucoma-related features from fundus images that are complementary to VCDR and not typically gradable by humans.

## Glaucoma prediction in the EPIC-Norfolk cohort

To further assess the utility of the ML-based elastic net VCDR PRS for prediction of glaucoma, we classified the status of EPIC-Norfolk participants (*n*=5,868) for POAG (175 cases and 5,693 controls) using previously described criteria (Methods). We additionally sub-categorized POAG cases into high-tension glaucoma (HTG; 98 cases) and normal-tension glaucoma (NTG; 77 cases; see Methods). Given the enrichment of the VCDR PRS for variants associated with neuronal development and function, we hypothesized that the PRS would be particularly associated with NTG. We fit a logistic regression model to predict POAG status using age, sex, and ML-based elastic net VCDR PRS as its three predictors.

The ML-based elastic net VCDR PRS was strikingly associated with POAG, and particularly NTG, in EPIC-Norfolk (Fig. 5). The ORs (95% CI) comparing the top risk decile with the bottom decile were 9.7 (3.4–27.6) for POAG, 7.4 (2.2–25.2) for HTG, and 16.5 (2.2–125.9) for NTG (Fig. 5). The overall prediction metrics were (AUC=0.74, 95% CI=0.70–0.77; AUPRC=0.08, 95% CI=0.06–0.11; prevalence=3.0%) for POAG, (AUC=0.73, 95% CI=0.68–0.78; AUPRC=0.05, 95% CI=0.03–0.08; prevalence=1.7%) for HTG, and (AUC=0.76, 95% CI=0.71–0.80; AUPRC=0.04, 95% CI=0.03-0.06; prevalence=1.3%) for NTG. The AUC and AUPRC show nominally significant improvements over those from an analogous model using the Craig *et al.* elastic net VCDR PRS for POAG (ΔAUC=0.014, 95% CI=0.0–0.03, $P$=0.03; ΔAUPRC=0.008, 95% CI=0.0–0.02, $P$=0.03, paired bootstrap test) and HTG (ΔAUC=0.014, 95% CI=0.0–0.03, $P$=0.04; ΔAUPRC=0.006, 95% CI=0.0–0.02, $P$=0.04, paired bootstrap test).

## Discussion

Large cohorts of genotyped and phenotyped individuals have enabled researchers to identify genetic influences of many traits. As methods to ascertain genetic variants in large cohorts continue to improve, we anticipate the major challenge for cohort generation to be accurate and deep phenotyping (Delude 2015) at scale. Here we demonstrated that ML-based phenotyping shows promise for improving both scalability to biobank-sized datasets and phenotyping accuracy. We predicted VCDR from all 175,337 UKB fundus images in less than one hour on a distributed computing system. Multiple lines of evidence indicate that the model-based VCDR predictions improve accuracy over manual labeling, including the reproduction of known VCDR-related biology, identification of plausible novel genetic associations, and generation of polygenic risk scores that better predict VCDR in multiple held-out datasets. Additional advantages of ML-based phenotyping over manual labeling are improved joint prediction accuracy for multiple correlated phenotypes and predicting liabilities instead of binary labels for binary phenotypes. By regressing out predicted VCDR from the predicted referrable glaucoma risk (i.e., whether the individual should seek further ophthalmologist care), we identified residual referrable risk not attributable to variation in VCDR.

The improvement of our model-based VCDR GWAS over the recent expert-labeled VCDR GWAS by Craig *et al.* is consistent with improved phenotyping accuracy by our model. The expert labels may include more noise or measurement error than the ML-based labels, as suggested by the inter-grader variability; the inter-grader Pearson's correlation between the two ophthalmologists as reported by Craig *et al.* for images graded multiple times was 0.75 (95% CI=0.72–0.77), whereas the ML model achieves a Pearson's correlation of 0.89 between the model predictions and adjudicated expert labels (95% CI=0.88–0.90). Noise or variability in human grading of VCDR can arise from difficulty in defining the cup-rim border of the optic disc. If the cup-rim border is sloping, rather than having vertical edges, defining it is challenging using two-dimensional images. In this situation, the average VCDR of multiple graders may be considered more accurate than a single grader's score. Our ML-based model was trained and tuned on images that were assessed by multiple graders, and may therefore be expected to outperform a single human grader, on average.

The 92 novel VCDR-associated loci discovered by ML-based phenotyping substantially expand our knowledge of the biological processes underlying optic nerve head morphology. While elevated IOP is an established cause of glaucoma (Chan et al. 2017), characterized by a pathologically enlarged VCDR, our results support the role of IOP contributing to variation in VCDR within the healthy range as well. Of particular note were common VCDR-associated variants in genes harboring mutations which cause inherited anterior segment dysgenesis that is well-characterized phenotypically. Our findings suggest these dysgenesis processes may also occur at subclinical levels and contribute to variation in the complex VCDR phenotype. Understanding the genotype-phenotype link in rare single-gene disorders can therefore improve our knowledge of some of the many contributory causes to complex traits. Our results also support an important role of neuronal development processes for VCDR. It remains uncertain whether these processes primarily influence VCDR during optic nerve development in early life, thereby reflecting population variation in baseline optic nerve head anatomy, or act later in life and reflect a pathological, glaucomatous change in VCDR over time. Interestingly, genes involved in developmental processes more broadly, including development of the cardiovascular and urogenital systems, were significantly enriched in our results (Supplementary Table 8). This may suggest early life processes are a major determinant of VCDR variation in adult populations.

This study also showed that a substantial proportion of VCDR variation can be predicted using a polygenic risk score. Improving VCDR prediction produces a concomitant improvement in glaucoma prediction, as we demonstrated by stratifying glaucoma prevalence using the VCDR PRS. While the UK National Screening Committee does not currently recommend population screening for glaucoma due to tests lacking sufficient positive predictive value (UK National Screening Committee 2019), using polygenic prediction to identify subsets of the general population that are at risk for glaucoma may enable effective screening. Notably, we identified a substantially higher POAG prevalence in the top decile of VCDR PRS and it may be that current screening tests would have sufficient positive predictive value if applied to this enriched population subset. Earlier detection and treatment of glaucoma, a disease that causes progressive and irreversible vision loss, is a key strategy outlined by the World Health Organization for the prevention of blindness worldwide (World Health Organization 2019).

While this study demonstrates the potential for ML-based phenotyping to expand our understanding of the genetic variation underlying complex traits, the method has important limitations that must be taken into account. Application of this technique relies on the trained model producing accurate predictions in the genomic discovery set. Here we showed strong generalizability of the model trained on non-UKB fundus images to the UKB fundus images used for genomic discovery by manually labeling a small subset of UKB fundus images and validating model predictions against these ground truth labels. Application to other phenotypes derived from fundus images, or other data modalities such as optical coherence tomography or magnetic resonance imaging, would require similar demonstrations of model generalizability. Additionally, the initial model training can be costly and time-intensive, as it requires manual labeling to be performed. While our ablation analysis showed that training on only 10% of the data still identified the majority of VCDR-associated loci, model performance did not appear to saturate even at the full training set size. Ongoing improvements to transfer learning may reduce future labeled data requirements (Kolesnikov et al. 2019), though the ability to extrapolate consumer imaging improvements to biomedical imaging is unclear (Raghu et al. 2019).

In summary, we have proposed a method for performing genomic discovery on biobank-scale datasets using machine learning algorithms for accurate phenotyping. A key benefit of the method is its ability to use a modest-sized biomedical dataset annotated with reasonable accuracy to train a model that identifies the underlying patterns and yields usable predictions. Extending the method to additional phenotypes and data modalities in large-scale biobanks could further expand our understanding of disease etiology and improve genetic risk modeling.

## URLs

1000 Genomes Project Phase 3:
ftp://ftp.1000genomes.ebi.ac.uk/vol1/ftp/release/20130502

AREDS dataset:
https://www.ncbi.nlm.nih.gov/projects/gap/cgi-bin/study.cgi?study_id=phs000001.v3.p1

BOLT-LMM software:
https://data.broadinstitute.org/alkesgroup/bolt-lmm

BaselineLD annotations:
https://data.broadinstitute.org/alkesgroup/ldscore

EPIC-Norfolk Study:
https://www.epic-norfolk.org.uk

EyePACS dataset:
http://www.eyepacs.com/research

FUMA:
https://fuma.ctglab.nl

GenomicRanges:
https://bioconductor.org/packages/release/bioc/html/genomicranges.html

GREAT:
http://great.stanford.edu

ImageNet dataset:
http://www.image-net.org

Inoveon:
http://www.inoveon.com

LocusZoom:
http://locuszoom.org

Meta-Soft:
http://genetics.cs.ucla.edu/meta_jemdoc

PLINK software:
https://www.cog-genomics.org/plink1.9

QCtools:
https://www.well.ox.ac.uk/~gav/qctool_v1

Scikit-learn:

https://scikit-learn.org/stable

Springelkamp *et al.* summary statistics:
https://academic.oup.com/hmg/article/26/2/438/2970289

TensorFlow:
https://www.tensorflow.org

TwoSampleMR:
https://github.com/mrcieu/twosamplemr

The UK Biobank Study:
https://www.ukbiobank.ac.uk

# Methods

## Model training and validation

We followed the procedure described previously by (Phene et al. 2019), modifying only to remove all UKB images. Briefly, we used 81,830 color fundus images from AREDS (Age-Related Eye Disease Study Research Group 1999), EyePACS (see URLs), Inoveon (see URLs) from United States, and two eye hospitals in India (Narayana Nethralaya and Sankara Nethralaya). We trained 10 independent multi-task Inception V3 (Szegedy et al. 2016) deep convolutional neural networks on the fundus images, using weights learned from the Image Net dataset (Deng et al. 2009) as pre-trained weights for the convolutional layers. Furthermore, we performed image augmentation (Shorten and Khoshgoftaar 2019) and early stopping (Prechelt 1998) based on mean squared error (MSE) for predicting VCDR on the tune dataset for picking the best model. The final prediction model is the average prediction of the 10 models in the ensemble.

## Phenotype calling in UK Biobank cohort

We included UK Biobank participants with color fundus images. After making predictions for 175,337 images, 21,400 were predicted to be ungradable and were removed. Individual-level VCDR values were computed as the average per-eye VCDR within a single visit, with preference for the initial visit (Supplementary Note).

## Genome-wide association study

We used BOLT-LMM v3.2 (Loh et al. 2018, 2015) to examine associations between genotype and ML-based VCDR in European individuals in UK Biobank, using the `--lmm` parameter to compute the Bayesian mixed model statistics. We used all genotyped variants with minor allele frequency > 0.001 to perform model-fitting and heritability estimation. ML-based VCDR was rank-based inverse normal (INT) transformed to increase the power for association discovery (McCaw et al. 2019). Finally, in our association study, we used sex, age at visit, visit number (i.e., 1 or 2 to indicate visit 1 or visit 2), number of eyes used to compute VCDR, genotyping array indicator, refractive error, average gradability scores of all fundus images included for each participant, and the top 15 genetic principal components as covariates.

## Detecting independent genome-wide significant loci

Genome-wide significant (GWS; $P \leq 5\times10^{-8}$) lead SNPs, independent at $R^2$=0.1, were identified using PLINK's `--clump` command (see URLs). The reference panel for LD calculation contained 10,000 unrelated subjects of European ancestry from the UK Biobank. Loci were formed around lead SNPs based on the span of reference panel SNPs in LD with the lead SNPs at $R^2 \geq 0.1$. Loci separated by fewer than 250 kb were subsequently merged.

## SNP-heritability estimates for ML-based VCDR

We computed the SNP heritability for ML-based VCDR by applying stratified LD score regression (Bulik-Sullivan et al. 2015) on the VCDR GWAS summary statistics while using the 75 baseline LD annotations provided by S-LDSC authors (see URLs).

## Replication of existing loci

Loci for ML-based VCDR and comparator studies were formed as described above, and the common reference panel of 10k randomly selected unrelated subjects from the UK Biobank. Replication was assessed via the proportion of ML-based VCDR loci that overlapped with comparators, and the proportion of comparator loci that overlapped with the ML-based VCDR loci. Thus, replication required that both studies had a GWS variant within a common genomic region, although not necessarily the same variant.

## Mendelian randomization and mediation analyses

We performed two sample Mendelian randomization analysis, implemented via TwoSampleMR (see URLs), to examine the causal association between IOP, as assessed by Khawaja *et al* (Khawaja et al. 2018), and ML-based VCDR. Per-SNP associations were meta-analyzed using Egger regression (Egger et al. 1997).

We performed mediation analysis to estimate the association between ML-based VCDR and glaucoma, as assessed by Gharakhani *et al.* (Gharahkhani et al. 2020). Mendelian randomization is in fact a special case of mediation analysis in which the instrumental variables (here, SNPs) have no effect on the outcome (here, glaucoma) other than through the mediator (here, ML-based VCDR). Our mediation analysis differs from Mendelian randomization in that, due to limited availability of summary statistics from Gharakhani *et al*, the SNP set was defined based on association with the mediator (ML-based VCDR) rather than the outcome (glaucoma). Among the 118 independent, significant glaucoma SNPs identified by Gharakhani *et al*, 116 remained after harmonizing with VCDR. To account for probable direct effects of the candidate SNPs on glaucoma odds, for example via IOP, the per-SNP associations were again meta-analyzed via Egger regression.

## VCDR polygenic risk score

We developed two PRS using the pruning and thresholding (P+T) (Chatterjee, Shi, and García-Closas 2016) and Elastic Net (Zou and Hastie 2005) methods. The UKB test cohort was graded using the same guidelines used in grading other datasets used in this study. The HRT-derived VCDR was examined and, for participants with good quality scans in both eyes, the mean value of right and left eyes was considered, as previously described (Khawaja et al. 2013). Genotyping was carried out on the Affymetrix UK Biobank Axiom array, as previously described (Khawaja et al. 2019).

In the P+T model, we used a set of variants common to the UKB and EPIC Norfolk cohorts. EPIC-Norfolk's imputation was performed using the HRC v1 panel and excludes indels (McCarthy et al. 2016); thus, to harmonize the variants we filtered out variants from Craig *et al.* and our ML-based GWAS not present in EPIC-Norfolk. This resulted in 58 variants from the 76 reported variants from Craig *et al.* GWAS (i.e., 18 variants were dropped) and 282 of the 299 variants from our ML-based GWAS (i.e., 17 fewer variants).

In the Elastic model, we used the ML-predicted VCDR as the target label from the 62,969 UKB training samples to train the elastic model. For Craig *et al,* we used 76 variants that included the 58 variants from the P+T model and 18 additional proxy variants that are in high LD ($R^2$≥0.6) with the 18 variants dropped from Craig *et al.* P+T model. The same set of 282 variants as used in P+T were used for the ML-based model. We performed 5-fold cross validation and used the L1-penalty ratios of [0.1, 0.5, 0.7, 0.9, 0.95, 0.99, 1.0].

## Glaucoma liability conditional analysis

We defined glaucoma risk liability as the logit transform of the highest-level of ML-based glaucoma probability ("likely glaucoma"; Supplementary Note) as:

$$g = \log\left(\frac{p}{1-p}\right)$$

where *p* and *g* denote ML-based glaucoma risk probability and liability, respectively. We performed conditional analysis on ML-based glaucoma risk liabilities using BOLT-LMM conditional on ML-based VCDR. In this conditional analysis, we additionally adjusted for the same covariates used in the primary ML-based VCDR GWAS.

## Glaucoma subtypes prediction in the EPIC-Norfolk cohort

We analyzed 5,868 participants from the EPIC-Norfolk Eye Study cohort who were genotyped using the Affymetrix UK Biobank Axiom Array, met inclusion criteria and quality control, and had scanning laser ophthalmoscopy VCDR measurements (Supplementary Note). Included participants had a mean age of 68 years (SD 7.7, range 48-90), 55% were women, and the mean VCDR was 0.34 (SD 0.23). Of the 5,868 samples, 175 were classified as POAG cases (see Supplementary Note for detailed POAG criteria), of which 98 were classified as high tension glaucoma (HTG; IOP > 21 mmHg) and 77 as normal tension glaucoma (NTG; IOP ≤ 21 mmHg) on the basis of the corneal-compensated IOP at the Eye Study assessment. Pre-treatment IOP was imputed by dividing by 0.7 for participants using glaucoma medication at the time of assessment, as previously described (Khawaja et al. 2018).

We extracted age, sex, POAG status, NTG status, and HTG status from all 5,868 samples. We fitted independent logistic regression models to predict POAG, HTG, and NTG statuses using VCDR PRS, age, and sex as predictors. We considered both the ML-based elastic net VCDR PRS and the Craig *et al.* elastic net PRS described above.

## Acknowledgements

We acknowledge and appreciate research participants in all datasets used for their dedication and contributions. We highly appreciate the contributions of members of both the Genomics and Ophthalmology teams in Google Health, and thank the labeling software team in Google Health for their assistance in data labeling. This research has been conducted using the UK Biobank Resource application 17643. We thank Inoveon, Aravind Eye Hospital, Sankara Nethralaya, and Narayana Nethralaya for providing de-identified data; Jorge Cuadros, OD, PhD, from EyePACS, for data access and helpful conversations; and the National Eye Institute Study of Age-Related Macular Degeneration Research Group and study participants for their valuable contribution to this research. The EPIC-Norfolk study (DOI 10.22025/2019.10.105.00004) has received funding from the Medical Research Council (MR/N003284/1 and MC-UU_12015/1) and Cancer Research UK (C864/A14136). The genetics work in the EPIC-Norfolk study was funded by the Medical Research Council (MC_PC_13048). We are grateful to all the participants who have been part of the project and to the many members of the study teams at the University of Cambridge who have enabled this research. Anthony Khawaja, MD, PhD was supported by a Moorfields Eye Charity Career Development Fellowship and a UK Research and Innovation Future Leaders Fellowship. Lastly, we highly appreciate all glaucoma specialists, ophthalmologists and optometrists who helped in labeling fundus images.

## Author contributions

BA, AWC, and CYM conceived the study. BA, FH, APK and CYM designed the study. BA, FH, BB, JC, ZRM, ES, APK and CYM performed experiments and analyzed results. LD, DS, SP, NH contributed to methodology and data collection. BA, FH, BB, ZRM, APK and CYM wrote the manuscript. All authors contributed to the final version of the manuscript.

## Competing interests

A.P.K. is an employee of the UCL Institute of Ophthalmology, London, UK. The remaining authors are employees of Google LLC and own Alphabet stock as part of the standard compensation package. This study was funded by Google LLC.

## Data availability

The UKB data are available through the UK Biobank Access Management System https://www.ukbiobank.ac.uk/. We will deposit the derived data fields and model predictions following UKB policy, which will be available through the UK Biobank Access Management System. The summary statistics, associated loci, and polygenic risk score coefficients will be made publicly available.

# Tables

**Table 1. Replicated loci of ML-based VCDR GWAS and meta-analysis at GWS level**. "ML-based 10% (VCDR)" denotes the GWAS performed on VCDR predictions of the ML model trained using only 10% of the training data. "ML-based + IGGC (VCDR)" denotes meta-analysis of ML-based and IGGC VCDR GWAS. Likewise, "Craig *et al.* + IGGC (VCDR)" denotes meta-analysis of Craig *et al.* VCDR and IGGC VCDR GWAS. Genetic correlation was only computed when the full set of summary statistics were available.

| **Discovery GWAS details** | | | **Number of loci replicated in ML-based VCDR GWAS** | **Number of loci replicated in ML-based + IGGC VCDR GWAS** | **S-LDSC-based genetic correlation with ML-based VCDR** |
|---|---|---|---|---|---|
| **Study (phenotype)** | **Number of participants** | **Loci** | | | |
| ML-based (VCDR) | 65,680 | 156 | – | 151 | – |
| ML-based 10% (VCDR) | 65,044 | 131 | 123 | 125 | 0.99 ($2.1x10^{-3}$) |
| ML-based + IGGC (VCDR) | 89,579 | 189 | 151 | – | 0.97 ($2.6x10^{-3}$) |
| IGGC (VCDR) | 23,899 | 22 | 22 | 22 | 0.95 (0.03) |
| Craig *et al.* (VCDR) | 67,040 | 65 | 62 | 63 | N/A |
| Craig *et al.* + IGGC (VCDR) | 90,939 | 90 | 82 | 85 | N/A |
| Khawaja *et al.* (IOP) | 139,555 | 107 | 14 | 22 | 0.19 (0.02) |
| Gharahkhani *et al.* (POAG) | 383,500 | 118 | 32 | 40 | N/A |

# Figures

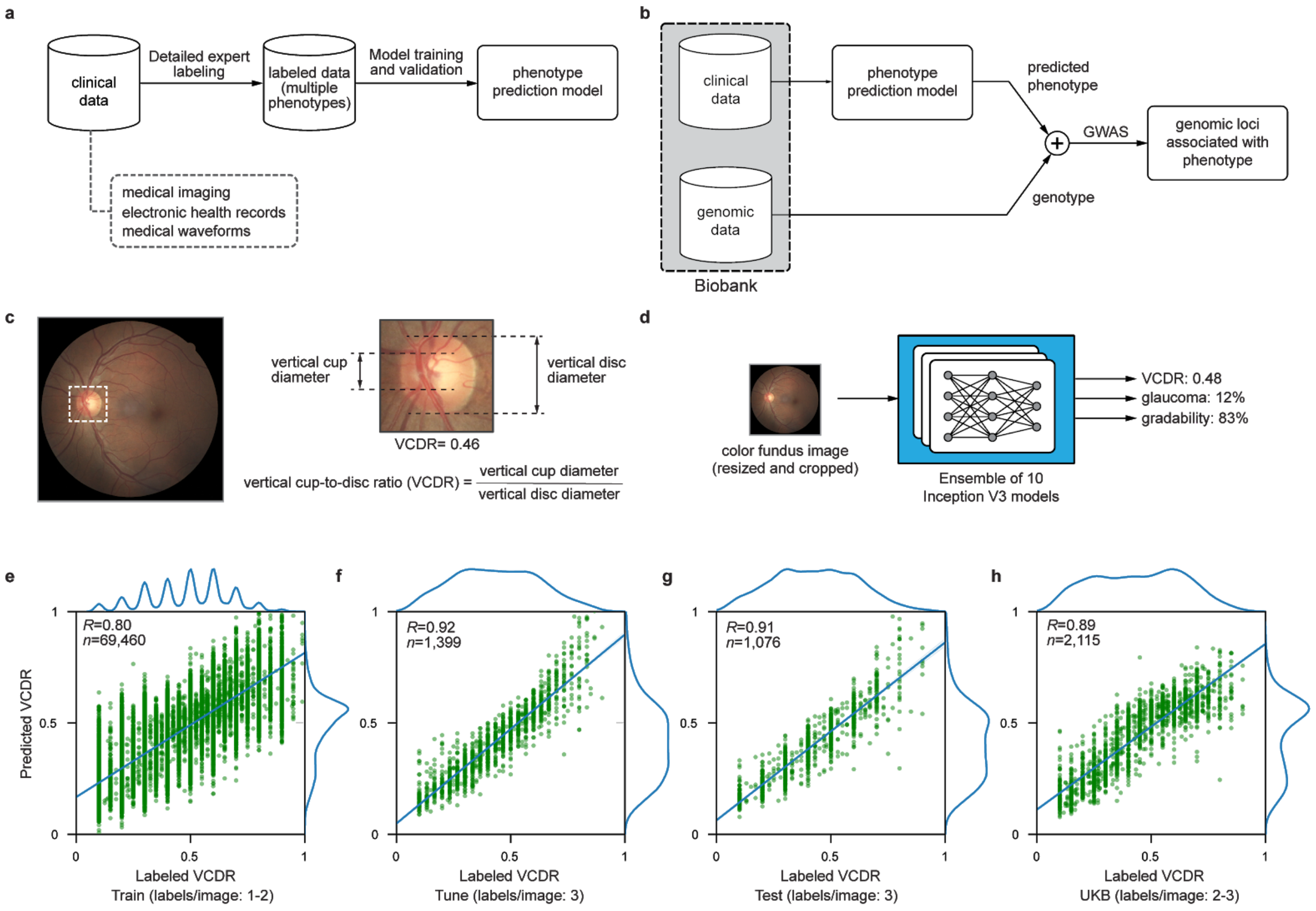

**Fig 1. ML-based phenotyping concept and its application to VCDR. a,** "Model training" phase in which a phenotype prediction model is trained using expert-labelled data. **b,** "Model application" phase in which the validated phenotype prediction model is applied to new, unlabelled data followed by genomic discovery. **c,** Definition of vertical cup-to-disc ratio (VCDR) in a real fundus image. **d,** Schematic of the multi-task ensemble model used in phenotype prediction. **e-h,** Scatter plots of the ML-based VCDR vs expert-labelled VCDR values for the train (**e**), tune (**f**), test (**g**), and UK Biobank (**h**) datasets. Number of grades per image is shown in parentheses.

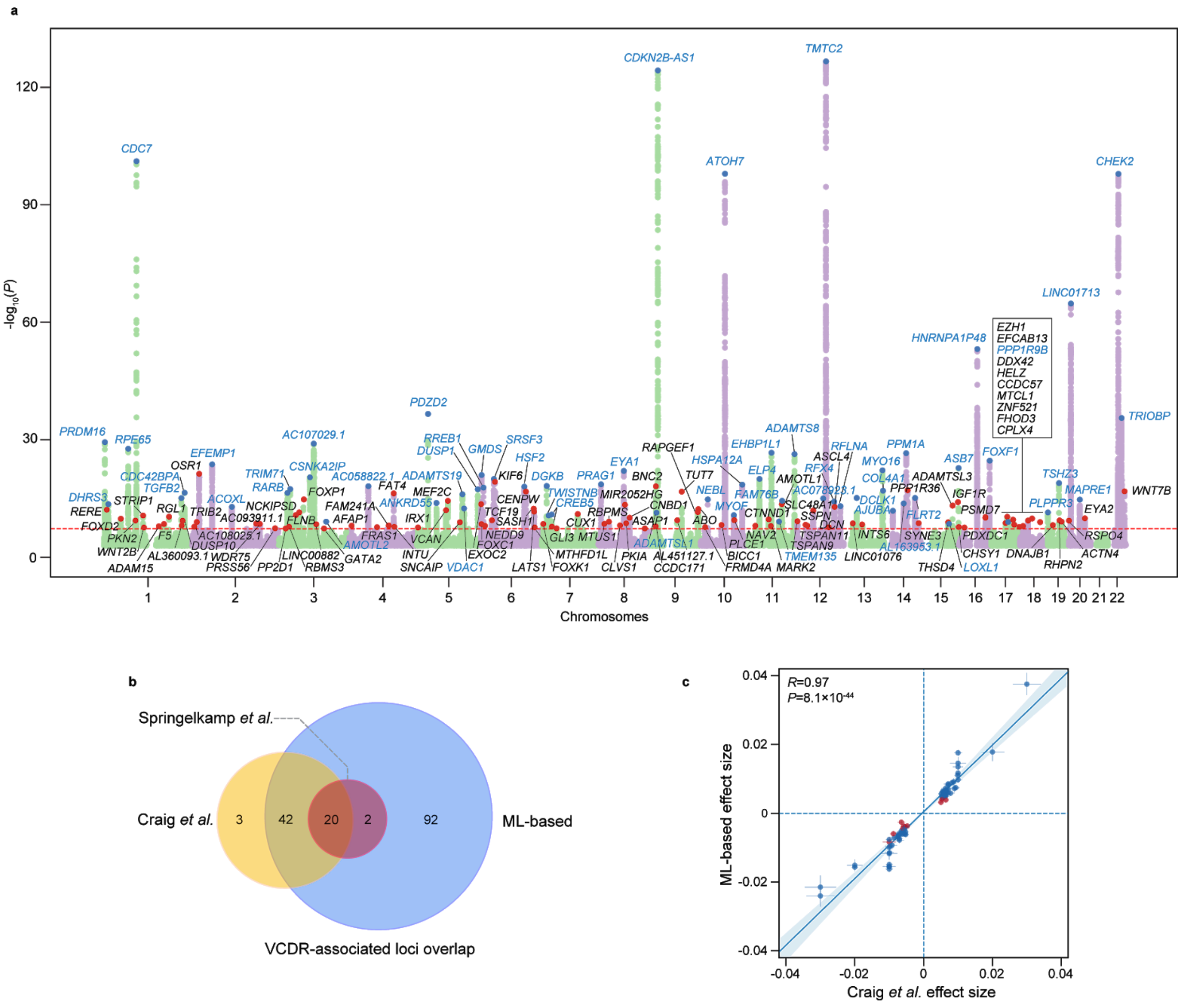

**Fig 2. ML-based VCDR GWAS results and comparison to known associations. a,** Manhattan plot depicting ML-based VCDR-associated *P* values from the BOLT-LMM analysis. There are 156 GWS (genome-wide significant) loci, representing 299 independent ($R^2$=0.1) GWS hits. For each locus, the closest gene is shown. Blue gene names and dots indicate loci also identified in the Craig *et al.*'s study and red dots and black gene names indicate novel loci. The dashed redline denotes the GWS *P* value, $5\times10^{-8}$. **b,** Venn diagram of loci overlap for three VCDR GWAS. ML-based GWAS replicates all 22 loci of the IGGC VCDR meta-analysis (Springelkamp *et al.*) and 62 of 65 loci identified by Craig *et al.*, while discovering 92 novel loci. **c,** Effect sizes for the 73 GWS hits shared by the Craig *et al.* and ML-based VCDR GWAS. The three Craig *et al.* hits not included failed the ML-based GWAS QC (rs61952219 for low imputation quality and rs7039467 and rs146055611 for violating Hardy-Weinberg equilibrium). Blue and red dots denote the SNP being more significant in the ML-based and Craig *et al.* GWAS, respectively. The banding in Craig *et al.* effect sizes is due to large effect sizes being reported in multiples of 0.01. The blue line is the best fit line and the shaded area shows the 95% confidence interval.

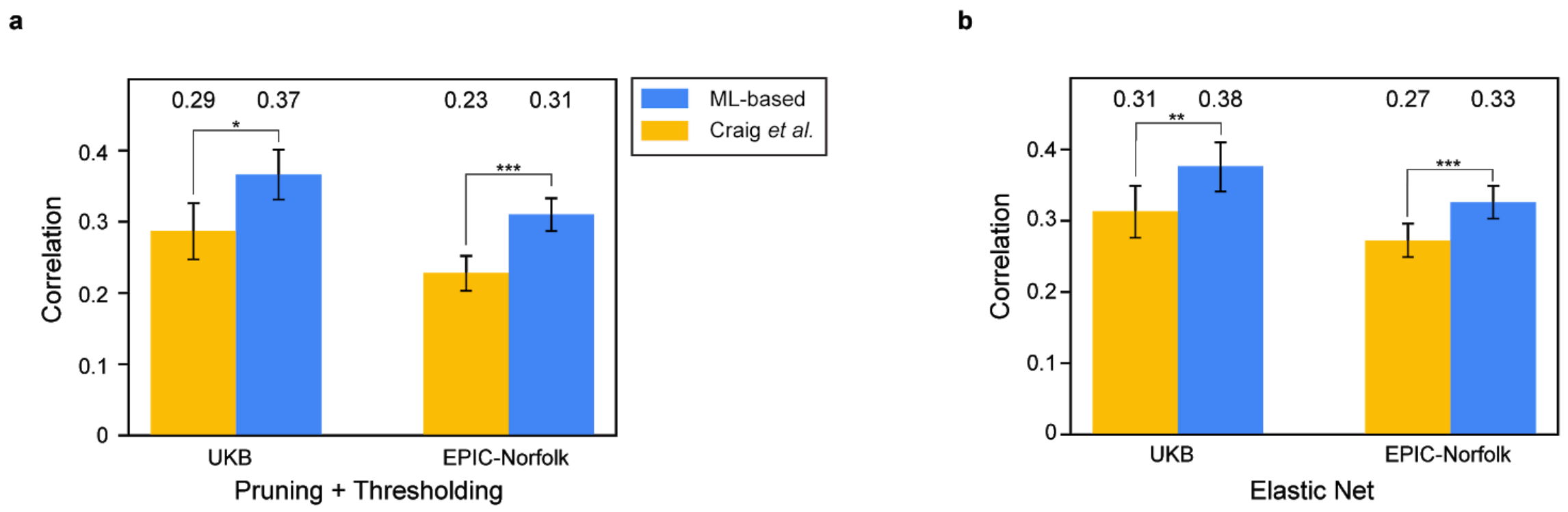

**Fig 3. VCDR polygenic risk score performance metrics.** Pearson's correlations between measured VCDR values and predictions of the pruning and thresholding (P+T) (**a**) and the Elastic Net models (**b**) are shown for the PRS learned from ML-based and Craig *et al.* hits. Error bars depict 95% confidence intervals. Numbers above bars are the observed Pearson's correlations. Indications of P value ranges: * $P \leq 0.05$, ** $P \leq 0.01$, *** $P \leq 0.001$. The Craig *et al.* P+T model uses 58 out of 76 hits. Measured VCDR values were obtained from adjudicated expert labeling of fundus images (UKB, $n$=2,076) and scanning laser ophthalmoscopy (HRT) (EPIC-Norfolk, $n$=5,868).

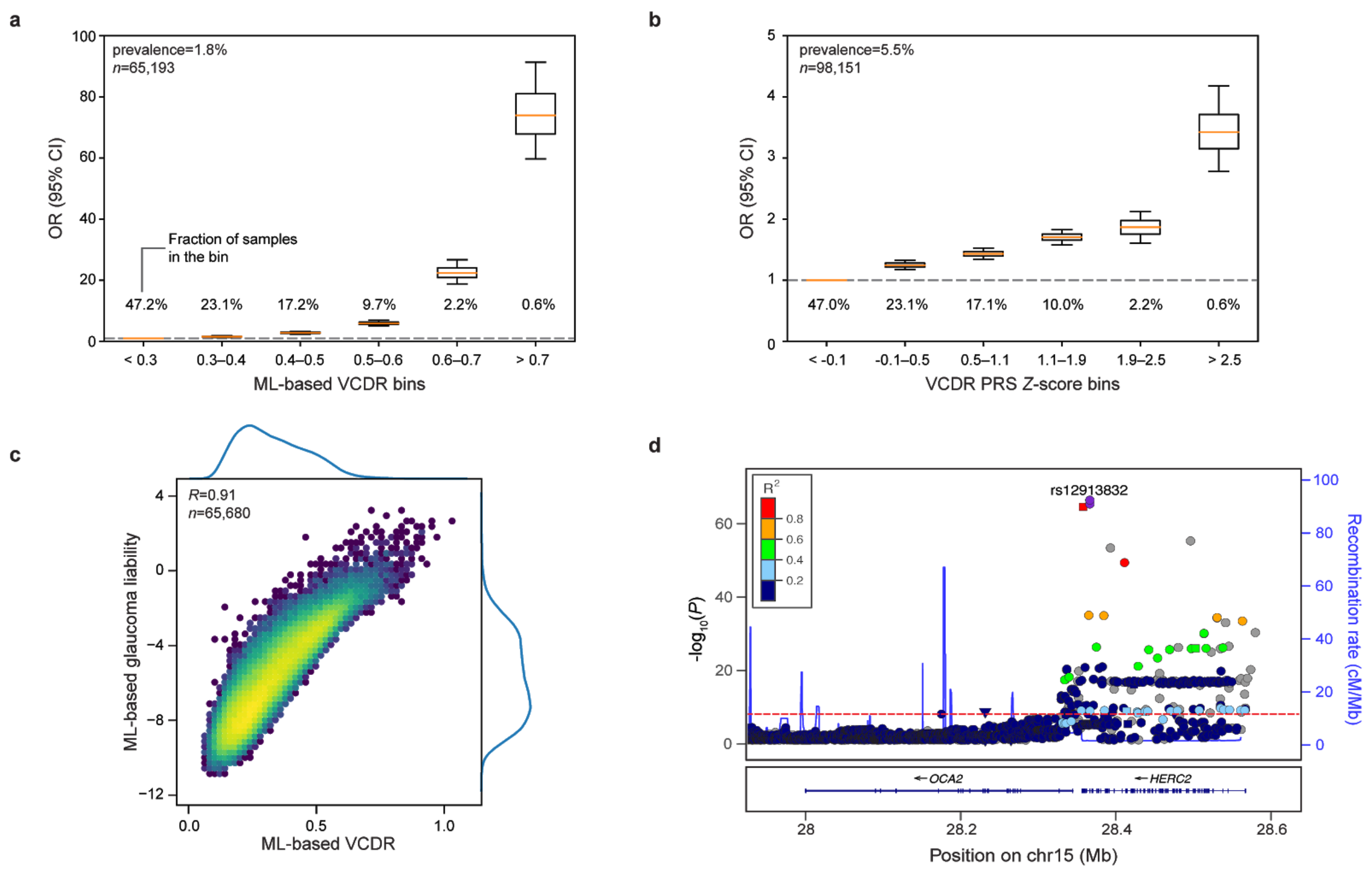

**Fig 4. Relationship between glaucoma and VCDR. a,** Glaucoma odds ratios for each ML-based VCDR bin vs. the bottom bin is shown. The fraction of individuals in each bin is shown ($n$=65,193). **b,** Glaucoma odds ratios for different VCDR elastic net PRS bins vs. the bottom bin for individuals with a glaucoma phenotype not used in the GWAS or developing the PRS ($n$=98,151). The fractions are selected to match those from **a**. **c,** A histogram of ML-based glaucoma liability vs. ML-based VCDR (Pearson's correlation $R$=0.91, $n$=65,680, $P<1\times10^{-300}$). **d,** LocusZoom for the strongest associated variant (rs12913832, $P=2.2\times10^{-66}$) in the ML-based glaucoma liability GWAS conditioned on the ML-based VCDR.

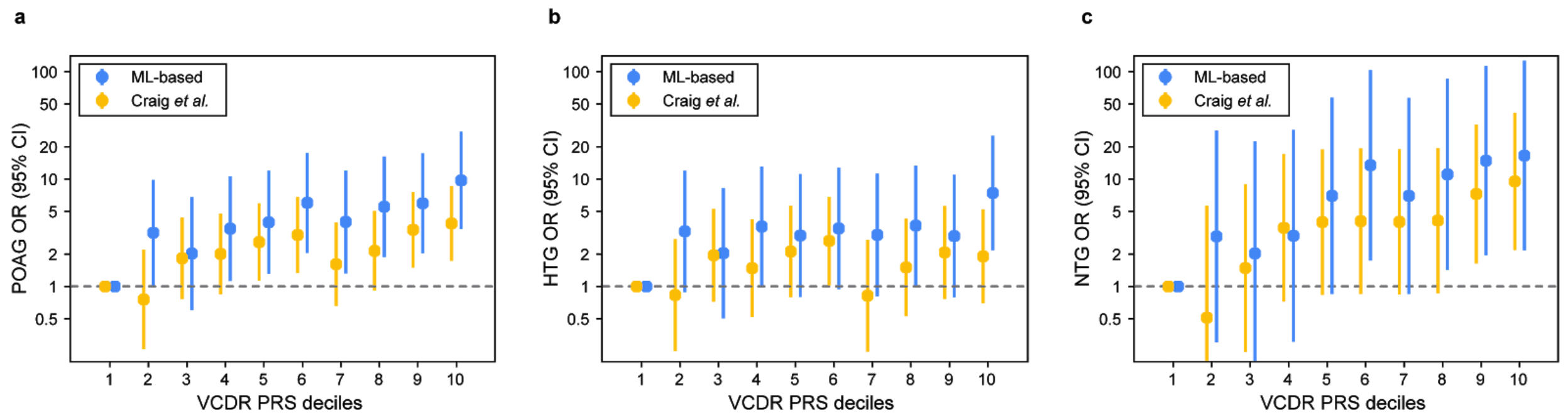

**Fig 5. Primary open-angle glaucoma (POAG) prediction in the EPIC-Norfolk cohort.** Odds ratios for POAG prevalence by decile of VCDR PRS; reference is decile 1. Results are from logistic regression models adjusted for age and sex for **a**, primary open-angle glaucoma (175 cases, 5,693 controls), **b**, high-tension glaucoma (HTG; 98 cases, 5,693 controls), and **c**, normal-tension glaucoma (NTG; 77 cases, 5,693 controls). Results are presented for the ML-based elastic net VCDR PRS (blue) and the Craig *et al*. elastic net VCDR PRS (yellow). Note the y-axis log scale.

# Supplementary Information - Large-scale machine learning-based phenotyping significantly improves genomic discovery for optic nerve head morphology

Babak Alipanahi[1,†,*], Farhad Hormozdiari[2,†], Babak Behsaz[2,†], Justin Cosentino[1,†], Zachary R. McCaw[1,†], Emanuel Schorsch[1], D Sculley[2], Elizabeth H. Dorfman[1], Sonia Phene[1], Naama Hammel[1], Andrew Carroll[1], Anthony P. Khawaja[3,4,‡], Cory Y. McLean[2,‡,*]

[1]Google, Health, Palo Alto, CA
[2]Google, Health, Cambridge, MA
[3]NIHR Biomedical Research Centre at Moorfields Eye Hospital & UCL Institute of Ophthalmology, London, UK
[4]Department of Public Health & Primary Care, University of Cambridge, Cambridge, UK
[†]These authors contributed equally to this work.
[‡]These authors contributed equally to this work.
[*]To whom correspondence should be addressed: cym@google.com or babaka@google.com.

## Table of Contents:

# Phenotype Prediction Model

## Data collection

Grading of images has been described in detail previously (Phene et al. 2019). In short, graders assessed each image for gradability, presence of various optic nerve head (ONH) features (including estimation of VCDR; the ratio between the vertical diameter of the cup and the vertical diameter of the disc) and referable glaucomatous optic neuropathy (GON). Gradability was measured based on image quality, blurring, media opacity, or any other confounding reason. If graders selected “ungradable” for a particular feature or referable GON, then no grade was collected for that aspect. To enable systematic training of graders, we developed grading guidelines and iterated on the guidelines with a panel of three fellowship-trained glaucoma specialists to increase inter-rater agreement; please refer to the Supplementary Table 1 in (Phene et al. 2019). Similar to clinical practice, for VCDR graders were asked to provide an estimate as a decimal between 0.0 and 1.0, with 0.1 increments ($0.0 < VCDR < 1.0$). For referable GON grading we developed guidelines for a four-point GON assessment ("non-glaucomatous", "low-risk glaucoma suspect", “high-risk glaucoma suspect”, and “likely glaucoma”) where the “high-risk glaucoma suspect” or “likely glaucoma” levels were considered referable, that is, the ONH appearance was worrisome enough to justify referral for comprehensive examination. Graders were asked to provide a referable GON grade after evaluating the image for the other ONH features.

## Model training and validation

Data processing and model training has been described previously (Phene et al. 2019). In short, we first remove all UK Biobank (UKB) samples from the "train", "tune", and "test" sets used by (Phene et al. 2019). We use 81,830 color fundus images from AREDS (age-related eye disease study) (Age-Related Eye Disease Study Research Group 1999), EyePACS (https://www.eyepacs.org/), Inoveon (http://www.inoveon.com/) from United States and two eye hospitals in India (Narayana Nethralaya and Sankara Nethralaya). In total, 69,460 of the 79,355 training images were gradable. All color fundus images are cropped to center the retinal image and resized to 587×587 pixels. The prediction model consists of ten independently trained multi-task Inception V3 (Szegedy et al. 2016) deep convolutional neural networks. To accelerate model training, convolutional layers were initialized using the weights learned from the Image Net dataset (Deng et al. 2009). We used image augmentation (Shorten and Khoshgoftaar 2019) (randomly changing brightness, hue, contrast, saturation and flipping the image horizontally and vertically) to regularize model training in TensorFlow (Abadi et al. 2016). Full set of hyperparameters is given in the "Model hyper-parameters" section. We used early stopping (Prechelt 1998) based on root mean squared error (RMSE) for predicting VCDR in the tune set for each model. The final prediction was the average prediction of the ten models in the ensemble. Model performance metrics are listed in Supplementary Table 1.

# Genomic Discovery

## UK Biobank cohort

The UK Biobank is a very large multisite cohort study established by the Medical Research Council, Department of Health, Wellcome Trust medical charity, Scottish Government and Northwest Regional Development Agency. Detailed study protocols are available online (http://www.ukbiobank.ac.uk/resources/ and http://biobank.ctsu.ox.ac.uk/crystal/docs.cgi). A baseline questionnaire, physical measurements, and biological samples were undertaken in 22 assessment centers across the UK between 2006 and 2010. All UK residents aged 40 to 69 years who were registered with the National Health Service (NHS) and living up to 25 miles from a study center were invited to participate. The study was conducted with the approval of the North-West Research Ethics Committee (ref 06/MRE08/65), in accordance with the principles of the Declaration of Helsinki, and all participants gave written informed consent. This research has been conducted using the UK Biobank Resource under Application Number 17643.

Ophthalmic assessment was not part of the original baseline assessment and was introduced as an enhancement in 2009 for 6 assessment centers which are spread across the UK (Liverpool and Sheffield in North England, Birmingham in the Midlands, Swansea in Wales, and Croydon and Hounslow in Greater London). Imaging of both eyes was performed using the Topcon 3D OCT- 1000 Mark II in a dark room without pupil dilation. The instrument takes a color photograph of the retina as well as an optical coherence tomography scan; we used the color photographs in the current study. The right eye was imaged first. Refractive status of both eyes was measured by autorefraction (Tomey RC5000; Erlangen-Tennenlohe). Spherical equivalent was calculated as the sphere + 0.5 * cylinder and participant-level refractive error was taken as the mean of right and left values.

## Phenotype calling

After predicting VCDR for all 175,337 fundus images from 85,665 individuals in UKB, we first remove the 21,400 images which are predicted as ungradable for VCDR. Recall that there are two imaging visits, called visit 1 and 2. We define the phenotype only based on one of these visits, because there is an approximate 5 years difference between the two visits and many factors such as age, medications, eye operations can be materially different between the two visits.

If an individual has any gradable image(s) from visit 1, we define the phenotype based on these images; otherwise, we define it based on visit 2 (a.k.a. first repeat imaging visit). For a specific visit, we first average the VCDRs of each eye and then average these per eye VCDRs if both eyes have gradable images. Moreover, to account for the impact of image gradability on the phenotype, we computed the average gradability score of all images used in defining an individual's phenotype. For the details and statistics of phenotype calling, please refer to Supplementary Fig. 1. To control for the small variations in phenotype calling, we add the visit number used (i.e., 1 or 2) and the number of eyes used in calling the phenotype (i.e., 1 or 2) as covariates. After subsetting to individuals of European ancestry and removing samples with excess heterozygosity or missingness, putative sex chromosome aneuplody, and missing refractive error report, we call the VCDR phenotype for 65,680 individuals.

## Genome-wide association study

We use linear mixed models as implemented in BOLT-LMM v3.2 (Loh et al. 2015) to account for population structure and cryptic relationships in UK Biobank, and to increase association power. We applied BOLT-LMM to all individuals of European ancestry with available VCDR who passed our sample QC and had non-missing covariates ($n$=65,680). We used sex, age at visit, visit number (i.e., 1 or 2 to indicate visit 1 or visit 2), number of eyes used to compute VCDR (i.e., 1 or 2 to indicate one eye or both eyes are used), genotyping array indicator, refractive error, average gradability scores of all fundus images used in phenotype calling and the top 15 genetic principal components as covariates. To increase association power and make the normality assumption more plausible, ML-based VCDR was rank-based inverse normal (INT; (McCaw et al. 2019)) transformed. We considered the autosomal chromosomes for our GWAS and filtered out variants with minor allele frequency (MAF) < 0.001, imputation INFO score < 0.8, or Hardy-Weinberg equilibrium (HWE) $P < 1\times10^{-10}$ in Europeans. Using these filters, 13,110,443 variants passed QC. To verify that our association results were not driven by population stratification, we applied LD score regression (Bulik-Sullivan et al. 2015).

## Identification and comparison of loci

Genome-wide significant (GWS; $P \leq 5\times10^{-8}$) lead SNPs, independent at $R^2$=0.1, were identified using the `plink --clump` command (v1.90b4). The reference panel comprised a random sample of 10,000 unrelated subjects of white European ancestry from the UK Biobank. Around each lead SNP, a locus was defined as the span of reference panel SNPs in LD with the lead SNP at $R^2 \geq 0.1$. For consistency with locus formation as implemented by FUMA (Watanabe et al. 2017), loci separated by fewer than 250 kb were merged, and the most significant, independent SNP in the merged locus was retained as the lead SNP. Gene context annotations were added from the GRCh37 version of GenCode v34 "comprehensive gene annotations." Only protein-coding genes and level 1 long noncoding RNAs (lncRNA) were considered.

For comparing loci across studies, loci were formed for each study using the common reference panel and procedure described above. Locus overlap metrics were calculated using the GenomicRanges package (Lawrence et al. 2013) in R (v3.2.3). In comparing loci form studies A and B, it is possible for a single locus from study A to overlap multiple loci from study B or conversely. To accommodate this, we report the maximum of the number of loci in study B *overlapped by* a locus from study A, and the number of loci in study A *overlapped by* a locus from study B.

## Fine-mapping

Fine-mapping of independent significant loci was performed via Sum of Single Effects Regression (SuSiE; v0.9.0) (Wang et al. 2020), as implemented in R. Briefly, SuSiE identifies the likely causal variants in a region using a variational approximation to Bayesian variable selection regression. A posterior inclusion probability (PIP) is assigned to each SNP in the locus, quantifying the probability that the SNP has a non-zero effect on the outcome. The sum of PIPs for SNPs in a locus is the posterior expectation of the number of causal variants in that locus. To estimate the total number of distinct genetic signals for ML-based VCDR detected in our analysis, PIPs were aggregated across all loci where SuSiE reported no more than the number of GWS variants in the locus. Loci where SuSiE reported more causal variants than GWS variants were considered potentially unreliable. The number of causal variants in such loci was conservatively estimated as the number of GWS variants in the locus, which is potentially an underestimate. Moreover, uncommon SNPs (those with minor allele frequencies below 5%) were removed from the fine-mapping analysis, some of which are likely causal. Nevertheless, the estimated number of genetic signals for VCDR detected by our analysis was 813.

## Ablation analysis

In order to assess the dependence of model quality on the training data size, we analyzed model performance when trained on progressively smaller subsets of the full training data. Predicted VCDR vs adjudicated VCDR correlations for different sets are depicted in Supplementary Fig. 3. In particular, when training only on 10% of the data (~7,900 samples), the Pearson's correlations (ratio with regard to the original correlation) were 0.87 (94%), 0.87 (96%) and 0.83 (93%), for the Tune, Test, and UKB Adjudicated cohorts.

We also performed a GWAS using the "10% model" predictions, which identified 131 genome-wide significant loci, replicating 123 of the 156 loci identified by the full model. The scatter plot of *P* values for the ML-based GWAS and the 10% ablation GWAS are presented in Supplementary Fig. 4.

## Genomic discovery power analysis

To assess how the power for genomic discovery varied with phenotyping quality, we followed the "Noisy Measurement Model" (Hormozdiari et al. 2016). Specifically, consider the following:

$$Y = X\beta + \epsilon \quad (1)$$

where $Y$ is the true VCDR, $X$ is genotype, and $\varepsilon$ is an environmental residual. Suppose $Y$ and $X$ have been standardized to mean zero and variance one. Let $h^2$ denote the per-SNP heritability, then the residual variance is $1-h^2$. We do not observe the true VCDR, but instead a mismeasured version $Y^*$, which is related to $Y$ via

$$Y^* = Y + \delta \quad (2)$$

where $\delta$ is mean-zero measurement error. Substituting (1) into (2) gives the variance component model

$$Y^* = X\beta + \epsilon + \delta \quad (3)$$

From model (3) we can derive the asymptotic non-centrality parameter (NCP) of the standard Wald $\chi^2$ test of association by considering a sequence of contiguous alternatives (Serfling 1980). The NCP for the $\chi^2$ test based on $Y^*$ takes the simple form

$$NCP = n \cdot \frac{\rho^2 h^2}{1 - \rho^2 h^2}$$

where *n* is the sample size, $\rho^2$ is the square of the correlation between the mismeasured *Y** and true *Y* phenotypes, and $h^2$ is the true heritability of *Y*. Power and Non-Centrality Curves as a function of per-SNP heritability, stratified by the correlation between the measured and true phenotypes are shown in Supplementary Fig. 5.

Applying the above model to compare the “10% model” and the model trained on the entire training set, at our GWAS sample size *n*=65,680, the difference in power between a GWAS where the correlation between the observed and true VCDR measurements is 0.89 and a GWAS where the correlation is 0.83 can reach as high as 15%.

## Replication slope analysis

To jointly test the ML-based hits for replication of the IGGC VCDR meta-analysis, we first scaled the effect size estimates of the ML-based GWAS results to account for winner's curse. Winner's curse correction was performed by fitting a two-component Gaussian mixture model, as described in supplemental section 5.3 of (Turley et al. 2018):

$$f(\hat{\beta}_j|\pi, \tau^2) = \pi \cdot N(\hat{\beta}_j|0, \sigma_j^2) + (1-\pi) \cdot N(\hat{\beta}_j|0, \sigma_j^2 + \tau^2).$$

Here $\hat{\beta}_j$ is the estimated effect size, $\pi$ is the prior probability of belonging to the null component, $\sigma_j{}^2$ is the sampling variance (i.e., squared standard error) of $\hat{\beta}_j$, and $\tau^2$ is the variance in effect sizes at non-null SNPs. Model parameters $(\pi, \tau^2)$ were estimated by maximum likelihood using the expectation maximization algorithm (McCaw, Julienne, and Aschard 2020; Meng and Rubin 1993). The observed effect sizes were shrunk to their posterior expectation via:

$$E[\beta_j|\hat{\beta}_j] = (1-\gamma_j)\frac{\tau^2}{\tau^2 + \sigma_j^2} \cdot \hat{\beta}_j$$

where $\gamma_j$ is the posterior responsibility of the null-component for SNP *j*:

$$\gamma_i = P[\beta_j = 0|\hat{\beta}_j] = \frac{\pi \cdot N(\hat{\beta}_j|0, \sigma_j^2)}{\pi \cdot N(\hat{\beta}_j|0, \sigma_j^2) + (1-\pi) \cdot N(\hat{\beta}_j|0, \sigma_j^2 + \tau^2)}.$$

Winner's curse correction was performed using all genotyped variants as input, with final model parameter estimates of $(\pi = 0.958, \tau^2 = 0.000427)$. We then identified 214 (of 299) ML-based hits additionally present in the IGGC VCDR meta-analysis and regressed the IGGC effect sizes on the winner's curse-corrected ML-based GWAS effect size estimates (Supplementary Fig. 7).

## Meta-analysis

GWAS summary statistics for ML-based VCDR were combined with summary statistics from a previous meta-analysis of VCDR by the International Glaucoma Genetics Consortium (IGGC) using Meta-Soft (Han and Eskin 2011). The following strategy was adopted for selecting the final *P* value. At each SNP, the $I^2$ statistic (Higgins and Thompson 2002) was calculated to quantify the proportion of total variation across studies that was attributable to effect size heterogeneity. For SNPs with $I^2 > 0$, a random effects meta-analysis was performed, using Han and Eskin's "RE2" model (see URLs), whereas for those SNPs with $I^2 = 0$, a fixed effects meta-analysis was performed. Selecting whether to perform random or fixed effects meta-analysis on the basis of $I^2$ is an effort to apply the most appropriate model for the observed effect sizes. Among the 8.6M SNPs present in both studies, 68% had $I^2 = 0$, while the remaining 32% had $I^2 > 0$. For the 4.5M SNPs present in our analysis but not in IGGC, the original P-value from the ML-based VCDR GWAS was retained. The S-LDSC intercept was 1.06 (s.e.m=0.01) and the SNP-heritability h2g was 0.37 (s.e.m=0.02).

## Functional analyses with FUMA and GREAT

Functional analyses were performed in FUMA (Watanabe et al. 2017). We assigned each variant to the nearest gene within 10kb using FUMA's "SNP2GENE" functionality, and performed gene-set enrichment analysis using FUMA's "GENE2FUNC" functionality. In both cases, we adopted the default parameter settings. We compared the *relative enrichment* of gene sets that were significant according to both the ML-based and the Craig *et al.* GWAS of VCDR. Specifically, enrichment refers to the odds that a gene in the gene-set was detected in a given GWAS, and relative enrichment is the odds ratio comparing our GWAS with the Craig et al. GWAS. Fisher's exact test was applied to determine whether enrichment differed significantly between the two studies. Those sets where the relative enrichment (odds ratio) exceeds 1 represent biologically interesting gene-sets where the ML-based GWAS captured more of the constituent genes.

GREAT enrichment analyses were performed on the human GRCh37 assembly using GREAT v4.0.4 (McLean et al. 2010). The default "basal+extension" region-gene association rule was used with 5 kb upstream, 1 kb downstream, 1000 kb extension, and curated regulatory domains included. Analyses were performed using the same loci as in the FUMA analyses described above; 65 loci from Craig *et al.* and 156 loci from the ML-derived GWAS. Terms were considered statistically significant if the Bonferroni-corrected *P*-values for both the region-based and gene-based tests were ≤ 0.05.

## VCDR-IOP Mendelian Randomization

Two sample Mendelian randomization (MR) for the association between intraocular pressure (IOP) and ML-based VCDR was performed using the TwoSampleMR (see **URLs**) package in R (4.0.2). Among the 187 independent significant SNPs for IOP from (Khawaja et al. 2018), 183 remained after harmonizing with ML-based VCDR. This provided 183 candidate instrumental variables for quantifying the association between IOP and ML-based VCDR. Based on Cochran's Q test, there was significant evidence of pleiotropy ($P<10^{-16}$). Therefore, per-SNP associations were meta-analyzed using Egger regression (Egger et al. 1997), which is robust to the exclusion restriction (Bowden et al. 2017). The Egger intercept did not differ from zero (intercept=0.001, $P$=0.69). The Egger slope of 0.07 ($P=4\times10^{-4}$) provided strong evidence of a directional association between IOP and ML-based VCDR. In a reversed analysis, regarding ML-based VCDR as the mediator and IOP as the outcome, the Egger slope was -0.03 ($P$=0.75), providing no significant evidence of association in the opposite direction.

# Polygenic VCDR Model

## Pruning and thresholding

Pruning and thresholding-based polygenic risk scores for VCDR were computed as the weighted sum of effect allele counts for independent genome-wide significant variants ($P\leq5\times10^{-8}$), where the weight of each variant was its estimated effect size from the GWAS (Chatterjee, Shi, and García-Closas 2016). To evaluate performance both within the UK Biobank and in the EPIC-Norfolk cohorts, index variants present in both cohorts were used in PRS creation, resulting in 58 of the 76 published variants from Craig *et al.* GWAS and 282 of the 299 index variants from the ML-based GWAS. The UK Biobank evaluation set consisted of adjudicated expert-annotated VCDR measurements in 2,076 individuals of European ancestry. The EPIC-Norfolk evaluation set consisted of scanning laser ophthalmoscopy (HRT)-measured VCDRs in 5,868 individuals.

## Elastic net

Elastic net-based polygenic risk scores for VCDR were trained using the ML-predicted VCDR as the target label in 62,969 individuals using scikit-learn (Pedregosa et al. 2011). The Craig *et al.* model used 76 variants (the 58 described in the pruning and thresholding section above, plus 18 proxy variants present in both UK Biobank and EPIC-Norfolk that were in highest linkage disequilibrium ($R^2\geq0.6$) with the 18 dropped Craig *et al.* variants) and the ML-based model used the same 282 variants as described above. Each model was trained with 5-fold cross-validation and L1-penalty ratios of [0.1, 0.5, 0.7, 0.9, 0.95, 0.99, 1.0]. Model evaluation was performed in the same evaluation sets as described above. Both the UK Biobank and EPIC-Norfolk test sets were scored using the `plink --score` command and the correlations were computed using the scores in the resulting `*.profile` files.

## Permutation *P*-values

A permutation test was applied to assess whether a polygenic risk score (PRS) trained using summary statistics from the ML-based GWAS significantly outperformed a PRS trained using summary statistics from the Craig *et al.* GWAS for predicting VCDR in the UK Biobank and EPIC-Norfolk cohorts. Phenotypic predictions were generated from both PRS. The test statistic was the difference in Pearson correlations between the observed and predicted phenotypes, comparing ML-based with Craig *et al.* A value exceeding zero indicates better performance by the ML-based PRS. Under the null hypothesis, the predictions from both PRS are exchangeable. To obtain a realization from the null distribution, for each subject, the predictions of the ML-based and Craig *et al.* PRS were randomly swapped, and the difference in correlations was recalculated. This procedure was repeated $10^5$ times to obtain the null distribution. The one-sided $P$ value is given by the proportion of realizations from the null distribution that were as or more extreme than the observed difference in correlations.

# Glaucoma Association

## Mediation Analysis

A mediation analysis was performed to estimate the association between ML-derived VCDR and glaucoma, as assessed by Gharakhani *et al* (Gharahkhani et al. 2020). MR is a special case of mediation analysis in which the SNPs have no direct effect on the outcome; that is, the effect of genotype on the phenotype passes entirely through the mediator. Our mediation analysis differs from MR in that, due to limited availability of summary statistics from Gharakhani *et al*, the SNP set was defined based on association with the mediator (ML-based VCDR) rather than the outcome (glaucoma). Among the 118 independent, significant glaucoma SNPs identified by Gharakhani *et al*, 116 remained after harmonizing with the VCDR summary statistics available from our study. As expected, Cochran's Q test provided strong evidence of pleiotropy ($P<10^{-16}$), and the Egger intercept of 0.04 ($P=7\times10^{-7}$) suggested that variants with tended to increase VCDR also tended to increase the odds of glaucoma via an alternative pathway. The Egger slope was 5.7 ($P=3\times10^{-3}$; Supplementary Figure 7), which is interpreted as a log odds ratio, provides substantial evidence that increased VCDR was associated with increased glaucoma odds. This estimate of the association between VCDR and glaucoma remains valid, despite the presence of pleiotropy, since Egger regression is robust to the exclusion restriction (Bowden et al. 2017). In a reversed analysis using the same set of candidate SNPs, but regarding glaucoma as the mediator and VCDR as the outcome, the Egger intercept was 0.00 ($P$=0.09), and the Egger slope was 0.02 ($P$=0.07), providing no strong evidence of association in the opposite direction.

## Glaucoma liability conditional analysis

One of the main advantages of the ML-based model is that we can apply our ML-based model to different phenotypes without additional cost. We computed the glaucoma liability (ML-based glaucoma) for the same set of individuals in UK Biobank for whom we had calculated VCDR as described above. We performed GWAS on glaucoma liability (logit scale of glaucoma probability), using BOLT-LMM, conditional on ML-based VCDR and all covariates included in the ML-based VCDR GWAS. The LD score regression intercept was 1.00 (SE=0.001), with a SNP-heritability of 0.06 (0.01). Moreover, QQ-plot is depicted in Supplementary Figure 9.

## UK Biobank glaucoma phenotype

UK Biobank participants who underwent an ophthalmic examination also completed an ophthalmic touchscreen questionnaire and were considered to have POAG if they responded "Glaucoma" to the question "Has a doctor told you that you have any of the following problems with your eyes?". Participants were also considered to have POAG if they had a recorded hospital episode statistic ICD 10 code for POAG (H40.1). Controls were defined as participants who underwent the ophthalmic touchscreen questionnaire but did not meet the criteria to be a case. Additionally, we excluded participants with an ICD 9/10 hospital episode statistic code for types of glaucoma types other than POAG (ICD 9: 365.*; ICD 10: H40.0, H40.2, H40.3, H40.4, H40.5, H40.6, H40.8, H40.9, H42.*), participants meeting the case criteria but reporting an age of glaucoma onset prior to 30 years, and participants reporting glaucoma laser treatment or eye surgery but not reporting glaucoma on the touchscreen questionnaire. Applying these criteria, there were 7,654 cases and 182,726 controls.

## EPIC-Norfolk cohort

The European Prospective Investigation into Cancer (EPIC) study is a pan-European prospective cohort study designed to investigate the etiology of major chronic diseases (Riboli and Kaaks 1997). EPIC-Norfolk, one of the UK arms of EPIC, recruited and examined 25,639 participants between 1993 and 1997 for the baseline examination (Day et al. 1999). Recruitment was via general practices in the city of Norwich and the surrounding small towns and rural areas, and methods have been described in detail previously (Hayat et al. 2014). Since virtually all residents in the UK are registered with a general practitioner through the National Health Service, general practice lists serve as population registers. Ophthalmic assessment formed part of the third health examination and this has been termed the EPIC-Norfolk Eye Study (Khawaja et al. 2013).

In total, 8,623 participants were seen for the Eye Study between 2004 and 2011. Ophthalmic examination included tonometry (Ocular Response Analyzer; Reichert, New York, USA; software V.3.01), optic disc photography (Nikon D80 camera; Nikon Corporation, Tokyo, Japan), scanning laser ophthalmoscopy (Heidelberg Retinal Tomograph 3; Heidelberg Engineering, Heidelberg, Germany) and nerve fiber layer assessment (GDx-VCC; Zeiss, Dublin, California, USA). Participants meeting pre-defined criteria and an additional 1:10 participants underwent automated visual field testing (Humphrey 750i Visual Field Analyzer; Carl Zeiss Meditech Ltd, Welwyn Garden City, UK). 99.7% of EPIC-Norfolk are of European descent. The EPIC-Norfolk Eye Study was carried out following the principles of the Declaration of Helsinki and the Research Governance Framework for Health and Social Care. The study was approved by the Norfolk Local Research Ethics Committee (05/Q0101/191) and East Norfolk & Waveney NHS Research Governance Committee (2005EC07L). All participants gave written, informed consent.

Ascertainment of POAG in the EPIC Norfolk third health examination has been described previously (Chan et al. 2017). In brief, participants with study results suspicious of glaucoma (using pre-defined criteria) were referred for further examination by a glaucoma specialist at the regional University Hospital (Khawaja et al. 2013). Additionally, a diagnosis refinement process was undertaken by a second glaucoma specialist who independently reviewed the test results of all participants classified as glaucoma and a proportion of participants who were not classified as having glaucoma. POAG was defined as the presence of a glaucomatous optic disc together with either a corresponding visual field defect or otherwise unexplained non-specific visual field loss, open angles on gonioscopy, and absence of secondary causes of glaucoma. A glaucomatous disc was defined as one with focal or diffuse neuro-retinal rim thinning, and may possess, though not necessary for the definition, additional characteristic features such as bared circumlinear vessels, disc hemorrhages or nerve fiber layer defects. Pseudoexfoliative and pigmentary glaucoma were defined as secondary glaucoma in this study and therefore did not contribute to POAG cases. We defined controls as participants not meeting referral criteria for glaucoma on initial ophthalmic assessment and participants who attended the University Hospital for further examination and were not classified as having or being suspect for any type of glaucoma or ocular hypertension.

Initial genotyping on a small subset of EPIC-Norfolk was undertaken using the Affymetrix GeneChip Human Mapping 500K Array Set and 1,096 of these participants contributed to the IGGC meta-analysis (Springelkamp et al. 2017). Subsequently, the rest of the EPIC-Norfolk cohort were genotyped using the Affymetrix UK Biobank Axiom Array (the same array as used in UK Biobank); it is 5,868 of these participants (which includes no overlap with the 1,096 participants contributing to the IGGC meta-analysis) that contributed to the EPIC-Norfolk analyses in the current study. SNP exclusion criteria included: call rate < 95%, abnormal cluster pattern on visual inspection, plate batch effect evident by significant variation in minor allele frequency, and/or Hardy-Weinberg equilibrium $P < 10^{-7}$. Sample exclusion criteria included: DishQC < 0.82 (poor fluorescence signal contrast), sex discordance, sample call rate < 97%, heterozygosity outliers (calculated separately for SNPs with minor allele frequency >1% and <1%), rare allele count outlier, and impossible identity-by-descent values. We removed individuals with relatedness corresponding to third-degree relatives or closer across all genotyped participants. Following these exclusions, there were no ethnic outliers. Imputation was carried out using the HRC v1.

Quality control for HRT3 images included requiring a topography SD > 40 μm and checking of the manually drawn optic disc margin contours by an ophthalmologist (with redrawing if necessary). The mean HRT3 VCDR of right and left eyes was considered as the participant's VCDR if good quality scans were available for both eyes. If a good quality scan was only available for one eye, the VCDR value for that eye was considered for the participant.

# Tables

**Supplementary Table 1. Phenotype prediction model performance metrics.** For VCDR Pearson's correlation is reported. AUC, area under ROC curve; AUPRC, area under precision-recall curve, RMSE, root mean square error. The numbers in parentheses are 95% confidence intervals.

**Shared nomenclature for all GWAS results:**

CHR, chromosome; POS, base-pair variant position; EA, effect allele; NEA, non-effect allele; EAF, effect allele frequency; BETA, estimated effect size; SE, standard error; P, GWAS *P*-value; NUM_INDV, sample size for the variant; SRC, imputed or genotyped variant; INFO, imputation INFO score (set to 1 for genotyped variants); CRAIG, locus replicated in Craig *et al.* GWAS, CRAIG_META, locus replicated in Craig *et al.* meta-analysis; GENE_CONTEXT: genomic context of the variant, as explained below.

- Overlapping gene(s)
    - `[A]`: variant overlaps gene A
    - `[A,B]`: variant overlaps genes A and B
- Downstream genes
    - `[]A`: variant position is $0 < p \leq 10^3$ bp upstream of closest downstream gene A
    - `[]-A`: variant position is $10^3 < p \leq 10^4$ bp upstream of closest downstream gene A
    - `[]--A`: variant position is $10^4 < p \leq 10^5$ bp upstream of closest downstream gene A
    - `[]---A`: variant position is $10^5 < p \leq 10^6$ bp upstream of closest downstream gene A
    - `[]`: closest downstream gene is further than $10^6$ bp
- Upstream genes
    - The notation for upstream genes is similar, but gene A is on the left side, e.g., `B-[]` means variant position is $10^3 < p \leq 10^4$ bp downstream of closest gene B

For example, `FOXD2--[]---TRABD2B` indicates the variant is $10^4 < p \leq 10^5$ downstream of *FOXD2* and $10^5 < p \leq 10^6$ upstream of *TRABD2B*.

**Supplementary Table 2.** ML-based VCDR GWAS independent GWS hits ($R^2 \leq 0.1$, $P \leq 5\times10^{-8}$).

**Supplementary Table 3.** ML-based VCDR GWAS independent GWS loci ($R^2 \leq 0.1$, $P \leq 5\times10^{-8}$, distance between top hits > 250k).

**Supplementary Table 4.** SuSiE per-SNP results for all fine-mapped SNPs. PIP is the posterior probability the SNP is causal, higher being more likely; and LOCUS_IDX is a locus identifying index (as defined in Supplementary Table 3). SNPs with PIP = 0 are not shown. Not included due to large size; available upon request.

**Supplementary Table 5.** SuSiE results summarized per-locus. N_FINEMAPPED is the number of SNPs in loci with PIPs available. N_GWS is the number of genome-wide significant SNPs with MAF>0.05. N_CAUSAL is the sum of PIPs across SNPs in the locus. The estimated number of causal SNPs for a locus is min(N_GWS, N_CAUSAL).

**Supplementary Table 6.** ML-based + IGGC VCDR meta-analysis independent GWS hits ($R^2 \leq 0.1$, $P \leq 5\times10^{-8}$).

**Supplementary Table 7.** ML-based + IGGC VCDR meta-analysis independent GWS loci ($R^2 \leq 0.1$, $P \leq 5\times10^{-8}$, distance between top hits > 250k).

**Supplementary Table 8. All GREAT ontology terms significant for at least one of the two sets of loci.** All terms in the ontologies of Supplementary Figure 7 were tested. Abbreviations: ML *P-val*, the Bonferroni-corrected *P* value for the region-based test with the ML-based GWAS loci; Craig *P-val*, the Bonferroni-corrected *P* value for the region-based test with the Craig *et al.* GWAS loci; GOBP, Gene Ontology Biological Process; MP1KO, Mouse Phenotype Single Knockout; HP, Human Phenotype.

**Supplementary Table 9.** ML-based glaucoma risk conditioned on ML-based VCDR independent GWS hits ($R^2 \leq 0.1$, $P \leq 5 \times 10^{-8}$).

**Supplementary Table 10.** ML-based glaucoma risk conditioned on ML-based VCDR independent GWS loci ($R^2 \leq 0.1$, $P \leq 5 \times 10^{-8}$, distance between top hits > 250k).

# Figures

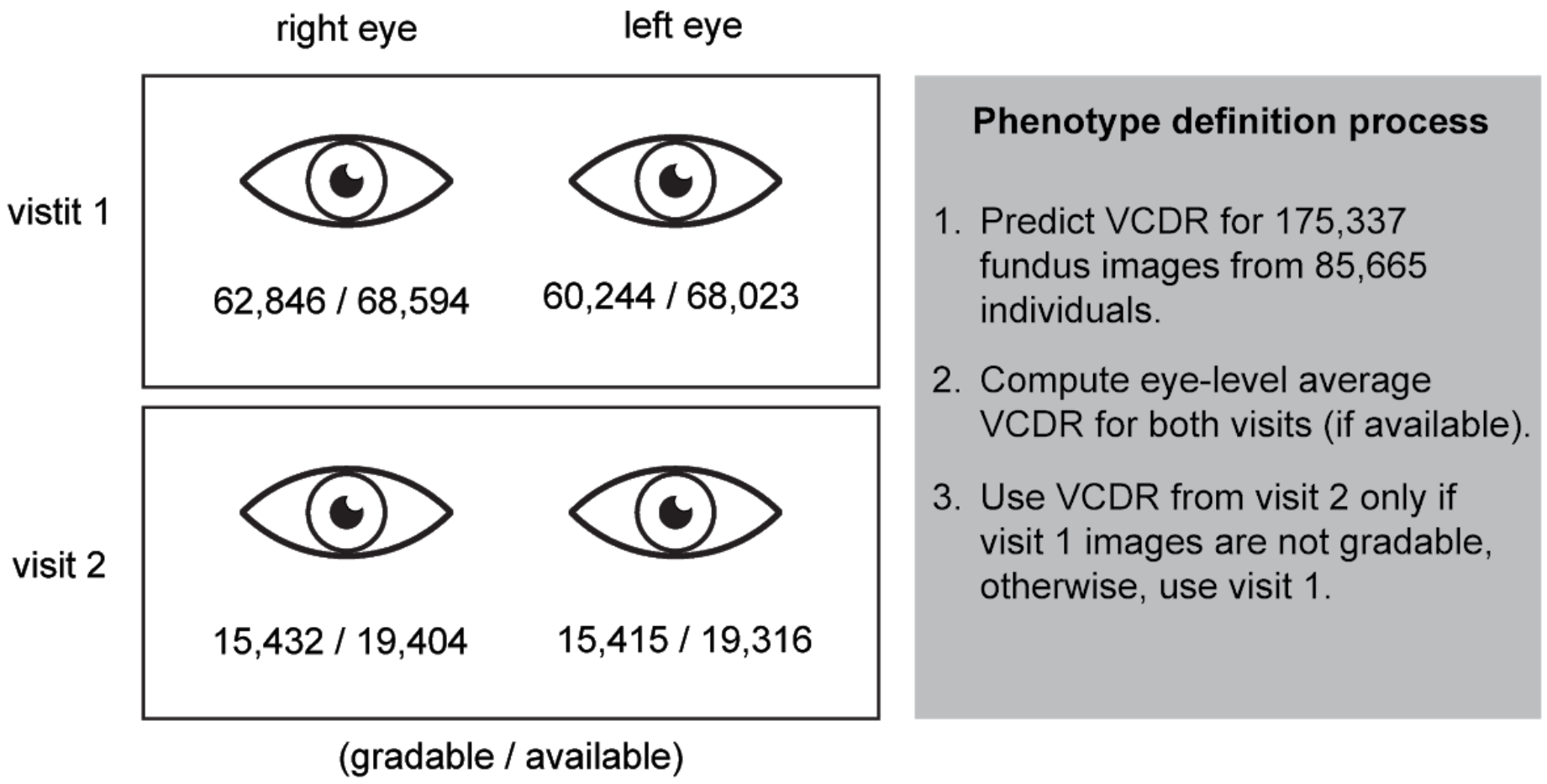

**Supplementary Figure 1. VCDR phenotype calling process.** The numbers below each eye indicate gradable / available images.

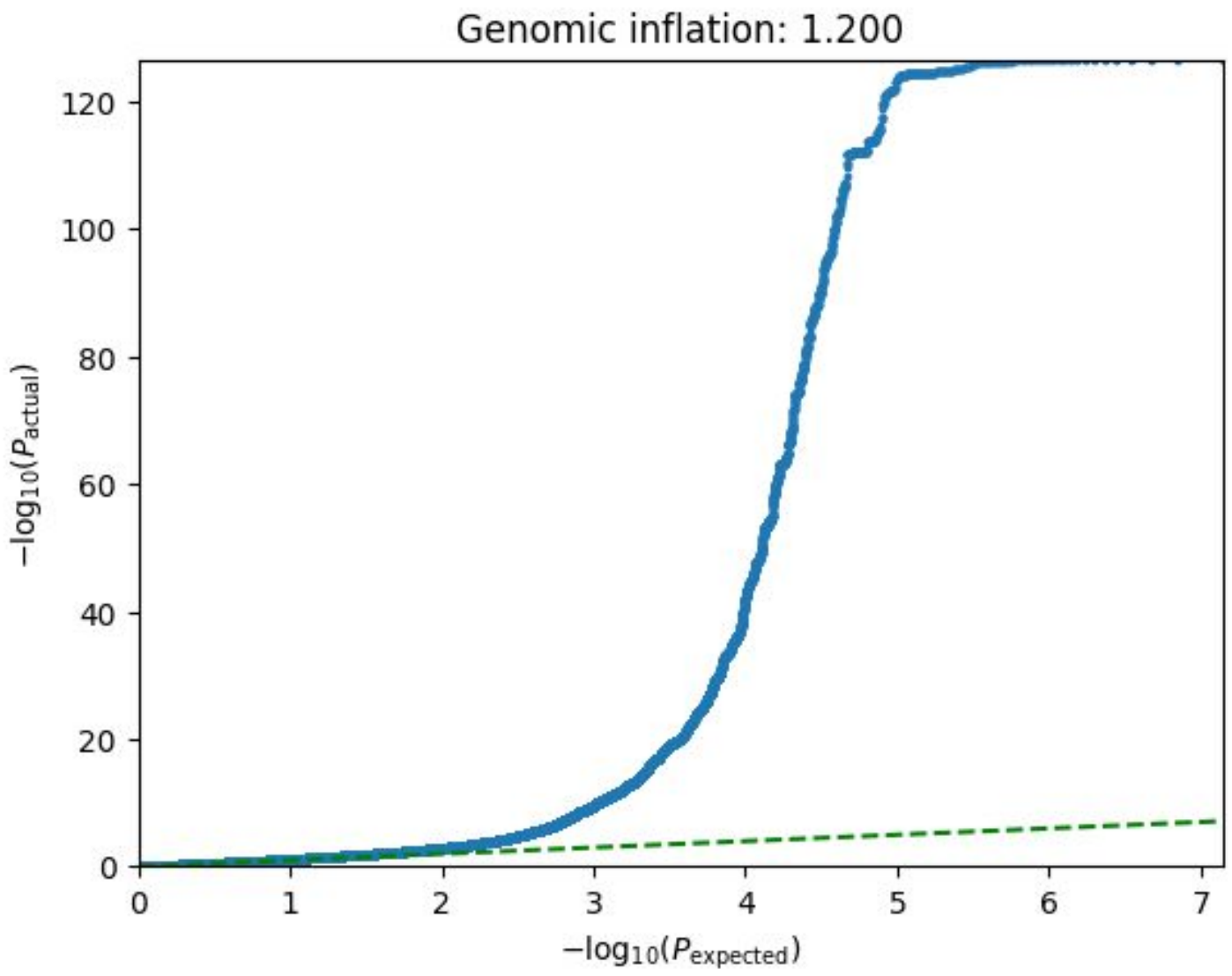

**Supplementary Figure 2. QQ-plot for the ML-based VCDR GWAS.** The expected *P* values are based on a uniform distribution.

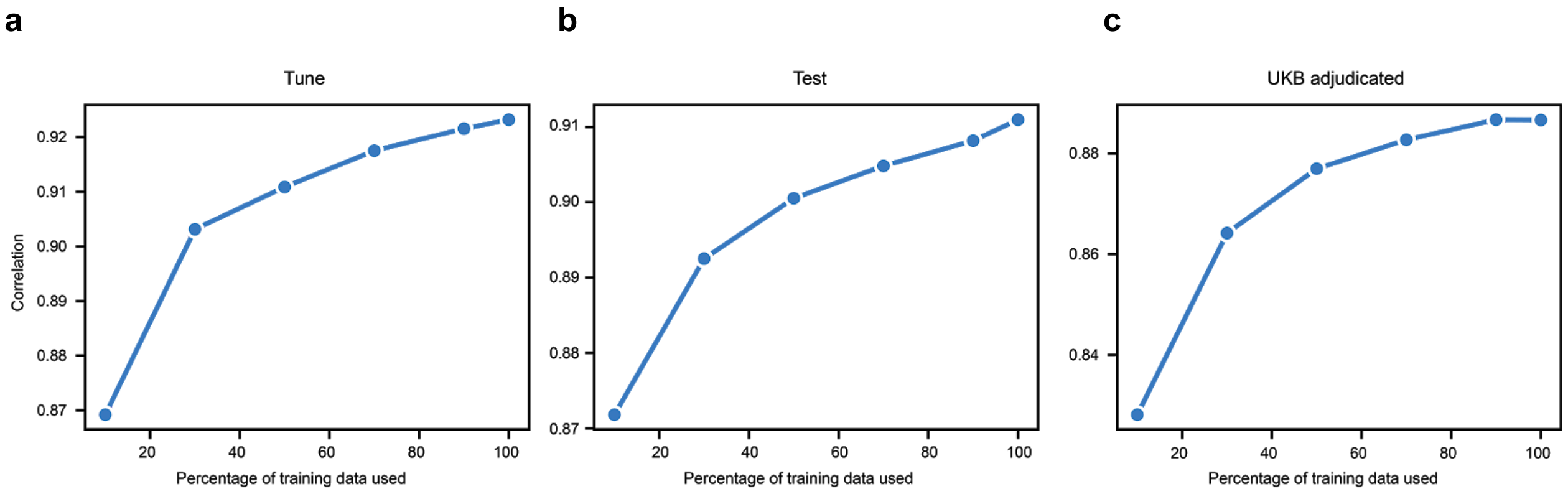

**Supplementary Figure 3. VCDR model performance as a function of the percentage of training data samples used to train the model.** Pearson's correlation between the model-predicted VCDR and the expert-labeled VCDR at training data percentages from 10% to 100% for **a,** the tune dataset, **b,** the test dataset, and **c,** the UKB adjudicated dataset. See **Model Training and Evaluation** section for detailed dataset definitions.

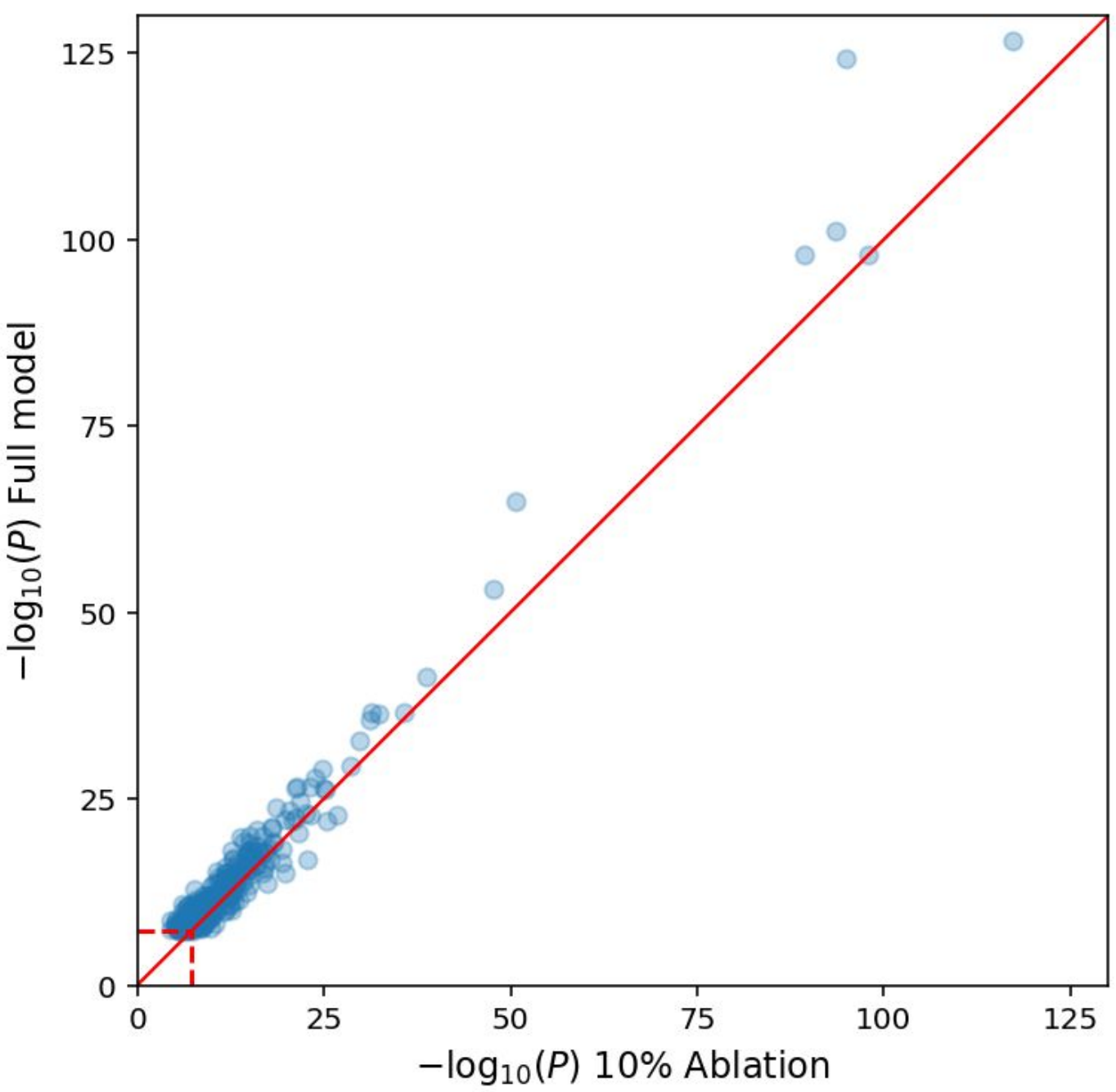

**Supplementary Figure 4. Comparison of the Full model *P* values with the 10% Ablation *P* values for 299 Full model hits.** The dashed red horizontal and vertical lines indicate the GWS level ($P<5\times10^{-8}$).

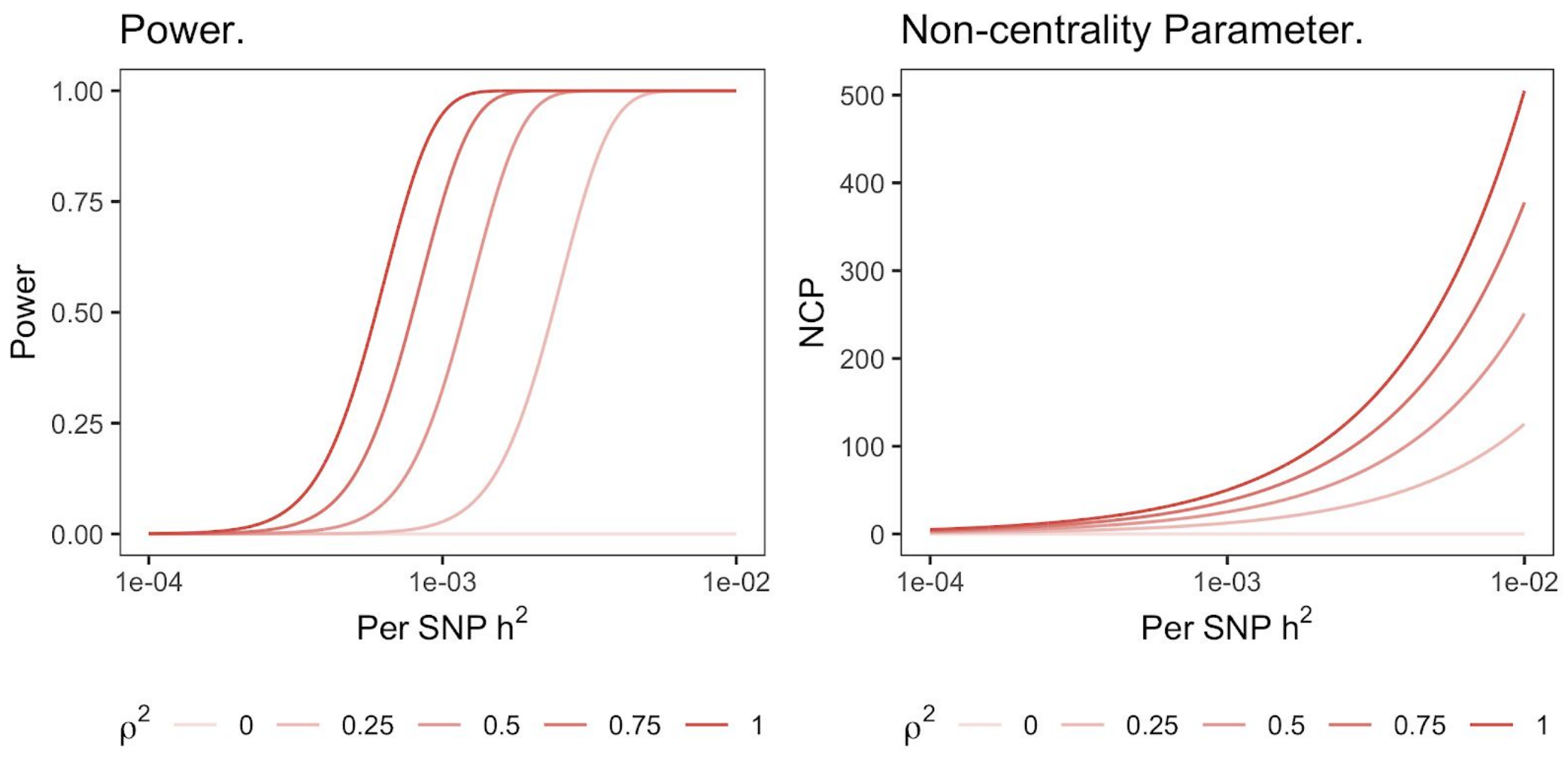

**Supplementary Figure 5. Power and non-centrality curves as a function of per-SNP heritability.** The curves are stratified by the correlation between the mismeasured and true phenotypes. Sample size was set to *n*=50,000; changing the sample sizes amounts to horizontally shifting the power curves. The range of per-SNP heritabilities was selected to demonstrate the inflection points of the power curves.

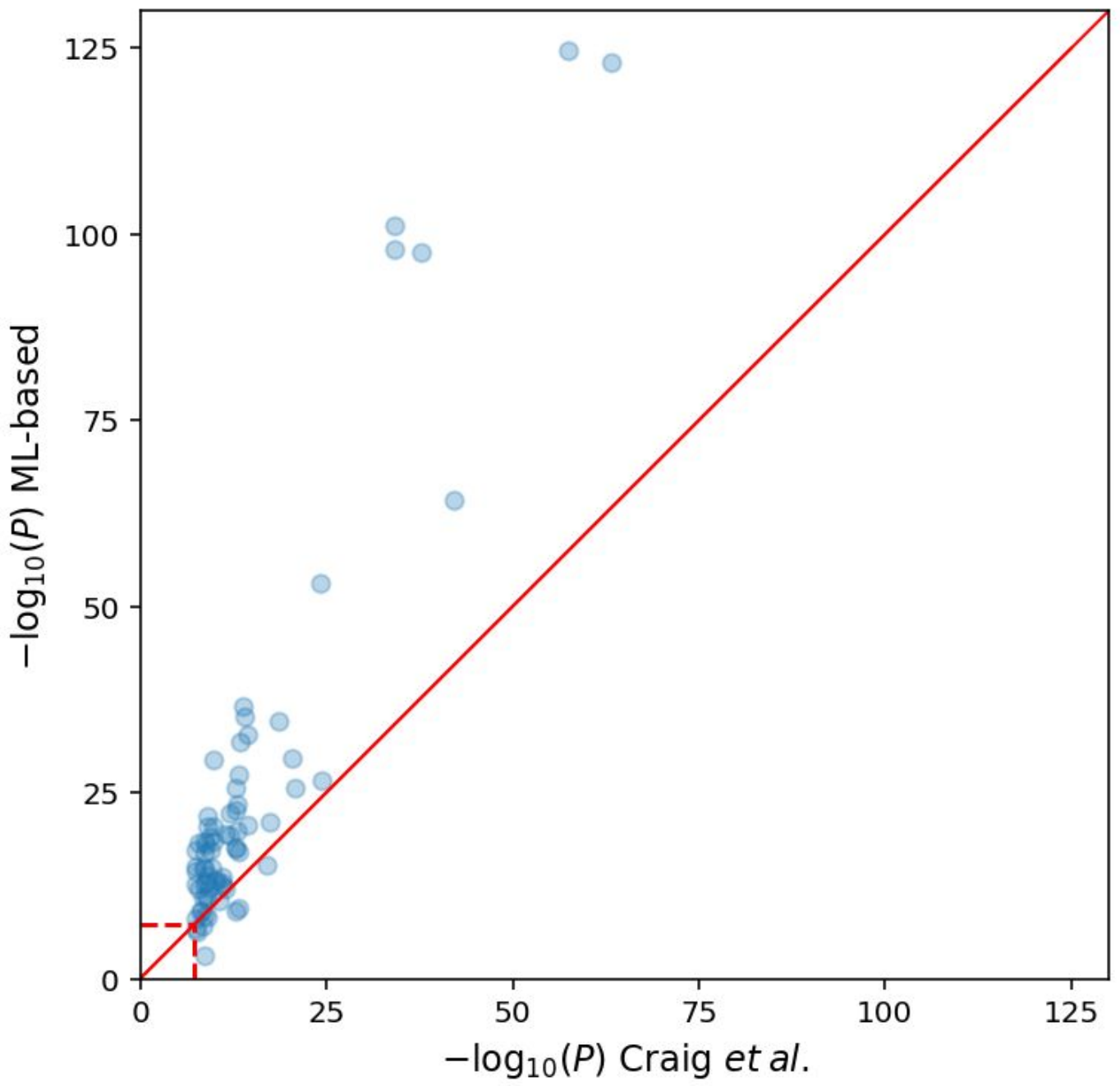

**Supplementary Figure 6. Comparison of ML-based VCDR *P* values with the Craig *et al.* *P* values for 73 Craig *et al.* hits.** The dashed red horizontal and vertical lines indicate the GWS level ($P<5\times10^{-8}$).

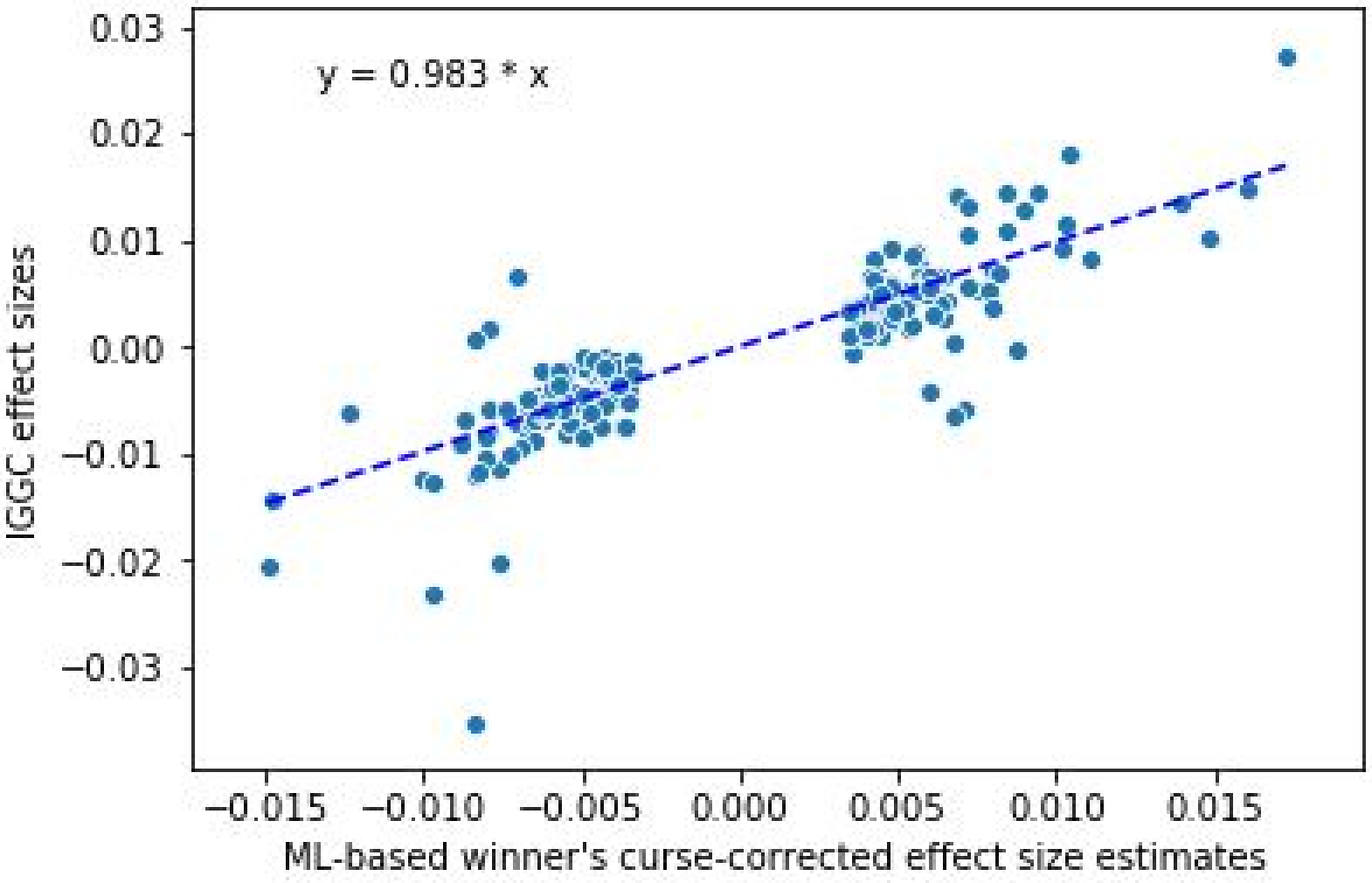

**Supplementary Figure 7. Regression of IGGC VCDR meta-analysis effect sizes on winner's curse-corrected ML-based VCDR effect size estimates.** Of 299 independent GWS hits, 214 were present in IGGC. Regression slope was computed with intercept fixed to zero. The dotted blue line shows the line of best fit.

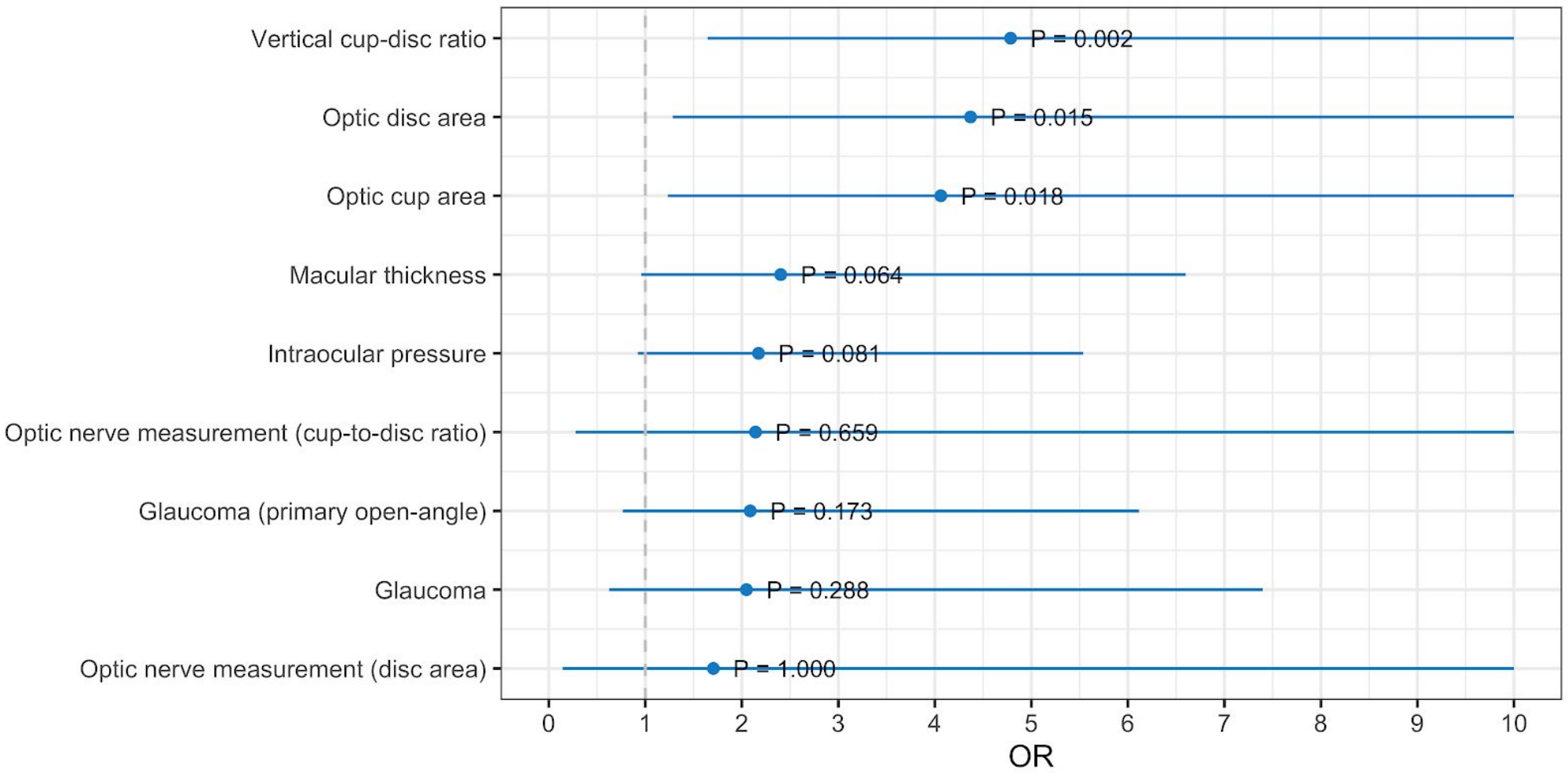

**Supplementary Figure 8. FUMA enrichment of eye-related gene sets from the ML-based VCDR GWAS versus the VCDR GWAS of Craig *et al.*** Enrichment is quantified via the odds ratio, with confidence interval and *P* value provided by Fisher's exact test.

a

b

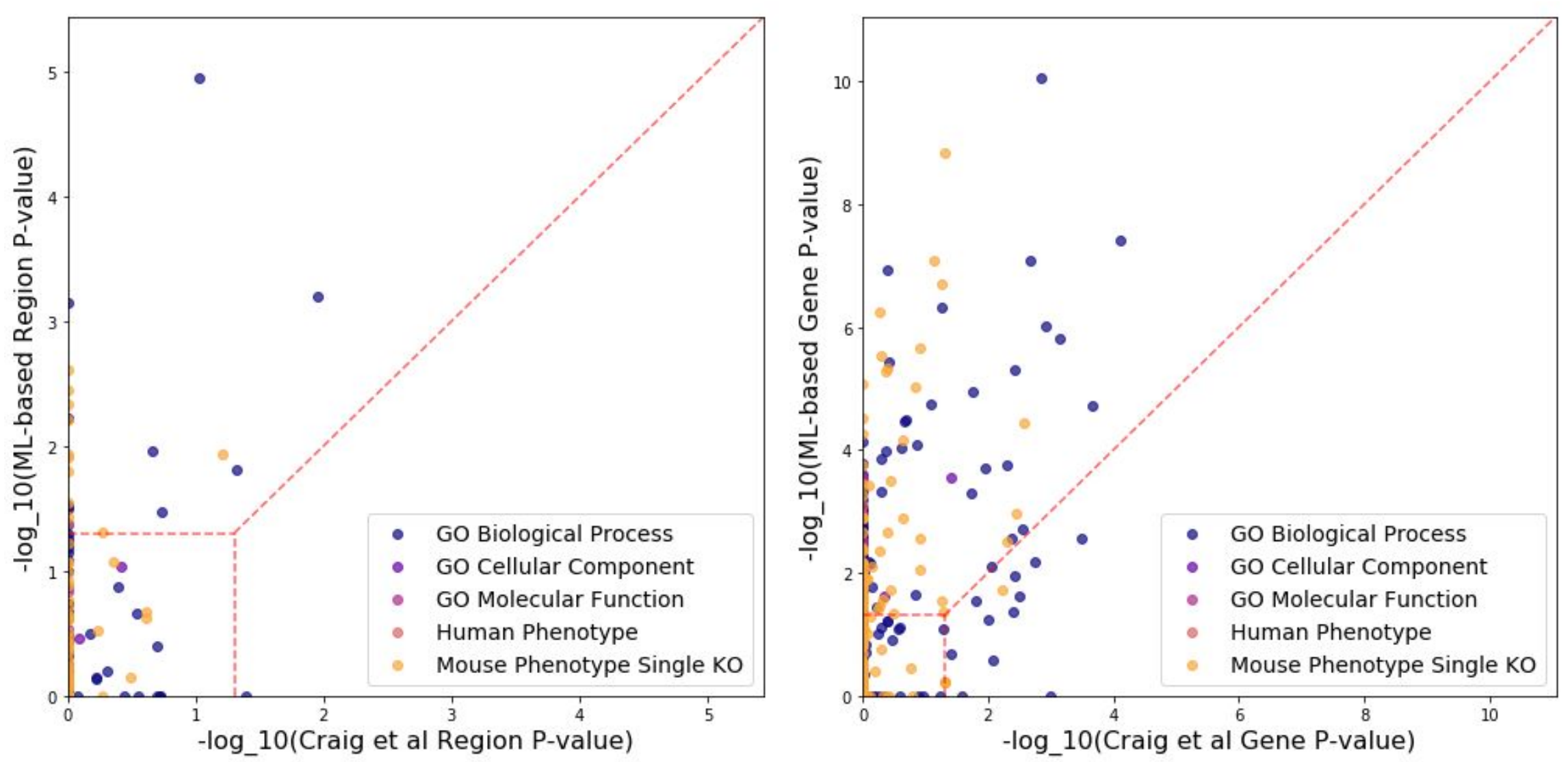

**Supplementary Figure 9. GREAT enrichment of loci from the ML-based VCDR GWAS vs the VCDR GWAS of Craig *et al.* a,** Comparison of Bonferroni-corrected *P* values for the region-based test reported by GREAT for the five listed ontologies. Dashed horizontal and vertical lines show the threshold for statistical significance at a Bonferroni-corrected $P \leq 0.05$, and the diagonal line indicates y=x. More ontology terms are statistically significant for the ML-derived GWAS than Craig *et al.* **b,** Comparison analogous to that in **a**, for the gene-based test.

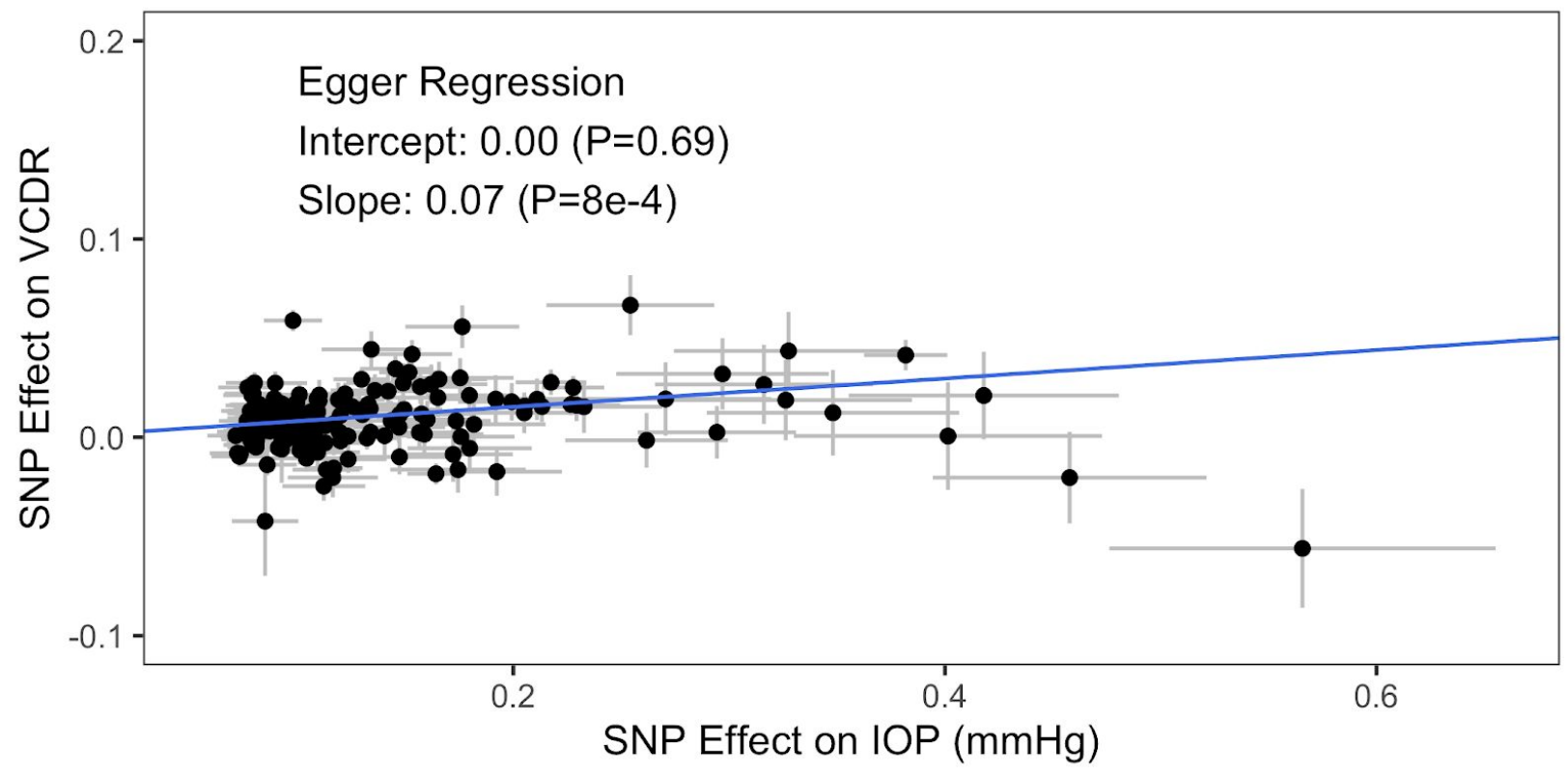

**Supplementary Figure 10. Egger Regression for Mendelian Randomization of the effect of IOP on ML-based VCDR.** Independent, significant IOP-associated SNPs were ascertained from Khawaja *et al.* and harmonized with GWAS results for ML-based VCDR.

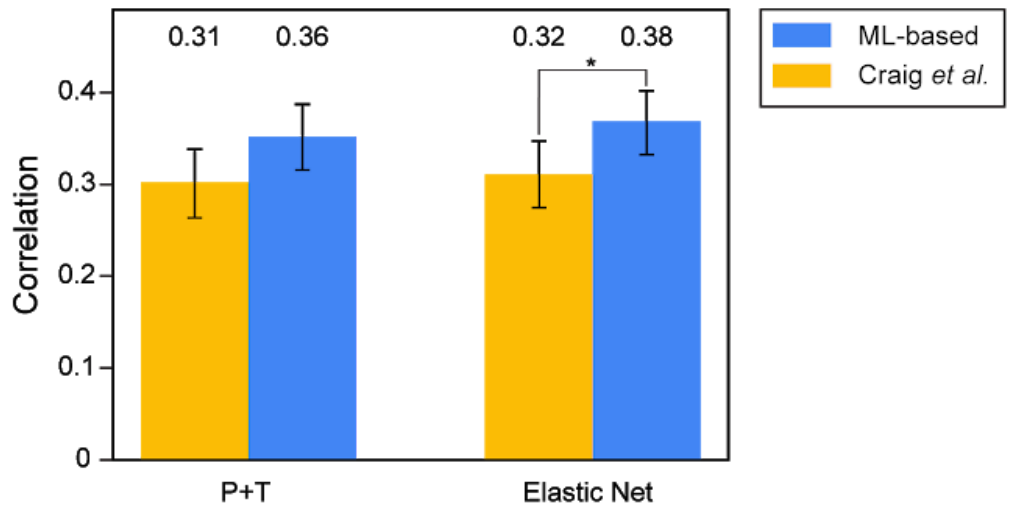

**Supplementary Figure 11. VCDR polygenic risk score performance metrics on the UKB imputation panel.** Pearson's correlations between measured VCDR values and predictions of the pruning and thresholding (P+T) and the Elastic Net models are shown for the PRS learned from ML-based and Craig *et al.* hits. Error bars depict 95% confidence intervals. Numbers above bars are the observed Pearson's correlations. Indications of P value ranges: * $P \leq 0.05$, ** $P \leq 0.01$, *** $P \leq 0.001$. Measured VCDR values were obtained from adjudicated expert labeling of CFPs (UKB, n=2,076).

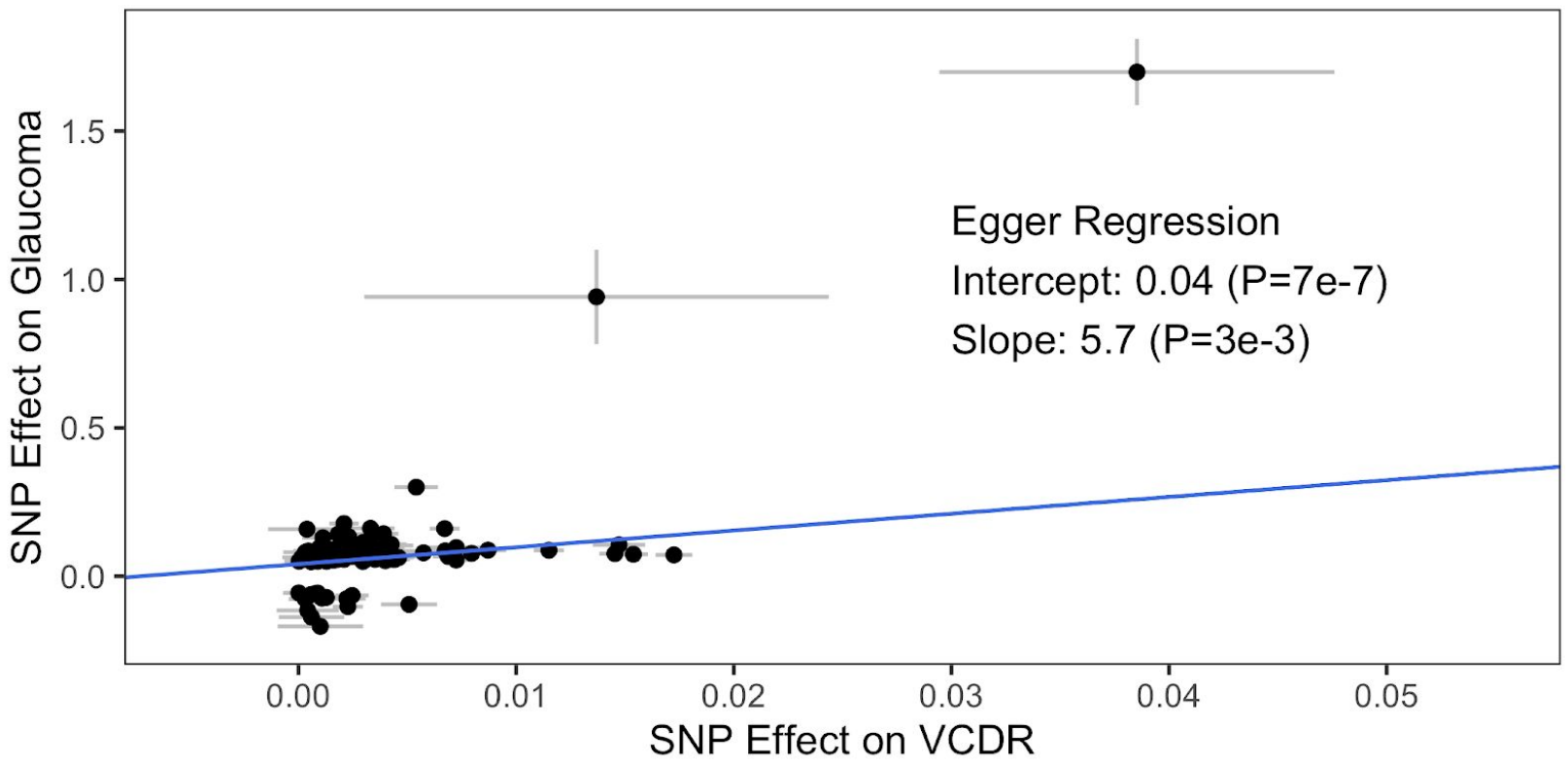

**Supplementary Figure 12. Egger Regression for Mendelian Randomization of the Effect of VCDR on Glaucoma log Odds.** Independent, significant Glaucoma risk SNPs were ascertained from Gharahkhani *et al.* and harmonized with GWAS results from ML-based VCDR.

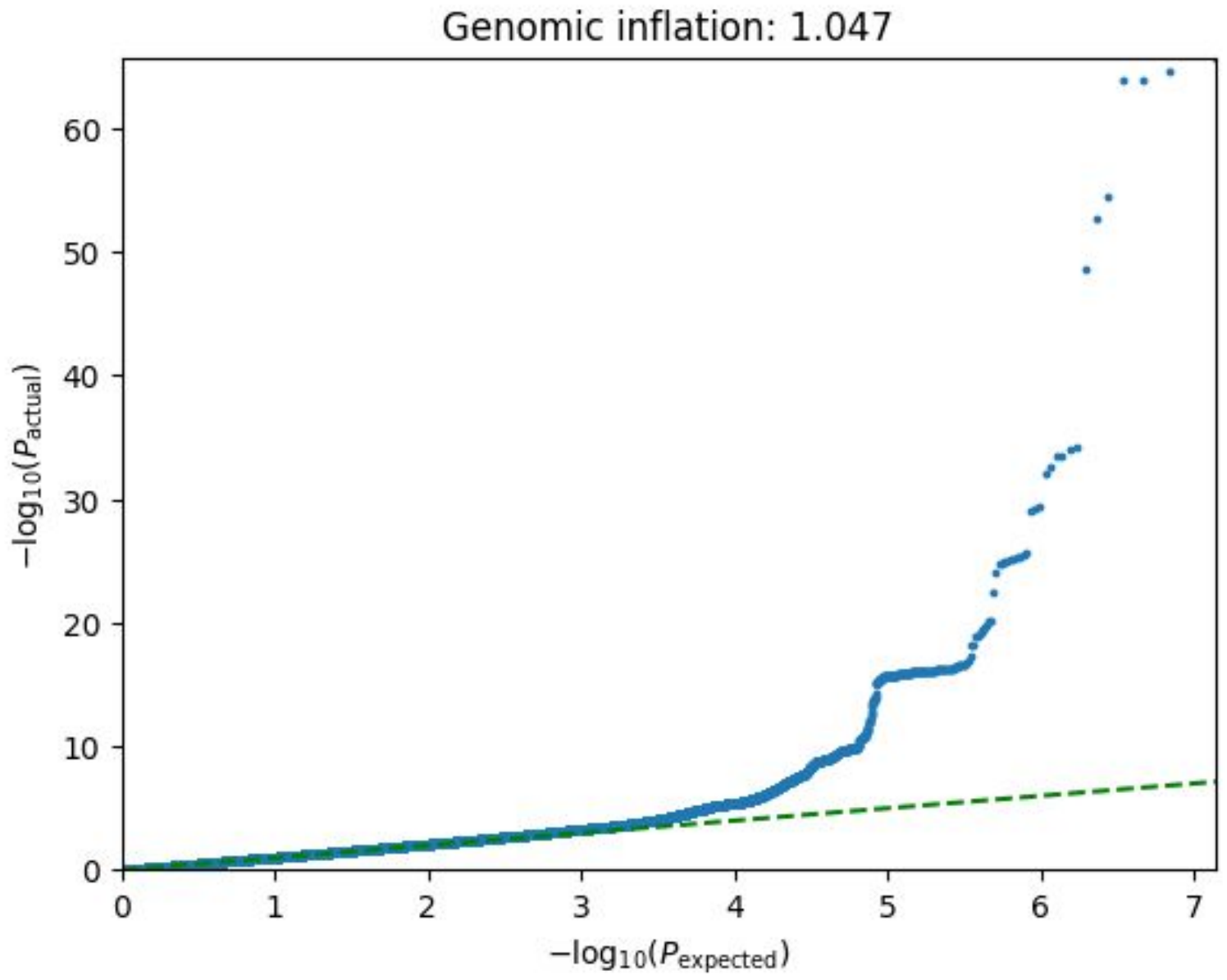

**Supplementary Figure 13. QQ-plot for the ML-based glaucoma liability GWAS conditional on ML-based VCDR.** The expected *P* values are based on a uniform distribution.

## Model hyper-parameters

The hyper-parameters used very closely follow (Krause et al. 2018; Phene et al. 2019):

- Inception V3 architecture
- Input image resolution: 587 x 587
- Learning rate: 0.001
- Batch size: 16
- Data augmentation:
    - Random horizontal reflections
    - Random vertical reflections
    - Random brightness changes (with a max delta of 0.1147528) [TensorFlow function `tf.image.random_brightness`]
    - Random saturation changes between 0.5597273 and 1.2748845 [`tf.image.random_saturation`]
    - Random hue changes (with a max delta of 0.0251488) [`tf.image.random_hue`]
    - Random contrast changes between 0.9996807 and 1.7704824 [`tf.image.random_constrast`]

Supplementary Information - Table 1. Model performance metrics

| Dataset | Gradable | Graders | VCDR (RMSE) | VCDR (correlation) | Glaucoma risk (AUC) | Glaucoma risk (AUPRC) | Glaucoma risk prevalence |
|---|---|---|---|---|---|---|---|
| Train | 69,460 | 1-2 | 0.106 (0.105-0.106) | 0.803 (0.800–0.806) | 0.896 (0.894–0.899) | 0.734 (0.727–0.741) | 22.6% |
| Tune | 1,399 | 3 | 0.073 (0.069-0.077) | 0.923 (0.915–0.931) | 0.964 (0.952–0.975) | 0.820 (0.768–0.866) | 11.5% |
| Test | 1,076 | 3 | 0.079 (0.074-0.085) | 0.911 (0.899–0.922) | 0.957 (0.944–0.969) | 0.860 (0.817–0.898) | 18.1% |
| UKB Adjudicated | 2,115 | 2-3 | 0.092 (0.088-0.096) | 0.886 (0.876–0.896) | 0.872 (0.854–0.890) | 0.506 (0.446–0.567) | 13.3% |

Supplementary Information - Table 2. ML-based VCDR (hits)

| CHR | POS | SNP | EA | NEA | EAF | BETA | SE | P | NUM_INDV | SRC | INFO | GENE_CONTEXT |
|---|---|---|---|---|---|---|---|---|---|---|---|---|
| 1 | 3049362 | rs12024620 | C | T | 0.936 | -1.55E-02 | 1.38E-03 | 4.00E-30 | 65680 | Imputed | 0.995 | [PRDM16] |
| 1 | 3055876 | rs10797380 | A | G | 0.551 | -4.06E-03 | 6.99E-04 | 8.30E-10 | 65680 | Imputed | 0.936 | [PRDM16] |
| 1 | 8468278 | rs301792 | A | G | 0.662 | -4.62E-03 | 7.13E-04 | 6.80E-13 | 65680 | Imputed | 0.998 | [RERE] |
| 1 | 12614029 | rs6541032 | T | C | 0.424 | -5.11E-03 | 6.82E-04 | 2.60E-14 | 65680 | Imputed | 0.999 | VPS13D--[]--DHRS3 |
| 1 | 47923058 | rs767682581 | C | CT | 0.374 | -4.27E-03 | 7.07E-04 | 1.40E-10 | 65680 | Imputed | 0.983 | FOXD2--[]---TRABD2B |
| 1 | 68773910 | rs34151819 | C | T | 0.983 | 1.78E-02 | 2.61E-03 | 4.50E-13 | 65654 | Genotyped | 1 | WLS--[]---RPE65 |
| 1 | 68840797 | rs2209559 | A | G | 0.601 | -7.49E-03 | 6.91E-04 | 1.90E-28 | 65484 | Genotyped | 1 | WLS---[]--RPE65 |
| 1 | 89253357 | rs786908 | A | G | 0.383 | -4.31E-03 | 6.95E-04 | 3.90E-10 | 65680 | Imputed | 0.997 | [PKN2] |
| 1 | 92016934 | rs61798047 | C | T | 0.486 | 4.77E-03 | 6.79E-04 | 6.50E-15 | 65680 | Imputed | 0.994 | CDC7--[]---TGFBR3 |
| 1 | 92024124 | rs75296423 | T | G | 0.98 | -1.78E-02 | 2.49E-03 | 5.20E-14 | 65680 | Imputed | 0.942 | CDC7--[]---TGFBR3 |
| 1 | 92030967 | rs7536573 | C | A | 0.675 | 5.05E-03 | 7.28E-04 | 2.30E-15 | 65680 | Imputed | 0.984 | CDC7--[]---TGFBR3 |
| 1 | 92070810 | rs77291384 | G | A | 0.986 | -1.76E-02 | 3.04E-03 | 4.10E-09 | 65680 | Imputed | 0.897 | CDC7--[]--TGFBR3 |
| 1 | 92077097 | rs1192415 | G | A | 0.188 | 1.76E-02 | 8.63E-04 | 7.60E-102 | 65610 | Genotyped | 1 | CDC7--[]--TGFBR3 |
| 1 | 92089160 | rs17569923 | G | A | 0.799 | 7.09E-03 | 8.52E-04 | 1.70E-18 | 65680 | Imputed | 0.981 | CDC7--[]--TGFBR3 |
| 1 | 92114938 | rs12046642 | A | G | 0.82 | -7.17E-03 | 8.81E-04 | 4.00E-16 | 65680 | Imputed | 0.99 | CDC7---[]--TGFBR3 |
| 1 | 110627923 | rs10857812 | T | A | 0.636 | 4.63E-03 | 7.03E-04 | 2.10E-11 | 65680 | Imputed | 0.993 | STRIP1--[]--UBL4B |
| 1 | 113045061 | rs351364 | A | T | 0.252 | -3.95E-03 | 7.81E-04 | 2.40E-08 | 65680 | Imputed | 0.984 | [WNT2B] |
| 1 | 155033308 | rs11589479 | G | A | 0.837 | 4.88E-03 | 9.17E-04 | 1.20E-08 | 65404 | Genotyped | 1 | [ADAM15] |
| 1 | 169551682 | rs6028 | T | C | 0.709 | -4.03E-03 | 7.44E-04 | 3.10E-09 | 65680 | Imputed | 0.996 | [F5] |
| 1 | 183849739 | rs41263652 | G | C | 0.896 | 7.20E-03 | 1.12E-03 | 4.20E-11 | 65680 | Imputed | 0.971 | [RGL1] |
| 1 | 218520995 | rs6658835 | A | G | 0.73 | -5.97E-03 | 7.64E-04 | 7.60E-16 | 65680 | Imputed | 0.992 | [TGFB2] |
| 1 | 219573841 | rs796959510 | C | CT | 0.491 | -4.02E-03 | 6.82E-04 | 7.40E-09 | 65680 | Imputed | 0.979 | AL360093.1---[]---ZC3H11B |
| 1 | 222014897 | rs11118873 | A | G | 0.485 | 3.74E-03 | 6.79E-04 | 4.20E-10 | 65680 | Imputed | 0.988 | DUSP10--[]---HHIPL2 |
| 1 | 227585983 | rs6670351 | G | A | 0.798 | -6.66E-03 | 8.42E-04 | 3.30E-17 | 65680 | Imputed | 0.993 | CDC42BPA--[]---ZNF678 |
| 2 | 5680539 | rs7575439 | C | A | 0.358 | 4.09E-03 | 7.27E-04 | 1.50E-08 | 62872 | Genotyped | 1 | LINC01249---[]--AC108025.1 |
| 2 | 12891476 | rs730126 | A | C | 0.587 | 3.98E-03 | 6.95E-04 | 9.50E-10 | 65680 | Imputed | 0.985 | TRIB2-[] |
| 2 | 19307666 | rs76455252 | C | G | 0.926 | -7.80E-03 | 1.30E-03 | 1.10E-09 | 65680 | Imputed | 0.99 | NT5C1B---[]---OSR1 |
| 2 | 19431423 | rs72778352 | C | T | 0.987 | 2.71E-02 | 2.97E-03 | 4.80E-22 | 65680 | Imputed | 0.982 | NT5C1B---[]---OSR1 |
| 2 | 19472235 | rs851359 | T | C | 0.443 | 5.55E-03 | 6.81E-04 | 8.00E-17 | 65680 | Imputed | 0.998 | NT5C1B---[]--OSR1 |
| 2 | 56072501 | rs1430202 | G | A | 0.798 | 8.29E-03 | 8.51E-04 | 1.80E-24 | 65680 | Imputed | 0.99 | PNPT1---[]--EFEMP1 |
| 2 | 56102744 | rs11899888 | A | G | 0.841 | -5.44E-03 | 9.34E-04 | 9.30E-10 | 65680 | Imputed | 0.987 | [EFEMP1] |
| 2 | 56210465 | rs11303332 | G | C | 0.786 | -4.83E-03 | 8.51E-04 | 6.90E-09 | 65680 | Imputed | 0.949 | EFEMP1--[]---CCDC85A |
| 2 | 111658010 | rs4849203 | A | G | 0.652 | 5.07E-03 | 7.12E-04 | 1.30E-13 | 65575 | Genotyped | 1 | [ACOXL] |
| 2 | 180196027 | rs12620141 | C | A | 0.636 | -4.06E-03 | 7.18E-04 | 2.50E-09 | 65680 | Imputed | 0.961 | AC093911.1--[]---ZNF385B |
| 2 | 190269957 | 2:190269957_CTTTT_C | CTTTT | C | 0.308 | 4.16E-03 | 7.40E-04 | 3.00E-09 | 65680 | Imputed | 0.99 | COL5A2---[]--WDR75 |
| 2 | 233389918 | rs2853447 | A | G | 0.297 | -3.75E-03 | 7.44E-04 | 3.80E-08 | 65680 | Imputed | 0.987 | [PRSS56] |
| 3 | 20061023 | rs35057657 | A | G | 0.327 | 4.06E-03 | 7.27E-04 | 4.30E-08 | 65680 | Imputed | 0.991 | PP2D1-[]--KAT2B |
| 3 | 25046463 | rs12490228 | C | T | 0.717 | -6.08E-03 | 7.54E-04 | 3.50E-17 | 65680 | Imputed | 0.996 | THRB-AS1---[]---RARB |
| 3 | 25187193 | rs73048443 | A | G | 0.822 | 7.10E-03 | 8.91E-04 | 1.00E-15 | 65680 | Imputed | 0.981 | THRB-AS1---[]--RARB |
| 3 | 25194626 | rs143822311 | A | AGTGT | 0.438 | -4.53E-03 | 7.01E-04 | 9.50E-11 | 65680 | Imputed | 0.956 | THRB-AS1---[]--RARB |
| 3 | 29493916 | rs9822629 | T | G | 0.626 | -3.71E-03 | 7.02E-04 | 1.50E-08 | 65680 | Imputed | 0.992 | [AC098650.1,RBMS3] |
| 3 | 32879823 | rs56131903 | A | T | 0.679 | 6.19E-03 | 7.34E-04 | 3.90E-18 | 65680 | Imputed | 0.983 | [TRIM71] |
| 3 | 48719638 | rs7633840 | T | C | 0.337 | 4.72E-03 | 7.39E-04 | 1.60E-11 | 65680 | Imputed | 0.949 | [NCKIPSD] |
| 3 | 57988028 | 3:57988028_AAAAT_A | AAAAT | A | 0.76 | -5.33E-03 | 7.94E-04 | 1.20E-11 | 65680 | Imputed | 0.998 | SLMAP--[]-FLNB |
| 3 | 58130168 | rs2362911 | A | G | 0.771 | -5.34E-03 | 8.06E-04 | 3.30E-12 | 65680 | Imputed | 0.991 | [FLNB] |
| 3 | 70061377 | rs190948281 | G | C | 0.997 | 4.43E-02 | 6.00E-03 | 1.60E-13 | 65680 | Imputed | 0.944 | MITF--[]---MDFIC2 |
| 3 | 71182447 | rs77877421 | A | T | 0.943 | -1.16E-02 | 1.50E-03 | 1.60E-15 | 65680 | Imputed | 0.964 | [FOXP1,AC097634.4] |
| 3 | 88380417 | rs9852080 | T | C | 0.448 | 6.19E-03 | 6.83E-04 | 3.70E-21 | 65680 | Imputed | 0.996 | C3orf38---[]-CSNKA2IP |
| 3 | 89603917 | rs373216501 | A | ATATT | 0.697 | 4.27E-03 | 7.72E-04 | 3.60E-08 | 65680 | Imputed | 0.91 | EPHA3--[] |
| 3 | 98943479 | rs13076500 | C | T | 0.461 | -6.73E-03 | 6.87E-04 | 9.90E-24 | 65680 | Imputed | 0.979 | DCBLD2---[]---AC107029.1 |
| 3 | 99078606 | rs1871794 | T | C | 0.748 | -8.53E-03 | 7.82E-04 | 9.40E-30 | 65680 | Imputed | 0.996 | DCBLD2---[]---AC107029.1 |
| 3 | 99313686 | 3:99313686_CAG_C | CAG | C | 0.953 | -1.11E-02 | 1.61E-03 | 1.70E-12 | 65680 | Imputed | 0.997 | AC107029.1--[]--COL8A1 |
| 3 | 99377999 | rs34578241 | T | C | 0.897 | 9.30E-03 | 1.11E-03 | 1.20E-18 | 65680 | Imputed | 0.999 | [COL8A1] |
| 3 | 99707371 | rs61144932 | A | AG | 0.322 | -3.95E-03 | 7.53E-04 | 9.00E-09 | 65680 | Imputed | 0.932 | [CMSS1,FILIP1L] |
| 3 | 100625703 | rs17398137 | G | A | 0.82 | 8.75E-03 | 8.90E-04 | 1.20E-23 | 65523 | Genotyped | 1 | [ABI3BP] |
| 3 | 106117209 | rs12637686 | G | A | 0.721 | -4.14E-03 | 7.57E-04 | 3.50E-09 | 65680 | Imputed | 0.993 | CBLB---[]--LINC00882 |
| 3 | 128196500 | rs2713594 | G | A | 0.583 | -3.68E-03 | 6.94E-04 | 4.20E-08 | 65680 | Imputed | 0.983 | DNAJB8--[]-GATA2 |
| 3 | 134089758 | rs143351962 | C | T | 0.99 | -2.15E-02 | 3.40E-03 | 7.40E-10 | 65680 | Imputed | 1 | [AMOTL2] |
| 4 | 7917204 | rs34939228 | T | TA | 0.617 | -4.17E-03 | 7.05E-04 | 8.30E-09 | 65680 | Imputed | 0.984 | [AFAP1] |
| 4 | 54979145 | rs1158402 | C | T | 0.378 | 5.88E-03 | 7.03E-04 | 6.80E-19 | 65680 | Imputed | 0.996 | [AC058822.1] |
| 4 | 55095682 | rs565335773 | G | GA | 0.792 | 5.84E-03 | 8.48E-04 | 9.80E-13 | 65680 | Imputed | 0.973 | [AC058822.1,PDGFRA] |
| 4 | 79122057 | rs372050829 | G | GTA | 0.775 | 4.46E-03 | 8.46E-04 | 2.40E-08 | 65680 | Imputed | 0.933 | [FRAS1] |
| 4 | 79396057 | 4:79396057_TC_T | TC | T | 0.367 | 4.02E-03 | 7.10E-04 | 2.00E-08 | 65680 | Imputed | 0.995 | [FRAS1] |
| 4 | 112399511 | rs2661764 | A | T | 0.638 | 4.03E-03 | 7.09E-04 | 7.30E-09 | 65680 | Imputed | 0.995 | PITX2---[]---FAM241A |
| 4 | 126239986 | rs1039808 | C | T | 0.547 | 4.23E-03 | 6.85E-04 | 5.90E-10 | 65550 | Genotyped | 1 | [FAT4] |
| 4 | 126407298 | rs532857051 | C | CTT | 0.708 | 6.06E-03 | 7.51E-04 | 4.70E-17 | 65680 | Imputed | 0.993 | [FAT4] |
| 4 | 128053375 | 4:128053375_AACAC_A | AACAC | A | 0.52 | 3.51E-03 | 6.96E-04 | 1.60E-08 | 65680 | Imputed | 0.953 | []---INTU |
| 5 | 3646121 | rs13165326 | C | T | 0.672 | -3.77E-03 | 7.25E-04 | 2.40E-08 | 65680 | Imputed | 0.991 | IRX1--[]---LINC02063 |

| CHR | POS | SNP | EA | NEA | EAF | BETA | SE | P | NUM_INDV | SRC | INFO | GENE_CONTEXT |
|---|---|---|---|---|---|---|---|---|---|---|---|---|
| 5 | 31952051 | rs72759609 | T | C | 0.898 | 1.35E-02 | 1.13E-03 | 2.40E-37 | 65680 | Imputed | 0.989 | [PDZD2] |
| 5 | 31966417 | rs10045838 | G | A | 0.585 | -3.68E-03 | 6.98E-04 | 1.80E-08 | 65680 | Imputed | 0.974 | [PDZD2] |
| 5 | 55578661 | rs158653 | G | A | 0.477 | 5.06E-03 | 6.85E-04 | 1.30E-14 | 65680 | Imputed | 0.992 | ANKRD55--[]---LINC01948 |
| 5 | 55701667 | rs140212185 | C | CTTTTTTT | 0.496 | -3.96E-03 | 6.86E-04 | 6.40E-10 | 65680 | Imputed | 0.983 | ANKRD55---[]--LINC01948 |
| 5 | 55744230 | rs30372 | T | C | 0.237 | -5.15E-03 | 8.07E-04 | 9.40E-12 | 65680 | Imputed | 0.978 | ANKRD55---[]-LINC01948 |
| 5 | 55781912 | rs12187324 | T | G | 0.918 | -7.70E-03 | 1.24E-03 | 8.60E-10 | 65680 | Imputed | 0.996 | LINC01948-[]--C5orf67 |
| 5 | 82770558 | rs11746859 | A | G | 0.539 | 4.77E-03 | 6.82E-04 | 8.90E-13 | 65680 | Imputed | 0.995 | [VCAN] |
| 5 | 87810199 | rs150221399 | G | A | 0.919 | -9.99E-03 | 1.28E-03 | 3.80E-15 | 65680 | Imputed | 0.95 | TMEM161B---[]---MEF2C |
| 5 | 121765728 | rs2570981 | T | C | 0.401 | 4.30E-03 | 6.93E-04 | 1.10E-09 | 65680 | Imputed | 0.994 | [SNCAIP] |
| 5 | 128931357 | rs7448395 | G | A | 0.204 | 6.62E-03 | 8.47E-04 | 8.00E-17 | 65680 | Imputed | 0.999 | [ADAMTS19] |
| 5 | 129100923 | rs75047204 | G | A | 0.885 | 5.54E-03 | 1.07E-03 | 4.60E-08 | 65680 | Imputed | 0.987 | [MINAR2] |
| 5 | 131466629 | rs3843503 | T | A | 0.55 | 4.25E-03 | 6.94E-04 | 1.90E-11 | 65680 | Imputed | 0.979 | CSF2--[]--AC063976.1 |
| 5 | 133393380 | 5:133393380_GA_G | GA | G | 0.844 | 7.22E-03 | 1.01E-03 | 3.30E-13 | 65680 | Imputed | 0.858 | VDAC1--[]--TCF7 |
| 5 | 172197790 | rs34013988 | C | T | 0.961 | 1.46E-02 | 1.75E-03 | 4.10E-18 | 65680 | Imputed | 0.998 | [AC022217.4,DUSP1] |
| 6 | 593289 | 6:593289_TG_T | TG | T | 0.869 | -7.47E-03 | 1.05E-03 | 2.60E-14 | 65680 | Imputed | 0.913 | [EXOC2] |
| 6 | 1548369 | rs2745572 | A | G | 0.666 | 4.35E-03 | 7.26E-04 | 3.10E-09 | 65680 | Imputed | 0.99 | FOXF2---[]--FOXC1 |
| 6 | 1983440 | rs6914444 | T | C | 0.866 | 9.30E-03 | 1.01E-03 | 9.70E-22 | 65680 | Imputed | 0.987 | [GMDS] |
| 6 | 7211818 | rs1334576 | G | A | 0.572 | -5.94E-03 | 6.92E-04 | 1.60E-18 | 65515 | Genotyped | 1 | [RREB1] |
| 6 | 11411838 | rs7742703 | C | T | 0.905 | 6.43E-03 | 1.16E-03 | 8.60E-09 | 65680 | Imputed | 0.996 | NEDD9--[]---TMEM170B |
| 6 | 31133577 | rs145919884 | A | AAAGCCC | 0.35 | 4.45E-03 | 7.18E-04 | 3.40E-10 | 65680 | Imputed | 0.996 | [TCF19,POU5F1] |
| 6 | 31158633 | rs9263861 | A | G | 0.879 | 6.21E-03 | 1.05E-03 | 4.40E-09 | 65680 | Imputed | 0.998 | POU5F1--[]-HCG27 |
| 6 | 36552592 | rs200252984 | G | A | 0.79 | -7.25E-03 | 8.47E-04 | 9.90E-21 | 65680 | Imputed | 0.973 | STK38--[]-SRSF3 |
| 6 | 39537880 | rs9369128 | T | C | 0.651 | -6.39E-03 | 7.15E-04 | 5.30E-20 | 65680 | Imputed | 0.998 | [KIF6] |
| 6 | 122392511 | rs2684249 | T | C | 0.593 | 6.38E-03 | 6.95E-04 | 9.00E-19 | 65680 | Imputed | 0.995 | GJA1---[]---HSF2 |
| 6 | 126767600 | rs1361108 | C | T | 0.543 | -6.05E-03 | 6.89E-04 | 1.60E-17 | 65542 | Genotyped | 1 | CENPW--[]---RSPO3 |
| 6 | 127289089 | rs4897198 | C | A | 0.548 | -4.56E-03 | 6.88E-04 | 2.70E-11 | 65680 | Imputed | 1 | CENPW---[]---RSPO3 |
| 6 | 148832343 | rs139973521 | A | ATGAG | 0.89 | -7.36E-03 | 1.10E-03 | 3.80E-13 | 65680 | Imputed | 0.987 | [SASH1] |
| 6 | 149989744 | 6:149989744_AT_A | AT | A | 0.648 | 4.99E-03 | 7.24E-04 | 2.10E-12 | 65680 | Imputed | 0.988 | [LATS1] |
| 6 | 151295133 | rs6900628 | A | G | 0.708 | 4.20E-03 | 7.55E-04 | 2.20E-08 | 65475 | Genotyped | 1 | [MTHFD1L] |
| 7 | 4767112 | rs6946034 | A | T | 0.52 | -3.68E-03 | 6.86E-04 | 3.00E-09 | 65680 | Imputed | 0.983 | [FOXK1] |
| 7 | 14237240 | rs10260511 | C | A | 0.842 | -7.78E-03 | 9.37E-04 | 5.90E-19 | 65597 | Genotyped | 1 | [DGKB] |
| 7 | 19612305 | rs2192476 | C | T | 0.372 | 4.26E-03 | 7.06E-04 | 2.00E-11 | 65680 | Imputed | 0.993 | FERD3L---[]---TWISTNB |
| 7 | 28393403 | rs7805378 | A | C | 0.559 | 4.60E-03 | 6.86E-04 | 1.20E-11 | 65680 | Imputed | 0.994 | [CREB5] |
| 7 | 28844815 | rs2282909 | T | G | 0.269 | 4.09E-03 | 7.70E-04 | 1.40E-08 | 65680 | Imputed | 0.999 | [CREB5] |
| 7 | 42108499 | rs2237417 | C | T | 0.593 | 3.69E-03 | 6.99E-04 | 4.00E-08 | 65680 | Imputed | 0.987 | [GLI3] |
| 7 | 101808020 | rs6976947 | A | G | 0.604 | 4.41E-03 | 6.97E-04 | 8.70E-12 | 65680 | Imputed | 0.997 | [CUX1] |
| 8 | 8254590 | rs2945880 | A | G | 0.113 | -9.34E-03 | 1.08E-03 | 2.30E-19 | 65680 | Imputed | 0.994 | PRAG1--[]---AC114550.3 |
| 8 | 17526359 | rs11203888 | C | T | 0.335 | -4.18E-03 | 7.24E-04 | 2.80E-09 | 65680 | Imputed | 0.995 | [MTUS1] |
| 8 | 30336017 | rs571194397 | A | AT | 0.817 | -5.07E-03 | 9.12E-04 | 1.70E-08 | 65680 | Imputed | 0.923 | [RBPMS] |
| 8 | 30386291 | rs7013873 | C | T | 0.784 | 4.49E-03 | 8.32E-04 | 7.90E-10 | 65680 | Imputed | 0.989 | [RBPMS] |
| 8 | 30510065 | 8:30510065_TA_T | TA | T | 0.578 | 3.54E-03 | 6.98E-04 | 2.30E-08 | 65680 | Imputed | 0.977 | [GTF2E2] |
| 8 | 61911070 | rs10957177 | A | G | 0.749 | 4.57E-03 | 7.94E-04 | 6.80E-09 | 65680 | Imputed | 0.978 | CHD7---[]--CLVS1 |
| 8 | 72278010 | rs12543430 | T | C | 0.39 | -6.08E-03 | 7.11E-04 | 1.10E-17 | 65680 | Imputed | 0.962 | [EYA1] |
| 8 | 72392687 | rs10093418 | A | G | 0.893 | 6.99E-03 | 1.12E-03 | 7.40E-11 | 65680 | Imputed | 0.977 | [EYA1] |
| 8 | 72579250 | rs10453110 | C | T | 0.874 | -9.94E-03 | 1.04E-03 | 8.50E-23 | 65680 | Imputed | 0.99 | EYA1--[]---AC104012.2 |
| 8 | 75522500 | rs10957731 | A | C | 0.402 | 5.12E-03 | 6.96E-04 | 3.70E-14 | 65680 | Imputed | 0.994 | [MIR2052HG] |
| 8 | 75532622 | rs2081498 | C | A | 0.936 | 8.58E-03 | 1.40E-03 | 2.40E-10 | 65680 | Imputed | 0.986 | [MIR2052HG] |
| 8 | 76025955 | rs11991447 | C | T | 0.756 | 4.78E-03 | 7.98E-04 | 2.80E-09 | 65680 | Imputed | 0.984 | CRISPLD1--[]---HNF4G |
| 8 | 78948855 | rs6999835 | T | C | 0.634 | 3.93E-03 | 7.10E-04 | 2.00E-09 | 65680 | Imputed | 0.996 | []---PKIA |
| 8 | 88664409 | rs200382882 | G | C | 0.485 | -4.07E-03 | 6.86E-04 | 7.60E-11 | 65680 | Imputed | 0.997 | CNBD1--[]---DCAF4L2 |
| 8 | 131606303 | 8:131606303_CTGTT_C | CTGTT | C | 0.647 | 3.90E-03 | 7.18E-04 | 4.20E-08 | 65680 | Imputed | 0.986 | ASAP1---[]---ADCY8 |
| 9 | 15912375 | rs10810475 | G | C | 0.564 | 3.67E-03 | 6.87E-04 | 1.20E-08 | 65680 | Imputed | 0.998 | [CCDC171] |
| 9 | 16619529 | rs13290470 | A | G | 0.603 | 5.52E-03 | 6.97E-04 | 7.20E-19 | 65680 | Imputed | 0.991 | [BNC2] |
| 9 | 18089275 | rs10738500 | C | A | 0.927 | -9.11E-03 | 1.31E-03 | 3.90E-12 | 65680 | Imputed | 0.99 | [ADAMTSL1] |
| 9 | 21951175 | rs117197971 | A | G | 0.945 | -1.03E-02 | 1.60E-03 | 1.30E-11 | 65680 | Imputed | 0.855 | [AL359922.1] |
| 9 | 22007330 | rs3217978 | C | A | 0.984 | 1.53E-02 | 2.84E-03 | 3.20E-08 | 65680 | Imputed | 0.914 | [AL359922.1,CDKN2B-AS1,CDKN2B] |
| 9 | 22044904 | rs74744824 | A | G | 0.984 | -1.85E-02 | 3.00E-03 | 2.30E-09 | 65680 | Imputed | 0.801 | [CDKN2B-AS1] |
| 9 | 22051670 | rs944801 | G | C | 0.428 | -1.57E-02 | 6.90E-04 | 5.30E-125 | 65680 | Imputed | 0.999 | [CDKN2B-AS1] |
| 9 | 22052068 | rs62560775 | A | G | 0.898 | -7.81E-03 | 1.13E-03 | 1.50E-12 | 65680 | Imputed | 0.993 | [CDKN2B-AS1] |
| 9 | 22082375 | rs1547705 | A | C | 0.877 | -8.34E-03 | 1.08E-03 | 8.50E-16 | 65680 | Imputed | 0.923 | [CDKN2B-AS1] |
| 9 | 22090416 | rs10965230 | C | T | 0.949 | 1.07E-02 | 1.62E-03 | 2.20E-11 | 65680 | Imputed | 0.903 | [CDKN2B-AS1] |
| 9 | 76622068 | rs11143754 | C | A | 0.573 | 4.29E-03 | 6.90E-04 | 3.40E-10 | 65680 | Imputed | 0.994 | AL451127.1---[]---AL355674.1 |
| 9 | 89252706 | rs10512176 | T | C | 0.722 | -6.49E-03 | 7.76E-04 | 1.70E-17 | 65680 | Imputed | 0.95 | TUT7---[]---GAS1 |
| 9 | 89380805 | rs1111066 | C | G | 0.516 | 3.70E-03 | 6.85E-04 | 7.80E-09 | 65680 | Imputed | 0.989 | TUT7---[]---GAS1 |
| 9 | 134563185 | rs11793533 | G | A | 0.719 | 5.28E-03 | 7.58E-04 | 2.80E-12 | 65419 | Genotyped | 1 | [RAPGEF1] |
| 9 | 136145414 | rs587611953 | C | A | 0.836 | -6.71E-03 | 1.02E-03 | 4.80E-13 | 65680 | Imputed | 0.824 | [ABO] |
| 10 | 14086579 | rs7099081 | A | G | 0.663 | 3.85E-03 | 7.20E-04 | 2.00E-08 | 65680 | Imputed | 0.993 | [FRMD4A] |

| CHR | POS | SNP | EA | NEA | EAF | BETA | SE | P | NUM_INDV | SRC | INFO | GENE_CONTEXT |
|---|---|---|---|---|---|---|---|---|---|---|---|---|
| 10 | 21462896 | 10:21462896_GGC_G | GGC | G | 0.987 | -2.40E-02 | 3.08E-03 | 1.60E-15 | 65680 | Imputed | 0.943 | [NEBL] |
| 10 | 60271824 | rs7069916 | G | A | 0.657 | -3.95E-03 | 7.16E-04 | 5.50E-09 | 65680 | Imputed | 0.997 | TFAM---[]BICC1 |
| 10 | 69838913 | rs117479359 | G | T | 0.917 | 8.56E-03 | 1.24E-03 | 2.70E-12 | 65680 | Imputed | 0.989 | HERC4-[]--MYPN |
| 10 | 69879366 | rs113337354 | A | G | 0.875 | 6.12E-03 | 1.03E-03 | 8.60E-10 | 65680 | Imputed | 0.995 | [MYPN] |
| 10 | 69902549 | rs3814180 | T | C | 0.395 | -3.92E-03 | 7.11E-04 | 1.40E-08 | 65680 | Imputed | 0.953 | [MYPN] |
| 10 | 69926319 | rs61854624 | C | A | 0.84 | 6.54E-03 | 9.31E-04 | 3.80E-14 | 65680 | Imputed | 1 | [MYPN] |
| 10 | 69974599 | 10:69974599_CG_C | CG | C | 0.71 | -6.42E-03 | 8.03E-04 | 9.90E-19 | 65680 | Imputed | 0.873 | MYPN-[]--ATOH7 |
| 10 | 69991853 | rs7916697 | A | G | 0.24 | -1.62E-02 | 7.98E-04 | 1.20E-98 | 65604 | Genotyped | 1 | [ATOH7] |
| 10 | 70026190 | rs112652264 | G | A | 0.977 | 1.74E-02 | 2.31E-03 | 9.90E-16 | 63682 | Genotyped | 1 | ATOH7--[]--PBLD |
| 10 | 70041695 | rs374085072 | G | A | 0.99 | 2.15E-02 | 3.81E-03 | 1.80E-09 | 65680 | Imputed | 0.828 | ATOH7--[]PBLD |
| 10 | 70206657 | rs367873689 | C | CA | 0.955 | 1.46E-02 | 1.71E-03 | 1.30E-18 | 65680 | Imputed | 0.932 | [DNA2] |
| 10 | 70312849 | rs4746777 | C | T | 0.154 | -5.48E-03 | 9.46E-04 | 2.00E-09 | 65680 | Imputed | 0.989 | SLC25A16--[]-TET1 |
| 10 | 70336105 | 10:70336105_CAA_C | CAA | C | 0.535 | -4.16E-03 | 6.85E-04 | 5.10E-10 | 65680 | Imputed | 0.994 | [TET1] |
| 10 | 70349416 | rs546745967 | C | T | 0.974 | 1.80E-02 | 2.35E-03 | 9.20E-17 | 65680 | Imputed | 0.825 | [TET1] |
| 10 | 70467778 | rs34117216 | C | A | 0.784 | 6.57E-03 | 8.26E-04 | 3.70E-18 | 65618 | Genotyped | 1 | TET1--[]--CCAR1 |
| 10 | 70775081 | rs2429022 | T | C | 0.833 | 4.80E-03 | 9.15E-04 | 2.00E-09 | 65680 | Imputed | 0.995 | [KIFBP] |
| 10 | 94974129 | rs6583871 | G | T | 0.264 | -5.05E-03 | 7.76E-04 | 1.70E-11 | 65680 | Imputed | 0.982 | CYP26A1---[]--MYOF |
| 10 | 96026184 | 10:96026184_CA_C | CA | C | 0.565 | -4.35E-03 | 6.91E-04 | 3.10E-10 | 65680 | Imputed | 0.995 | [PLCE1] |
| 10 | 96071561 | rs2077218 | G | A | 0.238 | 5.02E-03 | 8.02E-04 | 3.80E-10 | 65680 | Imputed | 0.992 | [PLCE1] |
| 10 | 118546046 | rs11197820 | G | A | 0.585 | -6.17E-03 | 6.94E-04 | 2.90E-19 | 65680 | Imputed | 0.985 | [HSPA12A] |
| 10 | 118562571 | 10:118562571_AT_A | AT | A | 0.067 | -7.68E-03 | 1.37E-03 | 1.20E-08 | 65680 | Imputed | 0.978 | [HSPA12A] |
| 10 | 118918956 | rs72840231 | A | T | 0.567 | -4.64E-03 | 6.95E-04 | 1.80E-11 | 65680 | Imputed | 0.975 | [MIR3663HG] |
| 11 | 19960147 | rs12807015 | G | T | 0.528 | -3.98E-03 | 6.97E-04 | 8.90E-09 | 65680 | Imputed | 0.96 | [NAV2] |
| 11 | 31719504 | 11:31719504_CT_C | CT | C | 0.233 | 7.13E-03 | 8.10E-04 | 9.60E-21 | 65680 | Imputed | 0.992 | [ELP4] |
| 11 | 57656794 | rs35328629 | T | TA | 0.561 | -4.29E-03 | 6.89E-04 | 2.00E-10 | 65680 | Imputed | 0.987 | CTNND1--[]---OR9Q1 |
| 11 | 58405674 | rs12799215 | C | T | 0.758 | 4.22E-03 | 7.96E-04 | 4.20E-08 | 65680 | Imputed | 0.996 | CNTF--[]-GLYAT |
| 11 | 63678128 | rs199826712 | T | TA | 0.926 | 6.78E-03 | 1.34E-03 | 9.90E-09 | 65680 | Imputed | 0.948 | [MARK2] |
| 11 | 65091708 | 11:65091708_AGTGT_A | AGTGT | A | 0.32 | 4.14E-03 | 7.57E-04 | 1.60E-08 | 65680 | Imputed | 0.931 | [AP000944.5] |
| 11 | 65240979 | 11:65240979_TAA_T | TAA | T | 0.961 | -9.80E-03 | 1.86E-03 | 1.30E-08 | 65680 | Imputed | 0.883 | NEAT1--[]--SCYL1 |
| 11 | 65343399 | rs547193816 | C | CCCCGCT | 0.797 | 9.19E-03 | 8.60E-04 | 2.00E-27 | 65680 | Imputed | 0.97 | FAM89B-[]EHBP1L1 |
| 11 | 86666906 | rs3758658 | T | C | 0.312 | 3.96E-03 | 7.36E-04 | 9.60E-09 | 65680 | Imputed | 0.987 | FZD4[]--TMEM135 |
| 11 | 86748437 | rs2445575 | T | C | 0.807 | 4.90E-03 | 8.62E-04 | 7.30E-10 | 65680 | Imputed | 0.997 | FZD4--[]TMEM135 |
| 11 | 94533444 | rs138059525 | G | A | 0.993 | 3.10E-02 | 4.07E-03 | 2.60E-14 | 65629 | Genotyped | 1 | [AMOTL1] |
| 11 | 95308854 | rs11021221 | T | A | 0.831 | 7.22E-03 | 9.09E-04 | 3.60E-15 | 65680 | Imputed | 0.994 | SESN3---[]---FAM76B |
| 11 | 130281735 | rs10894194 | T | C | 0.797 | 5.26E-03 | 8.54E-04 | 1.50E-10 | 65680 | Imputed | 0.983 | [ADAMTS8] |
| 11 | 130288797 | rs34248430 | A | ACCT | 0.282 | -7.71E-03 | 7.72E-04 | 4.50E-27 | 65680 | Imputed | 0.961 | [ADAMTS8] |
| 11 | 130307542 | rs56378410 | T | C | 0.974 | -1.23E-02 | 2.15E-03 | 1.30E-09 | 65680 | Imputed | 0.979 | ADAMTS8-[]--ADAMTS15 |
| 12 | 3353356 | rs73047017 | C | T | 0.937 | -8.15E-03 | 1.41E-03 | 6.00E-10 | 65680 | Imputed | 0.976 | [TSPAN9] |
| 12 | 26392080 | rs16930371 | A | G | 0.816 | 4.97E-03 | 8.80E-04 | 4.80E-09 | 65283 | Genotyped | 1 | [SSPN] |
| 12 | 31067490 | rs55710412 | C | T | 0.845 | 5.27E-03 | 9.65E-04 | 8.10E-09 | 65680 | Imputed | 0.948 | CAPRIN2---[]--TSPAN11 |
| 12 | 48157019 | rs12818241 | C | T | 0.84 | 6.29E-03 | 9.31E-04 | 1.10E-12 | 65680 | Imputed | 0.984 | [RAPGEF3,SLC48A1] |
| 12 | 76114872 | rs6582298 | G | A | 0.595 | 5.67E-03 | 7.23E-04 | 9.30E-16 | 65680 | Imputed | 0.901 | [AC078923.1] |
| 12 | 83762921 | 12:83762921_CA_C | CA | C | 0.358 | -5.37E-03 | 7.31E-04 | 4.20E-14 | 65680 | Imputed | 0.943 | TMTC2---[] |
| 12 | 83856540 | rs12826083 | G | A | 0.859 | -7.51E-03 | 1.00E-03 | 1.20E-14 | 65680 | Imputed | 0.948 | TMTC2---[] |
| 12 | 83858003 | rs76005250 | G | A | 0.941 | 1.06E-02 | 1.49E-03 | 4.70E-14 | 65680 | Imputed | 0.94 | TMTC2---[] |
| 12 | 83861261 | rs77725841 | A | G | 0.916 | 8.51E-03 | 1.24E-03 | 5.90E-14 | 65680 | Imputed | 0.969 | TMTC2---[] |
| 12 | 84076137 | rs55667441 | A | G | 0.558 | 1.58E-02 | 6.86E-04 | 2.50E-127 | 65680 | Imputed | 0.995 | TMTC2---[] |
| 12 | 84119063 | rs76513789 | C | A | 0.945 | 9.92E-03 | 1.50E-03 | 4.90E-12 | 65680 | Imputed | 0.994 | TMTC2---[] |
| 12 | 84132842 | rs1380758 | A | G | 0.088 | 7.41E-03 | 1.24E-03 | 2.20E-10 | 65680 | Imputed | 0.927 | TMTC2---[] |
| 12 | 84135457 | rs142121892 | C | CA | 0.666 | 7.49E-03 | 7.40E-04 | 6.30E-27 | 65680 | Imputed | 0.949 | TMTC2---[] |
| 12 | 84141754 | rs79301152 | A | G | 0.917 | -8.96E-03 | 1.28E-03 | 1.80E-13 | 65680 | Imputed | 0.924 | TMTC2---[] |
| 12 | 84154996 | rs71450946 | A | G | 0.927 | -8.72E-03 | 1.31E-03 | 9.30E-13 | 65545 | Genotyped | 1 | TMTC2---[] |
| 12 | 84225935 | rs7137822 | A | T | 0.654 | 7.24E-03 | 7.73E-04 | 8.50E-23 | 65680 | Imputed | 0.858 | TMTC2---[] |
| 12 | 84234482 | rs73154757 | T | A | 0.962 | 1.03E-02 | 1.80E-03 | 1.40E-09 | 65680 | Imputed | 0.97 | TMTC2---[] |
| 12 | 84237124 | 12:84237124_CTAA_C | CTAA | C | 0.948 | 9.14E-03 | 1.57E-03 | 4.50E-11 | 65680 | Imputed | 0.954 | TMTC2---[] |
| 12 | 84359892 | rs12423401 | C | A | 0.906 | -9.42E-03 | 1.19E-03 | 3.30E-16 | 65680 | Imputed | 0.961 | TMTC2---[]---SLC6A15 |
| 12 | 84404719 | rs4772012 | T | C | 0.769 | 4.64E-03 | 8.10E-04 | 2.90E-11 | 65680 | Imputed | 0.992 | TMTC2---[]---SLC6A15 |
| 12 | 84890667 | rs71445008 | C | CT | 0.561 | -5.27E-03 | 7.09E-04 | 4.10E-16 | 65680 | Imputed | 0.933 | []---SLC6A15 |
| 12 | 91816926 | rs147377344 | C | CTTTTACG | 0.4 | 4.03E-03 | 7.03E-04 | 2.10E-08 | 65680 | Imputed | 0.977 | DCN---[]---LINC01619 |
| 12 | 107073242 | 12:107073242_CA_C | CA | C | 0.653 | 5.36E-03 | 7.17E-04 | 6.00E-15 | 65680 | Imputed | 0.988 | [AC079385.1,RFX4] |
| 12 | 108165360 | rs2111281 | A | C | 0.636 | -5.30E-03 | 7.07E-04 | 1.30E-13 | 65680 | Imputed | 0.995 | PRDM4--[]-ASCL4 |
| 12 | 124665773 | rs11057488 | A | G | 0.435 | -4.85E-03 | 6.87E-04 | 8.50E-14 | 65680 | Imputed | 0.994 | [RFLNA] |
| 13 | 25778093 | rs17081940 | A | G | 0.858 | 5.44E-03 | 9.90E-04 | 2.10E-09 | 65680 | Imputed | 0.978 | [LINC01076] |
| 13 | 36683268 | rs9546383 | T | C | 0.754 | -5.96E-03 | 7.95E-04 | 6.70E-16 | 65680 | Imputed | 0.992 | [DCLK1] |
| 13 | 51945741 | rs9535652 | G | A | 0.834 | 5.41E-03 | 9.19E-04 | 4.10E-09 | 65680 | Imputed | 0.996 | [INTS6] |
| 13 | 109267985 | rs10162202 | T | C | 0.722 | 6.97E-03 | 7.68E-04 | 6.10E-23 | 65680 | Imputed | 0.982 | [MYO16] |
| 13 | 110778747 | 13:110778747_CCTTTT_C | CCTTTT | C | 0.641 | -6.06E-03 | 7.31E-04 | 9.90E-18 | 65680 | Imputed | 0.945 | IRS2---[]--COL4A1 |

Supplementary Information - Table 2. ML-based VCDR (hits)

| CHR | POS | SNP | EA | NEA | EAF | BETA | SE | P | NUM_INDV | SRC | INFO | GENE_CONTEXT |
|---|---|---|---|---|---|---|---|---|---|---|---|---|
| 14 | 23452128 | rs3811183 | C | G | 0.6 | -4.45E-03 | 7.01E-04 | 1.30E-12 | 65680 | Imputed | 0.987 | AJUBA[]-C14orf93 |
| 14 | 53991705 | rs2077940 | T | C | 0.668 | 5.45E-03 | 7.24E-04 | 1.70E-14 | 65680 | Imputed | 0.997 | DDHD1---[]--AL163953.1 |
| 14 | 60808553 | rs10162287 | C | G | 0.696 | -7.65E-03 | 7.48E-04 | 2.60E-27 | 65680 | Imputed | 0.979 | PPM1A--[]--C14orf39 |
| 14 | 60914325 | rs139811951 | G | A | 0.874 | 6.34E-03 | 1.04E-03 | 1.60E-10 | 65680 | Imputed | 0.976 | [C14orf39] |
| 14 | 61238781 | rs12147818 | G | A | 0.993 | 2.88E-02 | 4.26E-03 | 8.10E-13 | 65680 | Imputed | 0.955 | [MNAT1] |
| 14 | 65081054 | rs149761305 | G | GCT | 0.83 | -7.31E-03 | 9.14E-04 | 7.40E-18 | 65680 | Imputed | 0.993 | PPP1R36--[]--PLEKHG3 |
| 14 | 85668732 | rs11626115 | G | A | 0.949 | -1.05E-02 | 1.58E-03 | 1.10E-10 | 65680 | Imputed | 0.947 | []---FLRT2 |
| 14 | 85863058 | rs2145865 | T | A | 0.456 | 4.32E-03 | 6.97E-04 | 3.30E-10 | 65680 | Imputed | 0.967 | []---FLRT2 |
| 14 | 85922578 | rs1289426 | A | G | 0.767 | -6.91E-03 | 8.15E-04 | 6.80E-16 | 65680 | Imputed | 0.974 | []--FLRT2 |
| 14 | 86021748 | rs2018653 | A | G | 0.755 | -5.84E-03 | 8.00E-04 | 3.90E-13 | 65680 | Imputed | 0.986 | [FLRT2] |
| 14 | 95957694 | rs11160251 | T | G | 0.703 | 4.15E-03 | 7.47E-04 | 1.90E-09 | 65680 | Imputed | 0.991 | SYNE3--[]--GLRX5 |
| 15 | 71840327 | rs35194812 | T | C | 0.839 | -4.96E-03 | 9.39E-04 | 1.50E-09 | 65680 | Imputed | 0.975 | [THSD4] |
| 15 | 71882771 | rs4776562 | A | G | 0.554 | 3.95E-03 | 7.27E-04 | 1.10E-09 | 65680 | Imputed | 0.892 | [THSD4] |
| 15 | 74230660 | rs59755145 | G | A | 0.712 | -4.23E-03 | 7.57E-04 | 2.90E-09 | 65680 | Imputed | 0.992 | [LOXL1] |
| 15 | 84484384 | rs59199978 | A | G | 0.82 | -6.14E-03 | 8.91E-04 | 6.00E-14 | 65680 | Imputed | 0.995 | [ADAMTSL3] |
| 15 | 99458902 | rs28612945 | C | T | 0.794 | 6.20E-03 | 8.46E-04 | 7.80E-15 | 65547 | Genotyped | 1 | [IGF1R] |
| 15 | 101200873 | rs34222435 | C | T | 0.865 | -9.55E-03 | 1.00E-03 | 1.60E-23 | 65680 | Imputed | 0.991 | ASB7-[]---ALDH1A3 |
| 15 | 101200962 | rs11452536 | T | TA | 0.374 | -3.87E-03 | 7.12E-04 | 1.50E-09 | 65680 | Imputed | 0.98 | ASB7-[]---ALDH1A3 |
| 15 | 101751698 | rs8043304 | T | C | 0.752 | 4.35E-03 | 7.95E-04 | 1.80E-08 | 65680 | Imputed | 0.995 | [CHSY1] |
| 16 | 51183728 | rs11643654 | C | A | 0.389 | -4.74E-03 | 7.00E-04 | 1.50E-10 | 65680 | Imputed | 0.997 | [SALL1] |
| 16 | 51341412 | rs117537696 | T | G | 0.981 | -1.73E-02 | 2.47E-03 | 2.10E-13 | 65680 | Imputed | 1 | SALL1---[]---HNRNPA1P48 |
| 16 | 51400539 | rs57495036 | C | G | 0.972 | -1.57E-02 | 2.06E-03 | 1.10E-14 | 65680 | Imputed | 0.993 | SALL1---[]---HNRNPA1P48 |
| 16 | 51416004 | rs1111196 | A | T | 0.263 | -9.50E-03 | 7.80E-04 | 2.20E-37 | 65680 | Imputed | 0.985 | SALL1---[]---HNRNPA1P48 |
| 16 | 51469726 | rs8053277 | T | C | 0.303 | 1.11E-02 | 7.45E-04 | 6.90E-54 | 65680 | Imputed | 0.993 | SALL1---[]---HNRNPA1P48 |
| 16 | 51470928 | rs71386559 | C | T | 0.96 | 1.36E-02 | 1.78E-03 | 2.10E-16 | 65680 | Imputed | 0.954 | SALL1---[]---HNRNPA1P48 |
| 16 | 51526098 | rs12443961 | T | C | 0.494 | 5.00E-03 | 6.90E-04 | 2.90E-14 | 65680 | Imputed | 0.977 | SALL1---[]--HNRNPA1P48 |
| 16 | 51573106 | rs143024692 | C | A | 0.991 | 2.24E-02 | 3.68E-03 | 1.20E-10 | 65680 | Imputed | 0.953 | SALL1---[]--HNRNPA1P48 |
| 16 | 51590885 | rs116257475 | C | T | 0.915 | -1.50E-02 | 1.22E-03 | 4.40E-37 | 65680 | Imputed | 0.995 | [HNRNPA1P48] |
| 16 | 51649796 | 16:51649796_CTCTT_C | CTCTT | C | 0.644 | -4.71E-03 | 7.46E-04 | 8.90E-10 | 65680 | Imputed | 0.907 | [HNRNPA1P48] |
| 16 | 51667131 | rs8057507 | T | C | 0.533 | -6.92E-03 | 6.85E-04 | 3.00E-27 | 65680 | Imputed | 0.993 | [HNRNPA1P48] |
| 16 | 51855425 | rs2892063 | G | A | 0.371 | -4.62E-03 | 7.19E-04 | 1.20E-11 | 65680 | Imputed | 0.966 | HNRNPA1P48---[]---TOX3 |
| 16 | 74226221 | rs807293 | T | C | 0.636 | -4.41E-03 | 7.10E-04 | 7.70E-11 | 65680 | Imputed | 0.996 | ZFHX3---[]---PSMD7 |
| 16 | 74404812 | rs11865100 | C | T | 0.869 | 6.31E-03 | 1.07E-03 | 9.30E-10 | 65680 | Imputed | 0.875 | PSMD7--[]-NPIPB15 |
| 16 | 74458445 | rs77199494 | T | G | 0.967 | -1.09E-02 | 1.99E-03 | 3.30E-08 | 65680 | Imputed | 0.932 | [AC009053.4] |
| 16 | 86366644 | rs3762872 | C | T | 0.487 | 3.65E-03 | 6.86E-04 | 1.00E-08 | 65680 | Imputed | 0.99 | IRF8---[]---FOXF1 |
| 16 | 86380293 | rs1687628 | T | C | 0.09 | 1.21E-02 | 1.19E-03 | 2.10E-25 | 65680 | Imputed | 0.992 | IRF8---[]---FOXF1 |
| 16 | 86403160 | rs17178451 | A | G | 0.468 | -4.12E-03 | 6.84E-04 | 1.90E-09 | 65680 | Imputed | 0.996 | IRF8---[]---FOXF1 |
| 16 | 86439374 | rs12935509 | G | A | 0.919 | -8.16E-03 | 1.25E-03 | 1.50E-11 | 65680 | Imputed | 0.992 | IRF8---[]---FOXF1 |
| 16 | 86465590 | rs13332095 | G | A | 0.899 | -6.32E-03 | 1.13E-03 | 6.10E-09 | 65680 | Imputed | 0.991 | IRF8---[]--FOXF1 |
| 16 | 86513683 | 16:86513683_GCA_G | GCA | G | 0.224 | 5.95E-03 | 8.30E-04 | 7.00E-14 | 65680 | Imputed | 0.975 | IRF8---[]--FOXF1 |
| 17 | 40867365 | rs115818584 | C | G | 0.984 | 1.62E-02 | 2.73E-03 | 1.70E-09 | 65680 | Imputed | 0.984 | [EZH1] |
| 17 | 45438886 | rs769594276 | C | CAGTG | 0.548 | -4.05E-03 | 6.88E-04 | 4.40E-11 | 65680 | Imputed | 0.997 | [EFCAB13] |
| 17 | 48225686 | rs4794104 | C | G | 0.838 | -5.48E-03 | 9.29E-04 | 9.00E-10 | 65680 | Imputed | 0.995 | [PPP1R9B] |
| 17 | 61865670 | 17:61865670_CT_C | CT | C | 0.637 | -4.25E-03 | 7.28E-04 | 2.40E-10 | 65680 | Imputed | 0.947 | [DDX42] |
| 17 | 65073835 | rs577377763 | C | CA | 0.642 | -4.45E-03 | 7.38E-04 | 4.50E-09 | 65680 | Imputed | 0.925 | [HELZ] |
| 17 | 80169426 | rs796355894 | A | AT | 0.592 | 3.96E-03 | 7.09E-04 | 1.60E-08 | 65680 | Imputed | 0.956 | [CCDC57] |
| 18 | 8797487 | rs569735 | C | A | 0.269 | 4.57E-03 | 7.89E-04 | 1.00E-08 | 65680 | Imputed | 0.961 | [MTCL1] |
| 18 | 23063159 | rs766791666 | T | TATC | 0.415 | -4.02E-03 | 7.05E-04 | 4.00E-10 | 65680 | Imputed | 0.971 | ZNF521---[]---SS18 |
| 18 | 34289285 | rs61735998 | G | T | 0.974 | 1.37E-02 | 2.17E-03 | 1.20E-10 | 65680 | Imputed | 1 | [FHOD3] |
| 18 | 56943484 | rs77759734 | C | T | 0.951 | -8.73E-03 | 1.59E-03 | 1.00E-09 | 65493 | Genotyped | 1 | [CPLX4] |
| 19 | 817708 | rs7250902 | A | G | 0.681 | 5.21E-03 | 7.45E-04 | 3.40E-12 | 65680 | Imputed | 0.966 | [PLPPR3] |
| 19 | 14616371 | rs11882319 | C | A | 0.851 | 5.31E-03 | 9.65E-04 | 6.00E-09 | 65680 | Imputed | 0.982 | GIPC1-[]-DNAJB1 |
| 19 | 32027330 | rs8102936 | G | A | 0.658 | 6.80E-03 | 7.26E-04 | 1.10E-19 | 65680 | Imputed | 0.987 | TSHZ3---[]---ZNF507 |
| 19 | 33523197 | rs73039431 | A | G | 0.955 | -1.04E-02 | 1.66E-03 | 3.70E-10 | 65680 | Imputed | 0.984 | [RHPN2] |
| 19 | 39146780 | rs55876653 | G | C | 0.508 | -4.15E-03 | 6.84E-04 | 7.40E-10 | 65680 | Imputed | 0.997 | [ACTN4] |
| 20 | 1029686 | rs4816177 | A | G | 0.821 | -5.44E-03 | 9.01E-04 | 4.10E-10 | 64868 | Genotyped | 1 | RSPO4--[]--PSMF1 |
| 20 | 6155721 | rs751223591 | C | CT | 0.287 | 6.23E-03 | 7.63E-04 | 5.80E-17 | 65680 | Imputed | 0.984 | FERMT1--[]---LINC01713 |
| 20 | 6229367 | rs55708193 | C | A | 0.799 | -5.76E-03 | 8.58E-04 | 8.60E-12 | 65680 | Imputed | 0.99 | FERMT1---[]---LINC01713 |
| 20 | 6420731 | rs6117259 | T | C | 0.812 | -6.22E-03 | 8.75E-04 | 8.20E-13 | 65680 | Imputed | 0.996 | FERMT1---[]---LINC01713 |
| 20 | 6470094 | rs2326788 | G | A | 0.628 | 1.17E-02 | 7.07E-04 | 1.60E-65 | 65680 | Imputed | 0.992 | FERMT1---[]---LINC01713 |
| 20 | 6474916 | rs11483156 | C | CT | 0.889 | -7.26E-03 | 1.10E-03 | 4.80E-11 | 65680 | Imputed | 0.964 | FERMT1---[]---LINC01713 |
| 20 | 6514692 | rs6038531 | G | A | 0.016 | -1.93E-02 | 2.72E-03 | 4.80E-13 | 65680 | Imputed | 0.976 | FERMT1---[]---LINC01713 |
| 20 | 6535065 | rs78004679 | G | A | 0.941 | 1.16E-02 | 1.48E-03 | 1.50E-16 | 65680 | Imputed | 0.959 | FERMT1---[]---LINC01713 |
| 20 | 6544738 | rs73077173 | C | T | 0.959 | 9.85E-03 | 1.73E-03 | 8.70E-10 | 65593 | Genotyped | 1 | FERMT1---[]---LINC01713 |
| 20 | 6631055 | rs1358805 | A | G | 0.825 | -5.71E-03 | 9.02E-04 | 1.90E-10 | 65680 | Imputed | 0.992 | FERMT1---[]--LINC01713 |
| 20 | 6759115 | rs235768 | A | T | 0.391 | -5.08E-03 | 7.01E-04 | 2.60E-13 | 65595 | Genotyped | 1 | [BMP2] |
| 20 | 31157394 | rs4911242 | A | T | 0.678 | 4.21E-03 | 7.35E-04 | 4.20E-09 | 65680 | Imputed | 0.992 | [NOL4L] |

Supplementary Information - Table 2. ML-based VCDR (hits)

| CHR | POS | SNP | EA | NEA | EAF | BETA | SE | P | NUM_INDV | SRC | INFO | GENE_CONTEXT |
|---|---|---|---|---|---|---|---|---|---|---|---|---|
| 20 | 31438954 | rs4911268 | A | G | 0.816 | 6.64E-03 | 8.85E-04 | 1.80E-15 | 65680 | Imputed | 0.998 | MAPRE1[]-EFCAB8 |
| 20 | 45797259 | rs3091590 | C | T | 0.544 | -4.29E-03 | 6.87E-04 | 1.10E-10 | 65680 | Imputed | 0.991 | [EYA2] |
| 22 | 28208528 | rs11704137 | C | G | 0.745 | -6.84E-03 | 7.84E-04 | 6.10E-20 | 65680 | Imputed | 0.998 | MN1--[]--PITPNB |
| 22 | 28358946 | rs5762423 | A | G | 0.056 | -1.27E-02 | 1.53E-03 | 1.80E-16 | 65680 | Imputed | 0.942 | [TTC28-AS1] |
| 22 | 28501414 | rs77885044 | C | T | 0.985 | 3.49E-02 | 2.77E-03 | 3.70E-42 | 65624 | Genotyped | 1 | [TTC28] |
| 22 | 28629713 | rs16986177 | C | T | 0.84 | -9.02E-03 | 9.32E-04 | 3.80E-23 | 65680 | Imputed | 0.996 | [TTC28] |
| 22 | 28653727 | 22:28653727_CATAT_C | CATAT | C | 0.593 | 4.71E-03 | 7.37E-04 | 7.50E-11 | 65680 | Imputed | 0.886 | [TTC28] |
| 22 | 28925406 | rs34320674 | C | A | 0.951 | 9.15E-03 | 1.61E-03 | 2.70E-09 | 65680 | Imputed | 0.964 | [TTC28] |
| 22 | 28990300 | rs117456789 | C | T | 0.948 | 9.66E-03 | 1.59E-03 | 2.40E-09 | 65680 | Imputed | 0.938 | [TTC28] |
| 22 | 29063037 | rs542574575 | C | T | 0.538 | 4.59E-03 | 7.21E-04 | 2.00E-11 | 65680 | Imputed | 0.902 | [TTC28] |
| 22 | 29115066 | rs4822983 | C | T | 0.677 | 1.49E-02 | 7.29E-04 | 1.30E-98 | 65680 | Imputed | 0.999 | [CHEK2] |
| 22 | 29127402 | rs8184952 | T | C | 0.158 | 8.04E-03 | 9.38E-04 | 9.80E-20 | 65680 | Imputed | 0.996 | [CHEK2] |
| 22 | 29162191 | rs145346186 | A | C | 0.955 | 1.20E-02 | 1.67E-03 | 7.30E-15 | 65680 | Imputed | 0.956 | HSCB-[]-CCDC117 |
| 22 | 29405956 | rs4580479 | T | G | 0.452 | -4.89E-03 | 6.92E-04 | 9.80E-14 | 65680 | Imputed | 0.98 | [ZNRF3] |
| 22 | 30592487 | rs713875 | C | G | 0.446 | 6.54E-03 | 6.86E-04 | 4.50E-24 | 65568 | Genotyped | 1 | [AC002378.1] |
| 22 | 30606986 | rs9614164 | C | T | 0.781 | 4.98E-03 | 8.29E-04 | 4.00E-09 | 65680 | Imputed | 0.985 | AC002378.1-[]--LIF |
| 22 | 30620627 | rs6006405 | A | G | 0.843 | -7.50E-03 | 9.41E-04 | 1.70E-16 | 65680 | Imputed | 0.995 | AC002378.1--[]--LIF |
| 22 | 37908462 | rs113605227 | A | AC | 0.827 | -1.15E-02 | 9.37E-04 | 1.40E-33 | 65680 | Imputed | 0.917 | [CARD10] |
| 22 | 37911770 | rs6000761 | T | C | 0.301 | -4.60E-03 | 7.54E-04 | 5.90E-11 | 65680 | Imputed | 0.97 | [CARD10] |
| 22 | 37916113 | rs143643697 | T | C | 0.964 | -1.15E-02 | 1.89E-03 | 2.10E-08 | 65680 | Imputed | 0.93 | CARD10[]--CDC42EP1 |
| 22 | 37923956 | rs372494932 | C | T | 0.99 | -2.16E-02 | 3.63E-03 | 4.60E-10 | 65680 | Imputed | 0.871 | CARD10-[]--CDC42EP1 |
| 22 | 37925332 | rs549756240 | A | T | 0.912 | -7.43E-03 | 1.30E-03 | 1.90E-08 | 65680 | Imputed | 0.859 | CARD10-[]--CDC42EP1 |
| 22 | 37939510 | rs9610795 | G | T | 0.652 | -6.47E-03 | 7.19E-04 | 1.70E-21 | 65680 | Imputed | 0.989 | CARD10--[]--CDC42EP1 |
| 22 | 38057338 | rs9622678 | A | G | 0.781 | -4.98E-03 | 9.29E-04 | 1.90E-09 | 65680 | Imputed | 0.801 | [Z83844.3,PDXP] |
| 22 | 38123259 | rs748088318 | G | GT | 0.165 | 4.93E-03 | 9.31E-04 | 6.90E-09 | 65680 | Imputed | 0.975 | [TRIOBP] |
| 22 | 38180407 | 22:38180407_CAA_C | CAA | C | 0.677 | -8.77E-03 | 7.31E-04 | 2.70E-36 | 65680 | Imputed | 0.992 | TRIOBP-[]--H1-0 |
| 22 | 38594126 | rs147906180 | C | CAAAAA | 0.675 | -6.61E-03 | 7.32E-04 | 1.20E-20 | 65680 | Imputed | 0.983 | [PLA2G6] |
| 22 | 38662396 | rs12485196 | T | C | 0.804 | -4.63E-03 | 8.61E-04 | 3.20E-08 | 65680 | Imputed | 0.998 | [TMEM184B] |
| 22 | 46375611 | rs62654723 | C | T | 0.832 | -5.54E-03 | 9.23E-04 | 5.40E-09 | 65680 | Imputed | 0.976 | WNT7B-[]--LINC00899 |
| 22 | 46383612 | rs73175083 | C | T | 0.687 | 6.37E-03 | 7.39E-04 | 1.40E-17 | 65680 | Imputed | 0.995 | WNT7B--[]--LINC00899 |

| CHR | POS | SNP | EA | NEA | EAF | BETA | SE | P | NUM_INDV | SRC | INFO | GENE_CONTEXT | CRAIG | CRAIG_META |
|---|---|---|---|---|---|---|---|---|---|---|---|---|---|---|
| 1 | 3049362 | rs12024620 | C | T | 0.936 | -1.55E-02 | 1.38E-03 | 4.00E-30 | 65680 | Imputed | 0.995 | [PRDM16] | TRUE | TRUE |
| 1 | 8468278 | rs301792 | A | G | 0.662 | -4.62E-03 | 7.13E-04 | 6.80E-13 | 65680 | Imputed | 0.998 | [RERE] | FALSE | TRUE |
| 1 | 12614029 | rs6541032 | T | C | 0.424 | -5.11E-03 | 6.82E-04 | 2.60E-14 | 65680 | Imputed | 0.999 | VPS13D--[]--DHRS3 | TRUE | TRUE |
| 1 | 47923058 | rs767682581 | C | CT | 0.374 | -4.27E-03 | 7.07E-04 | 1.40E-10 | 65680 | Imputed | 0.983 | FOXD2--[]---TRABD2B | FALSE | FALSE |
| 1 | 68840797 | rs2209559 | A | G | 0.601 | -7.49E-03 | 6.91E-04 | 1.90E-28 | 65484 | Genotyped | 1 | WLS---[]--RPE65 | TRUE | TRUE |
| 1 | 89253357 | rs786908 | A | G | 0.383 | -4.31E-03 | 6.95E-04 | 3.90E-10 | 65680 | Imputed | 0.997 | [PKN2] | FALSE | TRUE |
| 1 | 92077097 | rs1192415 | G | A | 0.188 | 1.76E-02 | 8.63E-04 | 7.60E-102 | 65610 | Genotyped | 1 | CDC7--[]--TGFBR3 | TRUE | TRUE |
| 1 | 110627923 | rs10857812 | T | A | 0.636 | 4.63E-03 | 7.03E-04 | 2.10E-11 | 65680 | Imputed | 0.993 | STRIP1--[]--UBL4B | FALSE | FALSE |
| 1 | 113045061 | rs351364 | A | T | 0.252 | -3.95E-03 | 7.81E-04 | 2.40E-08 | 65680 | Imputed | 0.984 | [WNT2B] | FALSE | FALSE |
| 1 | 155033308 | rs11589479 | G | A | 0.837 | 4.88E-03 | 9.17E-04 | 1.20E-08 | 65404 | Genotyped | 1 | [ADAM15] | FALSE | FALSE |
| 1 | 169551682 | rs6028 | T | C | 0.709 | -4.03E-03 | 7.44E-04 | 3.10E-09 | 65680 | Imputed | 0.996 | [F5] | FALSE | TRUE |
| 1 | 183849739 | rs41263652 | G | C | 0.896 | 7.20E-03 | 1.12E-03 | 4.20E-11 | 65680 | Imputed | 0.971 | [RGL1] | FALSE | FALSE |
| 1 | 218520995 | rs6658835 | A | G | 0.73 | -5.97E-03 | 7.64E-04 | 7.60E-16 | 65680 | Imputed | 0.992 | [TGFB2] | TRUE | TRUE |
| 1 | 219573841 | rs796959510 | C | CT | 0.491 | -4.02E-03 | 6.82E-04 | 7.40E-09 | 65680 | Imputed | 0.979 | AL360093.1---[]---ZC3H11B | FALSE | FALSE |
| 1 | 222014897 | rs11118873 | A | G | 0.485 | 3.74E-03 | 6.79E-04 | 4.20E-10 | 65680 | Imputed | 0.988 | DUSP10--[]---HHIPL2 | FALSE | FALSE |
| 1 | 227585983 | rs6670351 | G | A | 0.798 | -6.66E-03 | 8.42E-04 | 3.30E-17 | 65680 | Imputed | 0.993 | CDC42BPA--[]---ZNF678 | TRUE | TRUE |
| 2 | 5680539 | rs7575439 | C | A | 0.358 | 4.09E-03 | 7.27E-04 | 1.50E-08 | 62872 | Genotyped | 1 | LINC01249---[]--AC108025.1 | FALSE | FALSE |
| 2 | 12891476 | rs730126 | A | C | 0.587 | 3.98E-03 | 6.95E-04 | 9.50E-10 | 65680 | Imputed | 0.985 | TRIB2-[] | FALSE | FALSE |
| 2 | 19431423 | rs72778352 | C | T | 0.987 | 2.71E-02 | 2.97E-03 | 4.80E-22 | 65680 | Imputed | 0.982 | NT5C1B---[]---OSR1 | FALSE | TRUE |
| 2 | 56072501 | rs1430202 | G | A | 0.798 | 8.29E-03 | 8.51E-04 | 1.80E-24 | 65680 | Imputed | 0.99 | PNPT1---[]--EFEMP1 | TRUE | TRUE |
| 2 | 111658010 | rs4849203 | A | G | 0.652 | 5.07E-03 | 7.12E-04 | 1.30E-13 | 65575 | Genotyped | 1 | [ACOXL] | TRUE | TRUE |
| 2 | 180196027 | rs12620141 | C | A | 0.636 | -4.06E-03 | 7.18E-04 | 2.50E-09 | 65680 | Imputed | 0.961 | AC093911.1--[]---ZNF385B | FALSE | FALSE |
| 2 | 190269957 | 2:190269957_CTTTT_C | CTTTT | C | 0.308 | 4.16E-03 | 7.40E-04 | 3.00E-09 | 65680 | Imputed | 0.99 | COL5A2---[]--WDR75 | FALSE | FALSE |
| 2 | 233389918 | rs2853447 | A | G | 0.297 | -3.75E-03 | 7.44E-04 | 3.80E-08 | 65680 | Imputed | 0.987 | [PRSS56] | FALSE | FALSE |
| 3 | 20061023 | rs35057657 | A | G | 0.327 | 4.06E-03 | 7.27E-04 | 4.30E-08 | 65680 | Imputed | 0.991 | PP2D1-[]--KAT2B | FALSE | FALSE |
| 3 | 25046463 | rs12490228 | C | T | 0.717 | -6.08E-03 | 7.54E-04 | 3.50E-17 | 65680 | Imputed | 0.996 | THRB-AS1---[]---RARB | TRUE | TRUE |
| 3 | 29493916 | rs9822629 | T | G | 0.626 | -3.71E-03 | 7.02E-04 | 1.50E-08 | 65680 | Imputed | 0.992 | [AC098650.1,RBMS3] | FALSE | FALSE |
| 3 | 32879823 | rs56131903 | A | T | 0.679 | 6.19E-03 | 7.34E-04 | 3.90E-18 | 65680 | Imputed | 0.983 | [TRIM71] | TRUE | TRUE |
| 3 | 48719638 | rs7633840 | T | C | 0.337 | 4.72E-03 | 7.39E-04 | 1.60E-11 | 65680 | Imputed | 0.949 | [NCKIPSD] | FALSE | FALSE |
| 3 | 58130168 | rs2362911 | A | G | 0.771 | -5.34E-03 | 8.06E-04 | 3.30E-12 | 65680 | Imputed | 0.991 | [FLNB] | FALSE | TRUE |
| 3 | 71182447 | rs77877421 | A | T | 0.943 | -1.16E-02 | 1.50E-03 | 1.60E-15 | 65680 | Imputed | 0.964 | [FOXP1,AC097634.4] | FALSE | FALSE |
| 3 | 88380417 | rs9852080 | T | C | 0.448 | 6.19E-03 | 6.83E-04 | 3.70E-21 | 65680 | Imputed | 0.996 | C3orf38---[]-CSNKA2IP | TRUE | TRUE |
| 3 | 99078606 | rs1871794 | T | C | 0.748 | -8.53E-03 | 7.82E-04 | 9.40E-30 | 65680 | Imputed | 0.996 | DCBLD2---[]---AC107029.1 | TRUE | TRUE |
| 3 | 106117209 | rs12637686 | G | A | 0.721 | -4.14E-03 | 7.57E-04 | 3.50E-09 | 65680 | Imputed | 0.993 | CBLB---[]--LINC00882 | FALSE | FALSE |
| 3 | 128196500 | rs2713594 | G | A | 0.583 | -3.68E-03 | 6.94E-04 | 4.20E-08 | 65680 | Imputed | 0.983 | DNAJB8--[]-GATA2 | FALSE | FALSE |
| 3 | 134089758 | rs143351962 | C | T | 0.99 | -2.15E-02 | 3.40E-03 | 7.40E-10 | 65680 | Imputed | 1 | [AMOTL2] | TRUE | FALSE |
| 4 | 7917204 | rs34939228 | T | TA | 0.617 | -4.17E-03 | 7.05E-04 | 8.30E-09 | 65680 | Imputed | 0.984 | [AFAP1] | FALSE | FALSE |
| 4 | 54979145 | rs1158402 | C | T | 0.378 | 5.88E-03 | 7.03E-04 | 6.80E-19 | 65680 | Imputed | 0.996 | [AC058822.1] | TRUE | TRUE |
| 4 | 79396057 | 4:79396057_TC_T | TC | T | 0.367 | 4.02E-03 | 7.10E-04 | 2.00E-08 | 65680 | Imputed | 0.995 | [FRAS1] | FALSE | FALSE |
| 4 | 112399511 | rs2661764 | A | T | 0.638 | 4.03E-03 | 7.09E-04 | 7.30E-09 | 65680 | Imputed | 0.995 | PITX2---[]---FAM241A | FALSE | FALSE |
| 4 | 126407298 | rs532857051 | C | CTT | 0.708 | 6.06E-03 | 7.51E-04 | 4.70E-17 | 65680 | Imputed | 0.993 | [FAT4] | FALSE | TRUE |
| 4 | 128053375 | 4:128053375_AACAC_A | AACAC | A | 0.52 | 3.51E-03 | 6.96E-04 | 1.60E-08 | 65680 | Imputed | 0.953 | []---INTU | FALSE | FALSE |
| 5 | 3646121 | rs13165326 | C | T | 0.672 | -3.77E-03 | 7.25E-04 | 2.40E-08 | 65680 | Imputed | 0.991 | IRX1--[]---LINC02063 | FALSE | TRUE |
| 5 | 31952051 | rs72759609 | T | C | 0.898 | 1.35E-02 | 1.13E-03 | 2.40E-37 | 65680 | Imputed | 0.989 | [PDZD2] | TRUE | TRUE |
| 5 | 55578661 | rs158653 | G | A | 0.477 | 5.06E-03 | 6.85E-04 | 1.30E-14 | 65680 | Imputed | 0.992 | ANKRD55--[]---LINC01948 | TRUE | TRUE |
| 5 | 82770558 | rs11746859 | A | G | 0.539 | 4.77E-03 | 6.82E-04 | 8.90E-13 | 65680 | Imputed | 0.995 | [VCAN] | FALSE | TRUE |
| 5 | 87810199 | rs150221399 | G | A | 0.919 | -9.99E-03 | 1.28E-03 | 3.80E-15 | 65680 | Imputed | 0.95 | TMEM161B---[]---MEF2C | FALSE | TRUE |
| 5 | 121765728 | rs2570981 | T | C | 0.401 | 4.30E-03 | 6.93E-04 | 1.10E-09 | 65680 | Imputed | 0.994 | [SNCAIP] | FALSE | FALSE |
| 5 | 128931357 | rs7448395 | G | A | 0.204 | 6.62E-03 | 8.47E-04 | 8.00E-17 | 65680 | Imputed | 0.999 | [ADAMTS19] | TRUE | TRUE |
| 5 | 133393380 | 5:133393380_GA_G | GA | G | 0.844 | 7.22E-03 | 1.01E-03 | 3.30E-13 | 65680 | Imputed | 0.858 | VDAC1--[]--TCF7 | TRUE | TRUE |
| 5 | 172197790 | rs34013988 | C | T | 0.961 | 1.46E-02 | 1.75E-03 | 4.10E-18 | 65680 | Imputed | 0.998 | [AC022217.4,DUSP1] | TRUE | TRUE |
| 6 | 593289 | 6:593289_TG_T | TG | T | 0.869 | -7.47E-03 | 1.05E-03 | 2.60E-14 | 65680 | Imputed | 0.913 | [EXOC2] | FALSE | TRUE |
| 6 | 1548369 | rs2745572 | A | G | 0.666 | 4.35E-03 | 7.26E-04 | 3.10E-09 | 65680 | Imputed | 0.99 | FOXF2---[]--FOXC1 | FALSE | FALSE |
| 6 | 1983440 | rs6914444 | T | C | 0.866 | 9.30E-03 | 1.01E-03 | 9.70E-22 | 65680 | Imputed | 0.987 | [GMDS] | TRUE | TRUE |
| 6 | 7211818 | rs1334576 | G | A | 0.572 | -5.94E-03 | 6.92E-04 | 1.60E-18 | 65515 | Genotyped | 1 | [RREB1] | TRUE | TRUE |
| 6 | 11411838 | rs7742703 | C | T | 0.905 | 6.43E-03 | 1.16E-03 | 8.60E-09 | 65680 | Imputed | 0.996 | NEDD9--[]---TMEM170B | FALSE | FALSE |
| 6 | 31133577 | rs145919884 | A | AAAGCCC | 0.35 | 4.45E-03 | 7.18E-04 | 3.40E-10 | 65680 | Imputed | 0.996 | [TCF19,POU5F1] | FALSE | FALSE |
| 6 | 36552592 | rs200252984 | G | A | 0.79 | -7.25E-03 | 8.47E-04 | 9.90E-21 | 65680 | Imputed | 0.973 | STK38--[]-SRSF3 | TRUE | TRUE |
| 6 | 39537880 | rs9369128 | T | C | 0.651 | -6.39E-03 | 7.15E-04 | 5.30E-20 | 65680 | Imputed | 0.998 | [KIF6] | FALSE | TRUE |
| 6 | 122392511 | rs2684249 | T | C | 0.593 | 6.38E-03 | 6.95E-04 | 9.00E-19 | 65680 | Imputed | 0.995 | GJA1---[]---HSF2 | TRUE | TRUE |
| 6 | 126767600 | rs1361108 | C | T | 0.543 | -6.05E-03 | 6.89E-04 | 1.60E-17 | 65542 | Genotyped | 1 | CENPW--[]---RSPO3 | FALSE | FALSE |
| 6 | 148832343 | rs139973521 | A | ATGAG | 0.89 | -7.36E-03 | 1.10E-03 | 3.80E-13 | 65680 | Imputed | 0.987 | [SASH1] | FALSE | FALSE |
| 6 | 149989744 | 6:149989744_AT_A | AT | A | 0.648 | 4.99E-03 | 7.24E-04 | 2.10E-12 | 65680 | Imputed | 0.988 | [LATS1] | FALSE | FALSE |
| 6 | 151295133 | rs6900628 | A | G | 0.708 | 4.20E-03 | 7.55E-04 | 2.20E-08 | 65475 | Genotyped | 1 | [MTHFD1L] | FALSE | FALSE |
| 7 | 4767112 | rs6946034 | A | T | 0.52 | -3.68E-03 | 6.86E-04 | 3.00E-09 | 65680 | Imputed | 0.983 | [FOXK1] | FALSE | FALSE |
| 7 | 14237240 | rs10260511 | C | A | 0.842 | -7.78E-03 | 9.37E-04 | 5.90E-19 | 65597 | Genotyped | 1 | [DGKB] | TRUE | TRUE |
| 7 | 19612305 | rs2192476 | C | T | 0.372 | 4.26E-03 | 7.06E-04 | 2.00E-11 | 65680 | Imputed | 0.993 | FERD3L---[]---TWISTNB | TRUE | TRUE |
| 7 | 28393403 | rs7805378 | A | C | 0.559 | 4.60E-03 | 6.86E-04 | 1.20E-11 | 65680 | Imputed | 0.994 | [CREB5] | TRUE | TRUE |
| 7 | 28844815 | rs2282909 | T | G | 0.269 | 4.09E-03 | 7.70E-04 | 1.40E-08 | 65680 | Imputed | 0.999 | [CREB5] | FALSE | FALSE |
| 7 | 42108499 | rs2237417 | C | T | 0.593 | 3.69E-03 | 6.99E-04 | 4.00E-08 | 65680 | Imputed | 0.987 | [GLI3] | FALSE | FALSE |
| 7 | 101808020 | rs6976947 | A | G | 0.604 | 4.41E-03 | 6.97E-04 | 8.70E-12 | 65680 | Imputed | 0.997 | [CUX1] | FALSE | TRUE |
| 8 | 8254590 | rs2945880 | A | G | 0.113 | -9.34E-03 | 1.08E-03 | 2.30E-19 | 65680 | Imputed | 0.994 | PRAG1--[]---AC114550.3 | TRUE | TRUE |
| 8 | 17526359 | rs11203888 | C | T | 0.335 | -4.18E-03 | 7.24E-04 | 2.80E-09 | 65680 | Imputed | 0.995 | [MTUS1] | FALSE | FALSE |
| 8 | 30386291 | rs7013873 | C | T | 0.784 | 4.49E-03 | 8.32E-04 | 7.90E-10 | 65680 | Imputed | 0.989 | [RBPMS] | FALSE | FALSE |

| CHR | POS | SNP | EA | NEA | EAF | BETA | SE | P | NUM_INDV | SRC | INFO | GENE_CONTEXT | CRAIG | CRAIG_META |
|---|---|---|---|---|---|---|---|---|---|---|---|---|---|---|
| 8 | 61911070 | rs10957177 | A | G | 0.749 | 4.57E-03 | 7.94E-04 | 6.80E-09 | 65680 | Imputed | 0.978 | CHD7---[]--CLVS1 | FALSE | FALSE |
| 8 | 72579250 | rs10453110 | C | T | 0.874 | -9.94E-03 | 1.04E-03 | 8.50E-23 | 65680 | Imputed | 0.99 | EYA1--[]---AC104012.2 | TRUE | TRUE |
| 8 | 75522500 | rs10957731 | A | C | 0.402 | 5.12E-03 | 6.96E-04 | 3.70E-14 | 65680 | Imputed | 0.994 | [MIR2052HG] | FALSE | FALSE |
| 8 | 78948855 | rs6999835 | T | C | 0.634 | 3.93E-03 | 7.10E-04 | 2.00E-09 | 65680 | Imputed | 0.996 | []---PKIA | FALSE | FALSE |
| 8 | 88664409 | rs200382882 | G | C | 0.485 | -4.07E-03 | 6.86E-04 | 7.60E-11 | 65680 | Imputed | 0.997 | CNBD1--[]---DCAF4L2 | FALSE | FALSE |
| 8 | 131606303 | 8:131606303_CTGTT_C | CTGTT | C | 0.647 | 3.90E-03 | 7.18E-04 | 4.20E-08 | 65680 | Imputed | 0.986 | ASAP1---[]---ADCY8 | FALSE | FALSE |
| 9 | 15912375 | rs10810475 | G | C | 0.564 | 3.67E-03 | 6.87E-04 | 1.20E-08 | 65680 | Imputed | 0.998 | [CCDC171] | FALSE | FALSE |
| 9 | 16619529 | rs13290470 | A | G | 0.603 | 5.52E-03 | 6.97E-04 | 7.20E-19 | 65680 | Imputed | 0.991 | [BNC2] | FALSE | FALSE |
| 9 | 18089275 | rs10738500 | C | A | 0.927 | -9.11E-03 | 1.31E-03 | 3.90E-12 | 65680 | Imputed | 0.99 | [ADAMTSL1] | TRUE | TRUE |
| 9 | 22051670 | rs944801 | G | C | 0.428 | -1.57E-02 | 6.90E-04 | 5.30E-125 | 65680 | Imputed | 0.999 | [CDKN2B-AS1] | TRUE | TRUE |
| 9 | 76622068 | rs11143754 | C | A | 0.573 | 4.29E-03 | 6.90E-04 | 3.40E-10 | 65680 | Imputed | 0.994 | AL451127.1---[]---AL355674.1 | FALSE | FALSE |
| 9 | 89252706 | rs10512176 | T | C | 0.722 | -6.49E-03 | 7.76E-04 | 1.70E-17 | 65680 | Imputed | 0.95 | TUT7---[]---GAS1 | FALSE | TRUE |
| 9 | 134563185 | rs11793533 | G | A | 0.719 | 5.28E-03 | 7.58E-04 | 2.80E-12 | 65419 | Genotyped | 1 | [RAPGEF1] | FALSE | FALSE |
| 9 | 136145414 | rs587611953 | C | A | 0.836 | -6.71E-03 | 1.02E-03 | 4.80E-13 | 65680 | Imputed | 0.824 | [ABO] | FALSE | TRUE |
| 10 | 14086579 | rs7099081 | A | G | 0.663 | 3.85E-03 | 7.20E-04 | 2.00E-08 | 65680 | Imputed | 0.993 | [FRMD4A] | FALSE | FALSE |
| 10 | 21462896 | 10:21462896_GGC_G | GGC | G | 0.987 | -2.40E-02 | 3.08E-03 | 1.60E-15 | 65680 | Imputed | 0.943 | [NEBL] | TRUE | TRUE |
| 10 | 60271824 | rs7069916 | G | A | 0.657 | -3.95E-03 | 7.16E-04 | 5.50E-09 | 65680 | Imputed | 0.997 | TFAM---[]BICC1 | FALSE | FALSE |
| 10 | 69991853 | rs7916697 | A | G | 0.24 | -1.62E-02 | 7.98E-04 | 1.20E-98 | 65604 | Genotyped | 1 | [ATOH7] | TRUE | TRUE |
| 10 | 94974129 | rs6583871 | G | T | 0.264 | -5.05E-03 | 7.76E-04 | 1.70E-11 | 65680 | Imputed | 0.982 | CYP26A1---[]--MYOF | TRUE | TRUE |
| 10 | 96026184 | 10:96026184_CA_C | CA | C | 0.565 | -4.35E-03 | 6.91E-04 | 3.10E-10 | 65680 | Imputed | 0.995 | [PLCE1] | FALSE | TRUE |
| 10 | 118546046 | rs11197820 | G | A | 0.585 | -6.17E-03 | 6.94E-04 | 2.90E-19 | 65680 | Imputed | 0.985 | [HSPA12A] | TRUE | TRUE |
| 11 | 19960147 | rs12807015 | G | T | 0.528 | -3.98E-03 | 6.97E-04 | 8.90E-09 | 65680 | Imputed | 0.96 | [NAV2] | FALSE | FALSE |
| 11 | 31719504 | 11:31719504_CT_C | CT | C | 0.233 | 7.13E-03 | 8.10E-04 | 9.60E-21 | 65680 | Imputed | 0.992 | [ELP4] | TRUE | TRUE |
| 11 | 57656794 | rs35328629 | T | TA | 0.561 | -4.29E-03 | 6.89E-04 | 2.00E-10 | 65680 | Imputed | 0.987 | CTNND1--[]---OR9Q1 | FALSE | FALSE |
| 11 | 63678128 | rs199826712 | T | TA | 0.926 | 6.78E-03 | 1.34E-03 | 9.90E-09 | 65680 | Imputed | 0.948 | [MARK2] | FALSE | FALSE |
| 11 | 65343399 | rs547193816 | C | CCCCGCT | 0.797 | 9.19E-03 | 8.60E-04 | 2.00E-27 | 65680 | Imputed | 0.97 | FAM89B-[]EHBP1L1 | TRUE | TRUE |
| 11 | 86748437 | rs2445575 | T | C | 0.807 | 4.90E-03 | 8.62E-04 | 7.30E-10 | 65680 | Imputed | 0.997 | FZD4--[]TMEM135 | TRUE | TRUE |
| 11 | 94533444 | rs138059525 | G | A | 0.993 | 3.10E-02 | 4.07E-03 | 2.60E-14 | 65629 | Genotyped | 1 | [AMOTL1] | FALSE | FALSE |
| 11 | 95308854 | rs11021221 | T | A | 0.831 | 7.22E-03 | 9.09E-04 | 3.60E-15 | 65680 | Imputed | 0.994 | SESN3---[]---FAM76B | TRUE | TRUE |
| 11 | 130288797 | rs34248430 | A | ACCT | 0.282 | -7.71E-03 | 7.72E-04 | 4.50E-27 | 65680 | Imputed | 0.961 | [ADAMTS8] | TRUE | TRUE |
| 12 | 3353356 | rs73047017 | C | T | 0.937 | -8.15E-03 | 1.41E-03 | 6.00E-10 | 65680 | Imputed | 0.976 | [TSPAN9] | FALSE | FALSE |
| 12 | 26392080 | rs16930371 | A | G | 0.816 | 4.97E-03 | 8.80E-04 | 4.80E-09 | 65283 | Genotyped | 1 | [SSPN] | FALSE | TRUE |
| 12 | 31067490 | rs55710412 | C | T | 0.845 | 5.27E-03 | 9.65E-04 | 8.10E-09 | 65680 | Imputed | 0.948 | CAPRIN2---[]--TSPAN11 | FALSE | FALSE |
| 12 | 48157019 | rs12818241 | C | T | 0.84 | 6.29E-03 | 9.31E-04 | 1.10E-12 | 65680 | Imputed | 0.984 | [RAPGEF3,SLC48A1] | FALSE | TRUE |
| 12 | 76114872 | rs6582298 | G | A | 0.595 | 5.67E-03 | 7.23E-04 | 9.30E-16 | 65680 | Imputed | 0.901 | [AC078923.1] | TRUE | FALSE |
| 12 | 84076137 | rs55667441 | A | G | 0.558 | 1.58E-02 | 6.86E-04 | 2.50E-127 | 65680 | Imputed | 0.995 | TMTC2---[] | TRUE | TRUE |
| 12 | 91816926 | rs147377344 | C | CTTTTACG | 0.4 | 4.03E-03 | 7.03E-04 | 2.10E-08 | 65680 | Imputed | 0.977 | DCN---[]---LINC01619 | FALSE | FALSE |
| 12 | 107073242 | 12:107073242_CA_C | CA | C | 0.653 | 5.36E-03 | 7.17E-04 | 6.00E-15 | 65680 | Imputed | 0.988 | [AC079385.1,RFX4] | TRUE | TRUE |
| 12 | 108165360 | rs2111281 | A | C | 0.636 | -5.30E-03 | 7.07E-04 | 1.30E-13 | 65680 | Imputed | 0.995 | PRDM4--[]-ASCL4 | FALSE | FALSE |
| 12 | 124665773 | rs11057488 | A | G | 0.435 | -4.85E-03 | 6.87E-04 | 8.50E-14 | 65680 | Imputed | 0.994 | [RFLNA] | TRUE | TRUE |
| 13 | 25778093 | rs17081940 | A | G | 0.858 | 5.44E-03 | 9.90E-04 | 2.10E-09 | 65680 | Imputed | 0.978 | [LINC01076] | FALSE | FALSE |
| 13 | 36683268 | rs9546383 | T | C | 0.754 | -5.96E-03 | 7.95E-04 | 6.70E-16 | 65680 | Imputed | 0.992 | [DCLK1] | TRUE | TRUE |
| 13 | 51945741 | rs9535652 | G | A | 0.834 | 5.41E-03 | 9.19E-04 | 4.10E-09 | 65680 | Imputed | 0.996 | [INTS6] | FALSE | FALSE |
| 13 | 109267985 | rs10162202 | T | C | 0.722 | 6.97E-03 | 7.68E-04 | 6.10E-23 | 65680 | Imputed | 0.982 | [MYO16] | TRUE | TRUE |
| 13 | 110778747 | 13:110778747_CCTTTT_C | CCTTTT | C | 0.641 | -6.06E-03 | 7.31E-04 | 9.90E-18 | 65680 | Imputed | 0.945 | IRS2---[]--COL4A1 | TRUE | TRUE |
| 14 | 23452128 | rs3811183 | C | G | 0.6 | -4.45E-03 | 7.01E-04 | 1.30E-12 | 65680 | Imputed | 0.987 | AJUBA[]-C14orf93 | TRUE | TRUE |
| 14 | 53991705 | rs2077940 | T | C | 0.668 | 5.45E-03 | 7.24E-04 | 1.70E-14 | 65680 | Imputed | 0.997 | DDHD1---[]--AL163953.1 | TRUE | TRUE |
| 14 | 60808553 | rs10162287 | C | G | 0.696 | -7.65E-03 | 7.48E-04 | 2.60E-27 | 65680 | Imputed | 0.979 | PPM1A--[]--C14orf39 | TRUE | TRUE |
| 14 | 65081054 | rs149761305 | G | GCT | 0.83 | -7.31E-03 | 9.14E-04 | 7.40E-18 | 65680 | Imputed | 0.993 | PPP1R36--[]--PLEKHG3 | FALSE | TRUE |
| 14 | 85922578 | rs1289426 | A | G | 0.767 | -6.91E-03 | 8.15E-04 | 6.80E-16 | 65680 | Imputed | 0.974 | []--FLRT2 | TRUE | TRUE |
| 14 | 95957694 | rs11160251 | T | G | 0.703 | 4.15E-03 | 7.47E-04 | 1.90E-09 | 65680 | Imputed | 0.991 | SYNE3--[]--GLRX5 | FALSE | FALSE |
| 15 | 71882771 | rs4776562 | A | G | 0.554 | 3.95E-03 | 7.27E-04 | 1.10E-09 | 65680 | Imputed | 0.892 | [THSD4] | FALSE | TRUE |
| 15 | 74230660 | rs59755145 | G | A | 0.712 | -4.23E-03 | 7.57E-04 | 2.90E-09 | 65680 | Imputed | 0.992 | [LOXL1] | TRUE | TRUE |
| 15 | 84484384 | rs59199978 | A | G | 0.82 | -6.14E-03 | 8.91E-04 | 6.00E-14 | 65680 | Imputed | 0.995 | [ADAMTSL3] | FALSE | FALSE |
| 15 | 99458902 | rs28612945 | C | T | 0.794 | 6.20E-03 | 8.46E-04 | 7.80E-15 | 65547 | Genotyped | 1 | [IGF1R] | FALSE | TRUE |
| 15 | 101200873 | rs34222435 | C | T | 0.865 | -9.55E-03 | 1.00E-03 | 1.60E-23 | 65680 | Imputed | 0.991 | ASB7-[]---ALDH1A3 | TRUE | TRUE |
| 15 | 101751698 | rs8043304 | T | C | 0.752 | 4.35E-03 | 7.95E-04 | 1.80E-08 | 65680 | Imputed | 0.995 | [CHSY1] | FALSE | FALSE |
| 16 | 51469726 | rs8053277 | T | C | 0.303 | 1.11E-02 | 7.45E-04 | 6.90E-54 | 65680 | Imputed | 0.993 | SALL1---[]---HNRNPA1P48 | TRUE | TRUE |
| 16 | 74226221 | rs807293 | T | C | 0.636 | -4.41E-03 | 7.10E-04 | 7.70E-11 | 65680 | Imputed | 0.996 | ZFHX3---[]---PSMD7 | FALSE | FALSE |
| 16 | 86380293 | rs1687628 | T | C | 0.09 | 1.21E-02 | 1.19E-03 | 2.10E-25 | 65680 | Imputed | 0.992 | IRF8---[]---FOXF1 | TRUE | TRUE |
| 17 | 40867365 | rs115818584 | C | G | 0.984 | 1.62E-02 | 2.73E-03 | 1.70E-09 | 65680 | Imputed | 0.984 | [EZH1] | FALSE | FALSE |
| 17 | 45438886 | rs769594276 | C | CAGTG | 0.548 | -4.05E-03 | 6.88E-04 | 4.40E-11 | 65680 | Imputed | 0.997 | [EFCAB13] | FALSE | FALSE |
| 17 | 48225686 | rs4794104 | C | G | 0.838 | -5.48E-03 | 9.29E-04 | 9.00E-10 | 65680 | Imputed | 0.995 | [PPP1R9B] | TRUE | TRUE |
| 17 | 61865670 | 17:61865670_CT_C | CT | C | 0.637 | -4.25E-03 | 7.28E-04 | 2.40E-10 | 65680 | Imputed | 0.947 | [DDX42] | FALSE | FALSE |
| 17 | 65073835 | rs577377763 | C | CA | 0.642 | -4.45E-03 | 7.38E-04 | 4.50E-09 | 65680 | Imputed | 0.925 | [HELZ] | FALSE | TRUE |
| 17 | 80169426 | rs796355894 | A | AT | 0.592 | 3.96E-03 | 7.09E-04 | 1.60E-08 | 65680 | Imputed | 0.956 | [CCDC57] | FALSE | FALSE |
| 18 | 8797487 | rs569735 | C | A | 0.269 | 4.57E-03 | 7.89E-04 | 1.00E-08 | 65680 | Imputed | 0.961 | [MTCL1] | FALSE | FALSE |
| 18 | 23063159 | rs766791666 | T | TATC | 0.415 | -4.02E-03 | 7.05E-04 | 4.00E-10 | 65680 | Imputed | 0.971 | ZNF521---[]---SS18 | FALSE | FALSE |
| 18 | 34289285 | rs61735998 | G | T | 0.974 | 1.37E-02 | 2.17E-03 | 1.20E-10 | 65680 | Imputed | 1 | [FHOD3] | FALSE | FALSE |
| 18 | 56943484 | rs77759734 | C | T | 0.951 | -8.73E-03 | 1.59E-03 | 1.00E-09 | 65493 | Genotyped | 1 | [CPLX4] | FALSE | FALSE |
| 19 | 817708 | rs7250902 | A | G | 0.681 | 5.21E-03 | 7.45E-04 | 3.40E-12 | 65680 | Imputed | 0.966 | [PLPPR3] | TRUE | TRUE |
| 19 | 14616371 | rs11882319 | C | A | 0.851 | 5.31E-03 | 9.65E-04 | 6.00E-09 | 65680 | Imputed | 0.982 | GIPC1-[]-DNAJB1 | FALSE | FALSE |
| 19 | 32027330 | rs8102936 | G | A | 0.658 | 6.80E-03 | 7.26E-04 | 1.10E-19 | 65680 | Imputed | 0.987 | TSHZ3---[]---ZNF507 | TRUE | TRUE |
| 19 | 33523197 | rs73039431 | A | G | 0.955 | -1.04E-02 | 1.66E-03 | 3.70E-10 | 65680 | Imputed | 0.984 | [RHPN2] | FALSE | FALSE |

Supplementary Information - Table 3. ML-based VCDR (loci)

| CHR | POS | SNP | EA | NEA | EAF | BETA | SE | P | NUM_INDV | SRC | INFO | GENE_CONTEXT | CRAIG | CRAIG_META |
|---|---|---|---|---|---|---|---|---|---|---|---|---|---|---|
| 19 | 39146780 | rs55876653 | G | C | 0.508 | -4.15E-03 | 6.84E-04 | 7.40E-10 | 65680 | Imputed | 0.997 | [ACTN4] | FALSE | FALSE |
| 20 | 1029686 | rs4816177 | A | G | 0.821 | -5.44E-03 | 9.01E-04 | 4.10E-10 | 64868 | Genotyped | 1 | RSPO4--[]--PSMF1 | FALSE | FALSE |
| 20 | 6470094 | rs2326788 | G | A | 0.628 | 1.17E-02 | 7.07E-04 | 1.60E-65 | 65680 | Imputed | 0.992 | FERMT1---[]---LINC01713 | TRUE | TRUE |
| 20 | 31438954 | rs4911268 | A | G | 0.816 | 6.64E-03 | 8.85E-04 | 1.80E-15 | 65680 | Imputed | 0.998 | MAPRE1[]-EFCAB8 | TRUE | TRUE |
| 20 | 45797259 | rs3091590 | C | T | 0.544 | -4.29E-03 | 6.87E-04 | 1.10E-10 | 65680 | Imputed | 0.991 | [EYA2] | FALSE | FALSE |
| 22 | 29115066 | rs4822983 | C | T | 0.677 | 1.49E-02 | 7.29E-04 | 1.30E-98 | 65680 | Imputed | 0.999 | [CHEK2] | TRUE | TRUE |
| 22 | 38180407 | 22:38180407_CAA_C | CAA | C | 0.677 | -8.77E-03 | 7.31E-04 | 2.70E-36 | 65680 | Imputed | 0.992 | TRIOBP-[]--H1-0 | TRUE | TRUE |
| 22 | 46383612 | rs73175083 | C | T | 0.687 | 6.37E-03 | 7.39E-04 | 1.40E-17 | 65680 | Imputed | 0.995 | WNT7B--[]--LINC00899 | FALSE | FALSE |

| CHR | POS | SNP | P | LOCUS_IDX | N_FINEMAPPED | N_GWS | N_CAUSAL |
|---|---|---|---|---|---|---|---|
| 1 | 3049362 | rs12024620 | 4.00E-30 | 1 | 141 | 79 | 2 |
| 1 | 8468278 | rs301792 | 6.80E-13 | 2 | 508 | 56 | 1 |
| 1 | 12614029 | rs6541032 | 2.60E-14 | 3 | 160 | 51 | 1 |
| 1 | 47923058 | rs767682581 | 1.40E-10 | 4 | 505 | 20 | 1 |
| 1 | 68840797 | rs2209559 | 1.90E-28 | 5 | 1028 | 163 | 9 |
| 1 | 89253357 | rs786908 | 3.90E-10 | 6 | 683 | 127 | 1 |
| 1 | 92077097 | rs1192415 | 7.60E-102 | 7 | 1292 | 187 | 10 |
| 1 | 110627923 | rs10857812 | 2.10E-11 | 8 | 317 | 28 | 8 |
| 1 | 113045061 | rs351364 | 2.40E-08 | 9 | 568 | 2 | 7 |
| 1 | 155033308 | rs11589479 | 1.20E-08 | 10 | 924 | 1 | 9 |
| 1 | 169551682 | rs6028 | 3.10E-09 | 11 | 185 | 76 | 6 |
| 1 | 183849739 | rs41263652 | 4.20E-11 | 12 | 159 | 4 | 7 |
| 1 | 218520995 | rs6658835 | 7.60E-16 | 13 | 42 | 5 | 7 |
| 1 | 219573841 | rs796959510 | 7.40E-09 | 14 | 237 | 1 | 5 |
| 1 | 222014897 | rs11118873 | 4.20E-10 | 15 | 165 | 4 | 7 |
| 1 | 227585983 | rs6670351 | 3.30E-17 | 16 | 522 | 133 | 10 |
| 2 | 5680539 | rs7575439 | 1.50E-08 | 17 | 27 | 1 | 2 |
| 2 | 12891476 | rs730126 | 9.50E-10 | 18 | 155 | 4 | 8 |
| 2 | 19431423 | rs72778352 | 4.80E-22 | 19 | 445 | 46 | 9 |
| 2 | 56072501 | rs1430202 | 1.80E-24 | 20 | 1699 | 190 | 8 |
| 2 | 111658010 | rs4849203 | 1.30E-13 | 21 | 492 | 165 | 5 |
| 2 | 180196027 | rs12620141 | 2.50E-09 | 22 | 53 | 6 | 4 |
| 2 | 190269957 | 2:190269957_CTTTT_C | 3.00E-09 | 23 | 458 | 42 | 4 |
| 2 | 233389918 | rs2853447 | 3.80E-08 | 24 | 101 | 1 | 9 |
| 3 | 20061023 | rs35057657 | 4.30E-08 | 25 | 993 | 2 | 6 |
| 3 | 25046463 | rs12490228 | 3.50E-17 | 26 | 1530 | 386 | 7 |
| 3 | 29493916 | rs9822629 | 1.50E-08 | 27 | 392 | 5 | 6 |
| 3 | 32879823 | rs56131903 | 3.90E-18 | 28 | 217 | 54 | 6 |
| 3 | 48719638 | rs7633840 | 1.60E-11 | 29 | 2149 | 170 | 7 |
| 3 | 58130168 | rs2362911 | 3.30E-12 | 30 | 845 | 139 | 6 |
| 3 | 71182447 | rs77877421 | 1.60E-15 | 31 | 1597 | 4 | 8 |
| 3 | 88380417 | rs9852080 | 3.70E-21 | 32 | 4860 | 235 | 7 |
| 3 | 99078606 | rs1871794 | 9.40E-30 | 33 | 3389 | 909 | 6 |
| 3 | 106117209 | rs12637686 | 3.50E-09 | 34 | 444 | 4 | 9 |
| 3 | 128196500 | rs2713594 | 4.20E-08 | 35 | 11 | 0 | 7 |
| 3 | 134089758 | rs143351962 | 7.40E-10 | 36 | 125 | 1 | 9 |
| 4 | 7917204 | rs34939228 | 8.30E-09 | 37 | 619 | 64 | 7 |
| 4 | 54979145 | rs1158402 | 6.80E-19 | 38 | 532 | 225 | 8 |
| 4 | 79396057 | 4:79396057_TC_T | 2.00E-08 | 39 | 1201 | 29 | 7 |
| 4 | 112399511 | rs2661764 | 7.30E-09 | 40 | 181 | 6 | 5 |
| 4 | 126407298 | rs532857051 | 4.70E-17 | 41 | 205 | 39 | 6 |
| 4 | 128053375 | 4:128053375_AACAC_A | 1.60E-08 | 42 | 753 | 1 | 2 |
| 5 | 3646121 | rs13165326 | 2.40E-08 | 43 | 397 | 4 | 7 |
| 5 | 31952051 | rs72759609 | 2.40E-37 | 44 | 191 | 34 | 5 |
| 5 | 55578661 | rs158653 | 1.30E-14 | 45 | 786 | 124 | 10 |
| 5 | 82770558 | rs11746859 | 8.90E-13 | 46 | 107 | 28 | 7 |
| 5 | 87810199 | rs150221399 | 3.80E-15 | 47 | 412 | 26 | 9 |
| 5 | 121765728 | rs2570981 | 1.10E-09 | 48 | 169 | 3 | 5 |
| 5 | 128931357 | rs7448395 | 8.00E-17 | 49 | 1996 | 141 | 10 |
| 5 | 133393380 | 5:133393380_GA_G | 3.30E-13 | 50 | 228 | 37 | 8 |

| CHR | POS | SNP | P | LOCUS_IDX | N_FINEMAPPED | N_GWS | N_CAUSAL |
|---|---|---|---|---|---|---|---|
| 5 | 172197790 | rs34013988 | 4.10E-18 | 51 | 273 | 82 | 5 |
| 6 | 593289 | 6:593289_TG_T | 2.60E-14 | 52 | 425 | 39 | 7 |
| 6 | 1548369 | rs2745572 | 3.10E-09 | 53 | 187 | 1 | 7 |
| 6 | 1983440 | rs6914444 | 9.70E-22 | 54 | 784 | 48 | 6 |
| 6 | 7211818 | rs1334576 | 1.60E-18 | 55 | 788 | 49 | 7 |
| 6 | 11411838 | rs7742703 | 8.60E-09 | 56 | 277 | 7 | 6 |
| 6 | 31133577 | rs145919884 | 3.40E-10 | 57 | 19009 | 88 | 10 |
| 6 | 36552592 | rs200252984 | 9.90E-21 | 58 | 601 | 380 | 7 |
| 6 | 39537880 | rs9369128 | 5.30E-20 | 59 | 589 | 180 | 7 |
| 6 | 122392511 | rs2684249 | 9.00E-19 | 60 | 369 | 60 | 9 |
| 6 | 126767600 | rs1361108 | 1.60E-17 | 61 | 401 | 133 | 5 |
| 6 | 148832343 | rs139973521 | 3.80E-13 | 62 | 77 | 4 | 6 |
| 6 | 149989744 | 6:149989744_AT_A | 2.10E-12 | 63 | 532 | 297 | 8 |
| 6 | 151295133 | rs6900628 | 2.20E-08 | 64 | 132 | 5 | 3 |
| 7 | 4767112 | rs6946034 | 3.00E-09 | 65 | 57 | 12 | 5 |
| 7 | 14237240 | rs10260511 | 5.90E-19 | 66 | 113 | 51 | 6 |
| 7 | 19612305 | rs2192476 | 2.00E-11 | 67 | 1115 | 173 | 5 |
| 7 | 28393403 | rs7805378 | 1.20E-11 | 68 | 264 | 5 | 6 |
| 7 | 28844815 | rs2282909 | 1.40E-08 | 69 | 274 | 4 | 8 |
| 7 | 42108499 | rs2237417 | 4.00E-08 | 70 | 269 | 2 | 8 |
| 7 | 101808020 | rs6976947 | 8.70E-12 | 71 | 202 | 68 | 6 |
| 8 | 8254590 | rs2945880 | 2.30E-19 | 72 | 1543 | 5 | 7 |
| 8 | 17526359 | rs11203888 | 2.80E-09 | 73 | 738 | 27 | 7 |
| 8 | 30386291 | rs7013873 | 7.90E-10 | 74 | 917 | 21 | 6 |
| 8 | 61911070 | rs10957177 | 6.80E-09 | 75 | 114 | 1 | 6 |
| 8 | 72579250 | rs10453110 | 8.50E-23 | 76 | 906 | 74 | 7 |
| 8 | 75522500 | rs10957731 | 3.70E-14 | 77 | 1749 | 55 | 9 |
| 8 | 78948855 | rs6999835 | 2.00E-09 | 78 | 2040 | 124 | 9 |
| 8 | 88664409 | rs200382882 | 7.60E-11 | 79 | 1189 | 71 | 8 |
| 8 | 131606303 | 8:131606303_CTGTT_C | 4.20E-08 | 80 | 74 | 0 | 7 |
| 9 | 15912375 | rs10810475 | 1.20E-08 | 81 | 470 | 3 | 6 |
| 9 | 16619529 | rs13290470 | 7.20E-19 | 82 | 184 | 26 | 10 |
| 9 | 18089275 | rs10738500 | 3.90E-12 | 83 | 307 | 21 | 6 |
| 9 | 22051670 | rs944801 | 5.30E-125 | 84 | 220 | 58 | 7 |
| 9 | 76622068 | rs11143754 | 3.40E-10 | 85 | 606 | 9 | 8 |
| 9 | 89252706 | rs10512176 | 1.70E-17 | 86 | 530 | 32 | 6 |
| 9 | 134563185 | rs11793533 | 2.80E-12 | 87 | 735 | 111 | 8 |
| 9 | 136145414 | rs587611953 | 4.80E-13 | 88 | 668 | 78 | 7 |
| 10 | 14086579 | rs7099081 | 2.00E-08 | 89 | 451 | 4 | 7 |
| 10 | 21462896 | 10:21462896_GGC_G | 1.60E-15 | 90 | 645 | 0 | 9 |
| 10 | 60271824 | rs7069916 | 5.50E-09 | 91 | 89 | 15 | 6 |
| 10 | 69991853 | rs7916697 | 1.20E-98 | 92 | 1987 | 566 | 5 |
| 10 | 94974129 | rs6583871 | 1.70E-11 | 93 | 278 | 21 | 3 |
| 10 | 96026184 | 10:96026184_CA_C | 3.10E-10 | 94 | 943 | 22 | 10 |
| 10 | 118546046 | rs11197820 | 2.90E-19 | 95 | 457 | 59 | 6 |
| 11 | 19960147 | rs12807015 | 8.90E-09 | 96 | 109 | 1 | 7 |
| 11 | 31719504 | 11:31719504_CT_C | 9.60E-21 | 97 | 1255 | 555 | 8 |
| 11 | 57656794 | rs35328629 | 2.00E-10 | 98 | 2699 | 152 | 8 |
| 11 | 63678128 | rs199826712 | 9.90E-09 | 99 | 503 | 33 | 7 |
| 11 | 65343399 | rs547193816 | 2.00E-27 | 100 | 3174 | 192 | 7 |

| CHR | POS | SNP | P | LOCUS_IDX | N_FINEMAPPED | N_GWS | N_CAUSAL |
|---|---|---|---|---|---|---|---|
| 11 | 86748437 | rs2445575 | 7.30E-10 | 101 | 173 | 3 | 5 |
| 11 | 94533444 | rs138059525 | 2.60E-14 | 102 | 259 | 5 | 5 |
| 11 | 95308854 | rs11021221 | 3.60E-15 | 103 | 85 | 26 | 9 |
| 11 | 130288797 | rs34248430 | 4.50E-27 | 104 | 119 | 18 | 6 |
| 12 | 3353356 | rs73047017 | 6.00E-10 | 105 | 26 | 3 | 4 |
| 12 | 26392080 | rs16930371 | 4.80E-09 | 106 | 32 | 0 | 10 |
| 12 | 31067490 | rs55710412 | 8.10E-09 | 107 | 183 | 13 | 6 |
| 12 | 48157019 | rs12818241 | 1.10E-12 | 108 | 657 | 86 | 9 |
| 12 | 76114872 | rs6582298 | 9.30E-16 | 109 | 54 | 1 | 6 |
| 12 | 84076137 | rs55667441 | 2.50E-127 | 110 | 5946 | 2199 | 6 |
| 12 | 91816926 | rs147377344 | 2.10E-08 | 111 | 506 | 2 | 8 |
| 12 | 107073242 | 12:107073242_CA_C | 6.00E-15 | 112 | 1205 | 279 | 8 |
| 12 | 108165360 | rs2111281 | 1.30E-13 | 113 | 325 | 99 | 9 |
| 12 | 124665773 | rs11057488 | 8.50E-14 | 114 | 580 | 66 | 6 |
| 13 | 25778093 | rs17081940 | 2.10E-09 | 115 | 44 | 8 | 7 |
| 13 | 36683268 | rs9546383 | 6.70E-16 | 116 | 218 | 17 | 7 |
| 13 | 51945741 | rs9535652 | 4.10E-09 | 117 | 161 | 16 | 9 |
| 13 | 109267985 | rs10162202 | 6.10E-23 | 118 | 331 | 91 | 7 |
| 13 | 110778747 | 13:110778747_CCTTTT_ | 9.90E-18 | 119 | 285 | 30 | 7 |
| 14 | 23452128 | rs3811183 | 1.30E-12 | 120 | 295 | 119 | 5 |
| 14 | 53991705 | rs2077940 | 1.70E-14 | 121 | 641 | 120 | 6 |
| 14 | 60808553 | rs10162287 | 2.60E-27 | 122 | 2941 | 322 | 9 |
| 14 | 65081054 | rs149761305 | 7.40E-18 | 123 | 1229 | 74 | 6 |
| 14 | 85922578 | rs1289426 | 6.80E-16 | 124 | 1314 | 119 | 8 |
| 14 | 95957694 | rs11160251 | 1.90E-09 | 125 | 90 | 4 | 6 |
| 15 | 71882771 | rs4776562 | 1.10E-09 | 126 | 74 | 1 | 3 |
| 15 | 74230660 | rs59755145 | 2.90E-09 | 127 | 54 | 2 | 5 |
| 15 | 84484384 | rs59199978 | 6.00E-14 | 128 | 457 | 11 | 6 |
| 15 | 99458902 | rs28612945 | 7.80E-15 | 129 | 13 | 0 | 4 |
| 15 | 101200873 | rs34222435 | 1.60E-23 | 130 | 42 | 5 | 4 |
| 15 | 101751698 | rs8043304 | 1.80E-08 | 131 | 62 | 3 | 8 |
| 16 | 51469726 | rs8053277 | 6.90E-54 | 132 | 1806 | 514 | 6 |
| 16 | 74226221 | rs807293 | 7.70E-11 | 133 | 1687 | 55 | 5 |
| 16 | 86380293 | rs1687628 | 2.10E-25 | 134 | 782 | 166 | 8 |
| 17 | 40867365 | rs115818584 | 1.70E-09 | 135 | 313 | 2 | 6 |
| 17 | 45438886 | rs769594276 | 4.40E-11 | 136 | 330 | 105 | 4 |
| 17 | 48225686 | rs4794104 | 9.00E-10 | 137 | 46 | 0 | 7 |
| 17 | 61865670 | 17:61865670_CT_C | 2.40E-10 | 138 | 131 | 13 | 2 |
| 17 | 65073835 | rs577377763 | 4.50E-09 | 139 | 99 | 17 | 6 |
| 17 | 80169426 | rs796355894 | 1.60E-08 | 140 | 238 | 5 | 3 |
| 18 | 8797487 | rs569735 | 1.00E-08 | 141 | 59 | 1 | 7 |
| 18 | 23063159 | rs766791666 | 4.00E-10 | 142 | 128 | 19 | 4 |
| 18 | 34289285 | rs61735998 | 1.20E-10 | 143 | 492 | 1 | 6 |
| 18 | 56943484 | rs77759734 | 1.00E-09 | 144 | 62 | 1 | 7 |
| 19 | 817708 | rs7250902 | 3.40E-12 | 145 | 194 | 22 | 6 |
| 19 | 14616371 | rs11882319 | 6.00E-09 | 146 | 177 | 19 | 8 |
| 19 | 32027330 | rs8102936 | 1.10E-19 | 147 | 525 | 22 | 7 |
| 19 | 33523197 | rs73039431 | 3.70E-10 | 148 | 711 | 12 | 7 |
| 19 | 39146780 | rs55876653 | 7.40E-10 | 149 | 153 | 18 | 9 |
| 20 | 1029686 | rs4816177 | 4.10E-10 | 150 | 17 | 0 | 4 |

Supplementary Information - Table 5. SuSiE fine-mapping (loci)

| CHR | POS | SNP | P | LOCUS_IDX | N_FINEMAPPED | N_GWS | N_CAUSAL |
|---|---|---|---|---|---|---|---|
| 20 | 6470094 | rs2326788 | 1.60E-65 | 151 | 374 | 141 | 2 |
| 20 | 31438954 | rs4911268 | 1.80E-15 | 152 | 539 | 27 | 5 |
| 20 | 45797259 | rs3091590 | 1.10E-10 | 153 | 199 | 45 | 2 |
| 22 | 29115066 | rs4822983 | 1.30E-98 | 154 | 5927 | 1583 | 7 |
| 22 | 38180407 | 22:38180407_CAA_C | 2.70E-36 | 155 | 2128 | 745 | 8 |
| 22 | 46383612 | rs73175083 | 1.40E-17 | 156 | 233 | 36 | 8 |

| CHR | POS | SNP | EA | NEA | BETA | SE | P | GENE_CONTEXT |
|---|---|---|---|---|---|---|---|---|
| 1 | 3055876 | rs10797380 | A | G | -3.06E-02 | 5.27E-03 | 1.93E-09 | [PRDM16] |
| 1 | 3056222 | rs35271327 | A | AT | -1.18E-01 | 1.08E-02 | 2.10E-29 | [PRDM16] |
| 1 | 8486131 | rs302714 | A | C | -3.46E-02 | 5.37E-03 | 1.61E-14 | [RERE,RERE-AS1] |
| 1 | 12614029 | rs6541032 | T | C | -3.85E-02 | 5.14E-03 | 2.30E-18 | VPS13D--[]--DHRS3 |
| 1 | 47923058 | rs767682581 | C | CT | -3.22E-02 | 5.33E-03 | 1.40E-10 | FOXD2--[]---TRABD2B |
| 1 | 53565054 | rs58924052 | T | C | -6.34E-02 | 1.22E-02 | 3.70E-08 | [SLC1A7] |
| 1 | 56955386 | rs4638151 | C | T | -2.57E-02 | 5.51E-03 | 2.76E-08 | [AC119674.2] |
| 1 | 68773910 | rs34151819 | C | T | 1.34E-01 | 1.96E-02 | 6.27E-13 | WLS--[]---RPE65 |
| 1 | 68848681 | rs1925953 | A | T | -5.63E-02 | 5.20E-03 | 4.49E-34 | WLS---[]--RPE65 |
| 1 | 89295765 | rs786914 | C | A | -3.23E-02 | 5.23E-03 | 1.67E-11 | [PKN2] |
| 1 | 92021464 | rs12134245 | C | T | 3.61E-02 | 5.10E-03 | 1.21E-12 | CDC7--[]---TGFBR3 |
| 1 | 92024124 | rs75296423 | T | G | -1.34E-01 | 1.88E-02 | 3.69E-16 | CDC7--[]---TGFBR3 |
| 1 | 92039474 | rs11587658 | G | A | 3.74E-02 | 5.41E-03 | 1.61E-15 | CDC7--[]---TGFBR3 |
| 1 | 92070810 | rs77291384 | G | A | -1.33E-01 | 2.29E-02 | 4.10E-09 | CDC7--[]--TGFBR3 |
| 1 | 92077097 | rs1192415 | G | A | 1.33E-01 | 6.50E-03 | 8.55E-110 | CDC7--[]--TGFBR3 |
| 1 | 92089160 | rs17569923 | G | A | 5.34E-02 | 6.42E-03 | 1.26E-20 | CDC7--[]--TGFBR3 |
| 1 | 92114938 | rs12046642 | A | G | -5.40E-02 | 6.64E-03 | 3.28E-19 | CDC7---[]--TGFBR3 |
| 1 | 92195601 | rs10783002 | G | A | -2.44E-02 | 5.22E-03 | 4.74E-08 | [TGFBR3] |
| 1 | 103441814 | rs2061705 | A | G | 2.31E-02 | 5.13E-03 | 2.81E-08 | [COL11A1] |
| 1 | 110466338 | rs333970 | C | A | -2.79E-02 | 5.21E-03 | 3.13E-08 | [CSF1] |
| 1 | 110632536 | rs11102052 | A | T | 3.49E-02 | 5.32E-03 | 3.52E-12 | STRIP1--[]--UBL4B |
| 1 | 113046395 | rs351365 | T | C | -2.96E-02 | 5.89E-03 | 4.81E-09 | [WNT2B] |
| 1 | 155033308 | rs11589479 | G | A | 3.67E-02 | 6.91E-03 | 2.11E-08 | [ADAM15] |
| 1 | 165715300 | rs6426939 | C | T | 4.10E-02 | 7.68E-03 | 4.96E-09 | [TMCO1] |
| 1 | 167691909 | rs7548746 | A | G | 5.22E-02 | 1.08E-02 | 1.76E-08 | [MPZL1] |
| 1 | 169551682 | rs6028 | T | C | -3.04E-02 | 5.61E-03 | 1.64E-12 | [F5] |
| 1 | 183849739 | rs41263652 | G | C | 5.42E-02 | 8.45E-03 | 1.54E-11 | [RGL1] |
| 1 | 218520995 | rs6658835 | A | G | -4.50E-02 | 5.76E-03 | 5.42E-15 | [TGFB2] |
| 1 | 219451442 | rs7536147 | A | G | 2.89E-02 | 5.58E-03 | 3.26E-11 | AL360093.1--[]---ZC3H11B |
| 1 | 219648808 | rs2791546 | T | G | 2.29E-02 | 5.38E-03 | 3.92E-09 | AL360093.1---[]---ZC3H11B |
| 1 | 227585983 | rs6670351 | G | A | -5.02E-02 | 6.35E-03 | 6.82E-16 | CDC42BPA--[]---ZNF678 |
| 2 | 5680539 | rs7575439 | C | A | 3.08E-02 | 5.48E-03 | 5.18E-10 | LINC01249---[]--AC108025.1 |
| 2 | 12891476 | rs730126 | A | C | 3.00E-02 | 5.24E-03 | 1.58E-09 | TRIB2-[] |
| 2 | 19272693 | rs4380183 | G | A | -3.48E-02 | 7.18E-03 | 8.67E-09 | NT5C1B---[]---OSR1 |
| 2 | 19303748 | rs6755909 | C | T | -3.38E-02 | 6.20E-03 | 3.61E-10 | NT5C1B---[]---OSR1 |
| 2 | 19306891 | rs56202558 | C | T | -5.86E-02 | 9.80E-03 | 8.64E-11 | NT5C1B---[]---OSR1 |
| 2 | 19420060 | rs851321 | C | T | -1.98E-01 | 2.21E-02 | 6.99E-22 | NT5C1B---[]---OSR1 |
| 2 | 19479000 | rs1727198 | C | T | 4.16E-02 | 5.13E-03 | 2.49E-20 | NT5C1B---[]--OSR1 |
| 2 | 28383841 | rs10165930 | G | A | -2.90E-02 | 6.03E-03 | 1.80E-08 | [BABAM2] |
| 2 | 42509829 | rs6723361 | T | C | -3.00E-02 | 5.50E-03 | 8.98E-10 | [EML4] |
| 2 | 56013061 | rs17047234 | A | C | 5.97E-02 | 6.17E-03 | 1.34E-24 | PNPT1--[]--EFEMP1 |
| 2 | 56095994 | rs17278665 | C | G | -3.76E-02 | 6.79E-03 | 5.84E-11 | [EFEMP1] |
| 2 | 56205280 | rs6724872 | G | C | -3.09E-02 | 5.81E-03 | 1.31E-09 | EFEMP1--[]---CCDC85A |
| 2 | 111680818 | rs2880192 | A | G | 3.84E-02 | 5.37E-03 | 1.65E-16 | [ACOXL] |
| 2 | 180196027 | rs12620141 | C | A | -3.06E-02 | 5.41E-03 | 2.50E-09 | AC093911.1--[]---ZNF385B |
| 2 | 190269957 | 2:190269957_CTTTT_C | CTTTT | C | 3.13E-02 | 5.57E-03 | 3.00E-09 | COL5A2---[]--WDR75 |
| 2 | 239273949 | rs56330821 | C | T | 4.50E-02 | 9.14E-03 | 3.34E-09 | [TRAF3IP1] |
| 2 | 241931723 | rs12694992 | G | A | 2.87E-02 | 5.36E-03 | 9.14E-11 | [CROCC2] |
| 3 | 20059749 | rs2948098 | G | A | 2.99E-02 | 5.47E-03 | 2.00E-08 | PP2D1-[]--KAT2B |
| 3 | 25046477 | rs4547662 | T | G | -2.83E-02 | 5.13E-03 | 4.81E-12 | THRB-AS1---[]---RARB |
| 3 | 25059227 | rs1021702 | A | G | -3.73E-02 | 5.17E-03 | 7.86E-14 | THRB-AS1---[]---RARB |
| 3 | 25159627 | rs1604012 | A | T | -4.18E-02 | 5.14E-03 | 3.01E-20 | THRB-AS1---[]--RARB |

| CHR | POS | SNP | EA | NEA | BETA | SE | P | GENE_CONTEXT |
|---|---|---|---|---|---|---|---|---|
| 3 | 25290905 | rs34197182 | G | A | 4.74E-02 | 7.18E-03 | 3.28E-11 | [RARB] |
| 3 | 25381021 | rs11920003 | C | A | 2.60E-02 | 5.41E-03 | 1.55E-08 | [RARB] |
| 3 | 29493443 | rs1946825 | A | G | -2.78E-02 | 5.29E-03 | 2.71E-09 | [AC098650.1,RBMS3] |
| 3 | 32851635 | rs6773458 | G | A | -2.61E-02 | 5.20E-03 | 8.71E-09 | CNOT10--[]-TRIM71 |
| 3 | 32879823 | rs56131903 | A | T | 4.66E-02 | 5.53E-03 | 2.12E-21 | [TRIM71] |
| 3 | 48743342 | rs551116669 | T | TA | 3.29E-02 | 5.38E-03 | 2.80E-10 | [IP6K2] |
| 3 | 57990467 | 3:57990467_CTT_C | CTT | C | -3.09E-02 | 5.29E-03 | 2.00E-08 | SLMAP--[]-FLNB |
| 3 | 58035497 | rs12494328 | G | A | -4.28E-02 | 6.14E-03 | 5.89E-16 | [FLNB] |
| 3 | 58154671 | rs4681787 | T | C | -2.93E-02 | 5.12E-03 | 8.04E-10 | [FLNB,FLNB-AS1] |
| 3 | 70061377 | rs190948281 | G | C | 3.33E-01 | 4.52E-02 | 1.60E-13 | MITF--[]---MDFIC2 |
| 3 | 71182447 | rs77877421 | A | T | -8.71E-02 | 1.13E-02 | 1.06E-13 | [FOXP1,AC097634.4] |
| 3 | 88387796 | rs9879264 | A | G | 4.65E-02 | 5.15E-03 | 7.27E-24 | [CSNKA2IP] |
| 3 | 89603917 | rs373216501 | A | ATATT | 3.22E-02 | 5.81E-03 | 3.60E-08 | EPHA3--[] |
| 3 | 98943479 | rs13076500 | C | T | -5.07E-02 | 5.18E-03 | 1.89E-27 | DCBLD2---[]---AC107029.1 |
| 3 | 98982518 | rs114755946 | C | T | -8.57E-02 | 1.68E-02 | 4.24E-08 | DCBLD2---[]---AC107029.1 |
| 3 | 99086375 | rs34814291 | G | A | -6.42E-02 | 5.89E-03 | 2.37E-34 | DCBLD2---[]---AC107029.1 |
| 3 | 99324088 | rs140565493 | G | T | -8.33E-02 | 1.21E-02 | 6.11E-14 | AC107029.1--[]--COL8A1 |
| 3 | 99378077 | rs62281832 | G | A | 7.00E-02 | 8.39E-03 | 8.60E-20 | [COL8A1] |
| 3 | 99667565 | rs10936003 | G | A | -2.60E-02 | 5.12E-03 | 3.95E-09 | [CMSS1,FILIP1L] |
| 3 | 100625703 | rs17398137 | G | A | 6.59E-02 | 6.70E-03 | 3.22E-24 | [ABI3BP] |
| 3 | 100841979 | rs982312 | C | T | -2.38E-02 | 5.56E-03 | 1.90E-08 | ABI3BP---[]--IMPG2 |
| 3 | 106118371 | rs11424801 | C | CT | 3.08E-02 | 5.71E-03 | 5.30E-09 | CBLB---[]--LINC00882 |
| 3 | 126717964 | rs7644947 | A | G | -2.63E-02 | 5.59E-03 | 2.79E-08 | [PLXNA1] |
| 3 | 134089758 | rs143351962 | C | T | -1.62E-01 | 2.56E-02 | 2.01E-10 | [AMOTL2] |
| 4 | 7919903 | rs4696780 | A | G | -3.09E-02 | 5.32E-03 | 3.61E-11 | [AFAP1] |
| 4 | 53732586 | rs8287 | T | C | 3.21E-02 | 5.97E-03 | 2.29E-10 | [RASL11B] |
| 4 | 54979046 | rs1158401 | C | T | 4.43E-02 | 5.30E-03 | 3.68E-20 | [AC058822.1] |
| 4 | 55066077 | rs6554158 | G | A | 4.00E-02 | 6.05E-03 | 5.70E-15 | [AC058822.1] |
| 4 | 79122800 | rs17003043 | G | A | 2.89E-02 | 5.41E-03 | 4.66E-09 | [FRAS1] |
| 4 | 79396057 | 4:79396057_TC_T | TC | T | 3.03E-02 | 5.35E-03 | 2.00E-08 | [FRAS1] |
| 4 | 106911742 | rs13112725 | G | C | -3.17E-02 | 5.99E-03 | 2.96E-08 | [NPNT] |
| 4 | 112399511 | rs2661764 | A | T | 3.04E-02 | 5.34E-03 | 2.30E-09 | PITX2---[]---FAM241A |
| 4 | 112825940 | rs7677732 | A | G | 2.65E-02 | 5.44E-03 | 4.33E-08 | []---FAM241A |
| 4 | 126239986 | rs1039808 | C | T | 3.19E-02 | 5.16E-03 | 9.71E-12 | [FAT4] |
| 4 | 126399998 | rs77531977 | A | G | 4.48E-02 | 5.63E-03 | 2.19E-18 | [FAT4] |
| 4 | 128053375 | 4:128053375_AACAC_A | AACAC | A | 2.64E-02 | 5.25E-03 | 1.60E-08 | []---INTU |
| 4 | 166579647 | rs2611206 | G | A | -3.61E-02 | 7.32E-03 | 4.47E-09 | CPE---[]---TLL1 |
| 5 | 3645864 | rs13184559 | A | G | -2.82E-02 | 5.46E-03 | 4.57E-10 | IRX1--[]---LINC02063 |
| 5 | 15274048 | rs7709148 | C | T | 2.64E-02 | 5.20E-03 | 8.61E-09 | LINC02149-[]---FBXL7 |
| 5 | 31952051 | rs72759609 | T | C | 1.02E-01 | 8.52E-03 | 1.68E-40 | [PDZD2] |
| 5 | 55578661 | rs158653 | G | A | 3.81E-02 | 5.16E-03 | 1.26E-17 | ANKRD55--[]---LINC01948 |
| 5 | 55701667 | rs140212185 | C | CTTTTTTT | -2.98E-02 | 5.17E-03 | 6.40E-10 | ANKRD55---[]--LINC01948 |
| 5 | 55744230 | rs30372 | T | C | -3.88E-02 | 6.08E-03 | 9.60E-14 | ANKRD55---[]-LINC01948 |
| 5 | 55783715 | rs76051791 | C | A | -5.75E-02 | 9.32E-03 | 3.15E-11 | LINC01948-[]--C5orf67 |
| 5 | 82742118 | rs12188947 | A | C | 3.53E-02 | 5.23E-03 | 3.15E-17 | XRCC4--[]--VCAN |
| 5 | 87826536 | rs56755309 | T | C | -6.74E-02 | 8.98E-03 | 1.51E-19 | TMEM161B---[]---MEF2C |
| 5 | 121768585 | rs304380 | G | A | 3.19E-02 | 5.22E-03 | 5.47E-10 | [SNCAIP] |
| 5 | 125209500 | rs10044084 | A | T | -2.72E-02 | 5.45E-03 | 2.75E-09 | []---GRAMD2B |
| 5 | 125345974 | rs10075656 | A | C | -3.26E-02 | 6.03E-03 | 6.52E-10 | []---GRAMD2B |
| 5 | 128881629 | rs4276452 | C | T | -3.37E-02 | 5.43E-03 | 2.20E-11 | [ADAMTS19] |
| 5 | 129054770 | rs32819 | A | G | 8.94E-02 | 1.14E-02 | 2.79E-20 | [ADAMTS19] |
| 5 | 131466629 | rs3843503 | T | A | 3.20E-02 | 5.23E-03 | 1.57E-10 | CSF2--[]--AC063976.1 |

| CHR | POS | SNP | EA | NEA | BETA | SE | P | GENE_CONTEXT |
|---|---|---|---|---|---|---|---|---|
| 5 | 133384337 | rs247463 | T | G | 3.09E-02 | 6.16E-03 | 8.62E-10 | VDAC1--[]--TCF7 |
| 5 | 133411871 | rs187380 | C | T | 5.94E-02 | 8.63E-03 | 1.73E-15 | VDAC1--[]--TCF7 |
| 5 | 146925367 | rs7715946 | A | G | 3.91E-02 | 6.94E-03 | 3.11E-12 | DPYSL3--[]--JAKMIP2 |
| 5 | 172197790 | rs34013988 | C | T | 1.10E-01 | 1.32E-02 | 1.27E-23 | [AC022217.4,DUSP1] |
| 6 | 619600 | rs1150856 | A | C | -4.61E-02 | 6.81E-03 | 7.94E-17 | [EXOC2] |
| 6 | 1548369 | rs2745572 | A | G | 3.28E-02 | 5.47E-03 | 1.55E-13 | FOXF2---[]--FOXC1 |
| 6 | 1983440 | rs6914444 | T | C | 7.01E-02 | 7.59E-03 | 1.77E-22 | [GMDS] |
| 6 | 7205796 | rs4960295 | G | A | -4.37E-02 | 5.21E-03 | 3.53E-24 | [RREB1] |
| 6 | 11411838 | rs7742703 | C | T | 4.85E-02 | 8.73E-03 | 1.64E-11 | NEDD9--[]---TMEM170B |
| 6 | 31108306 | rs7771067 | G | A | 3.92E-02 | 6.82E-03 | 4.84E-09 | PSORS1C1[]-CCHCR1 |
| 6 | 31133577 | rs145919884 | A | AAAGCC( | 3.35E-02 | 5.41E-03 | 3.40E-10 | [TCF19,POU5F1] |
| 6 | 32627155 | rs9273402 | G | A | 2.87E-02 | 5.20E-03 | 3.24E-08 | HLA-DQA1--[]HLA-DQB1 |
| 6 | 36552592 | rs200252984 | G | A | -5.46E-02 | 6.38E-03 | 9.90E-21 | STK38--[]-SRSF3 |
| 6 | 39531474 | rs9369127 | T | A | -4.80E-02 | 5.42E-03 | 3.46E-20 | [KIF6] |
| 6 | 74850240 | rs7770032 | A | C | 3.49E-02 | 6.92E-03 | 2.34E-08 | CD109---[]---COL12A1 |
| 6 | 75348855 | rs2485070 | A | T | -3.17E-02 | 7.11E-03 | 4.65E-09 | CD109---[]---COL12A1 |
| 6 | 121939112 | rs10457423 | C | G | 3.71E-02 | 7.25E-03 | 8.59E-10 | GJA1---[]---HSF2 |
| 6 | 122392511 | rs2684249 | T | C | 4.81E-02 | 5.24E-03 | 7.95E-25 | GJA1---[]---HSF2 |
| 6 | 126730543 | rs576049 | T | G | -4.48E-02 | 5.19E-03 | 9.04E-19 | CENPW--[]---RSPO3 |
| 6 | 127298008 | rs769528910 | A | ACTG | -3.41E-02 | 5.19E-03 | 4.60E-11 | CENPW---[]---RSPO3 |
| 6 | 148832343 | rs139973521 | A | ATGAG | -5.54E-02 | 8.25E-03 | 3.80E-13 | [SASH1] |
| 6 | 149979416 | rs1125 | G | A | 3.76E-02 | 5.47E-03 | 7.56E-14 | [LATS1] |
| 6 | 151295133 | rs6900628 | A | G | 3.17E-02 | 5.69E-03 | 4.82E-09 | [MTHFD1L] |
| 7 | 4780514 | rs3087749 | G | T | -2.74E-02 | 5.16E-03 | 1.13E-08 | [FOXK1] |
| 7 | 14237240 | rs10260511 | C | A | -5.86E-02 | 7.06E-03 | 1.52E-23 | [DGKB] |
| 7 | 19624489 | rs574793622 | A | AT | -3.24E-02 | 5.37E-03 | 2.00E-11 | FERD3L---[]---TWISTNB |
| 7 | 28393403 | rs7805378 | A | C | 3.47E-02 | 5.17E-03 | 7.25E-14 | [CREB5] |
| 7 | 28854950 | rs6964597 | T | A | 2.97E-02 | 5.77E-03 | 7.04E-09 | [CREB5] |
| 7 | 42117040 | rs2072201 | A | T | 2.82E-02 | 5.28E-03 | 9.74E-11 | [GLI3] |
| 7 | 101777382 | rs201530 | A | G | 3.11E-02 | 5.14E-03 | 1.90E-14 | [CUX1] |
| 7 | 116140931 | rs28503222 | G | C | -3.03E-02 | 6.77E-03 | 1.64E-08 | [CAV2] |
| 7 | 117635382 | rs2188836 | C | T | -2.42E-02 | 5.24E-03 | 4.02E-08 | CTTNBP2---[]-AC003084.1 |
| 8 | 8254590 | rs2945880 | A | G | -7.03E-02 | 8.14E-03 | 1.81E-21 | PRAG1--[]---AC114550.3 |
| 8 | 17526359 | rs11203888 | C | T | -3.15E-02 | 5.45E-03 | 4.36E-10 | [MTUS1] |
| 8 | 30276920 | rs4338062 | C | T | -2.57E-02 | 5.22E-03 | 2.79E-08 | [RBPMS] |
| 8 | 30336017 | rs571194397 | A | AT | -3.82E-02 | 6.87E-03 | 1.70E-08 | [RBPMS] |
| 8 | 30445960 | rs79527387 | T | C | 3.95E-02 | 7.71E-03 | 9.11E-10 | [GTF2E2] |
| 8 | 61911070 | rs10957177 | A | G | 3.45E-02 | 5.98E-03 | 2.83E-12 | CHD7---[]--CLVS1 |
| 8 | 72278010 | rs12543430 | T | C | -4.58E-02 | 5.36E-03 | 5.21E-18 | [EYA1] |
| 8 | 72392687 | rs10093418 | A | G | 5.26E-02 | 8.42E-03 | 1.50E-11 | [EYA1] |
| 8 | 72579250 | rs10453110 | C | T | -7.49E-02 | 7.81E-03 | 7.27E-24 | EYA1--[]---AC104012.2 |
| 8 | 75517928 | 8:75517928_TC_T | TC | T | 5.85E-02 | 1.00E-02 | 1.20E-09 | [MIR2052HG] |
| 8 | 75519048 | 8:75519048_TTAAAA_T | TTAAAA | T | 3.73E-02 | 5.30E-03 | 1.18E-13 | [MIR2052HG] |
| 8 | 75927930 | rs28567420 | A | C | -3.21E-02 | 6.26E-03 | 2.36E-09 | [CRISPLD1] |
| 8 | 76025955 | rs11991447 | C | T | 3.61E-02 | 6.01E-03 | 1.37E-11 | CRISPLD1--[]---HNF4G |
| 8 | 78945804 | rs10646223 | A | AAC | 2.92E-02 | 5.36E-03 | 2.50E-09 | []---PKIA |
| 8 | 88761223 | rs12547416 | C | T | 2.97E-02 | 5.16E-03 | 2.53E-13 | CNBD1---[]---DCAF4L2 |
| 8 | 131636781 | rs4565471 | C | T | 2.58E-02 | 5.16E-03 | 1.30E-10 | ASAP1---[]---ADCY8 |
| 8 | 143765414 | rs2920293 | C | G | 2.24E-02 | 5.16E-03 | 1.67E-09 | PSCA-[]--LY6K |
| 9 | 16619529 | rs13290470 | A | G | 4.16E-02 | 5.25E-03 | 3.34E-16 | [BNC2] |
| 9 | 18089832 | rs78542921 | T | A | -8.83E-02 | 1.32E-02 | 2.43E-13 | [ADAMTSL1] |
| 9 | 21941315 | rs74903566 | G | A | 6.26E-02 | 1.27E-02 | 1.56E-08 | [AL359922.1] |

| CHR | POS | SNP | EA | NEA | BETA | SE | P | GENE_CONTEXT |
|---|---|---|---|---|---|---|---|---|
| 9 | 21951175 | rs117197971 | A | G | -7.73E-02 | 1.21E-02 | 2.50E-12 | [AL359922.1] |
| 9 | 22007330 | rs3217978 | C | A | 1.16E-01 | 2.14E-02 | 3.06E-09 | [AL359922.1,CDKN2B-AS1,CDKN2B] |
| 9 | 22044904 | rs74744824 | A | G | -1.39E-01 | 2.26E-02 | 2.09E-09 | [CDKN2B-AS1] |
| 9 | 22051670 | rs944801 | G | C | -1.18E-01 | 5.20E-03 | 5.11E-144 | [CDKN2B-AS1] |
| 9 | 22052068 | rs62560775 | A | G | -5.89E-02 | 8.52E-03 | 1.16E-12 | [CDKN2B-AS1] |
| 9 | 22082375 | rs1547705 | A | C | -6.28E-02 | 8.12E-03 | 1.15E-19 | [CDKN2B-AS1] |
| 9 | 22090416 | rs10965230 | C | T | 8.10E-02 | 1.22E-02 | 9.25E-11 | [CDKN2B-AS1] |
| 9 | 76020565 | rs80355279 | G | A | 3.19E-02 | 6.66E-03 | 1.21E-08 | ANXA1---[]---AL451127.1 |
| 9 | 76622068 | rs11143754 | C | A | 3.23E-02 | 5.20E-03 | 1.18E-09 | AL451127.1---[]---AL355674.1 |
| 9 | 89252706 | rs10512176 | T | C | -4.89E-02 | 5.85E-03 | 3.32E-20 | TUT7---[]---GAS1 |
| 9 | 89380805 | rs1111066 | C | G | 2.79E-02 | 5.16E-03 | 4.04E-09 | TUT7---[]---GAS1 |
| 9 | 89819377 | rs11141703 | T | C | -3.54E-02 | 6.66E-03 | 1.44E-09 | GAS1---[]---DAPK1 |
| 9 | 134572638 | rs35424590 | A | G | 3.88E-02 | 5.69E-03 | 5.82E-13 | [RAPGEF1] |
| 9 | 136145414 | rs587611953 | C | A | -5.06E-02 | 7.69E-03 | 4.80E-13 | [ABO] |
| 10 | 21437861 | rs190927291 | C | G | -1.60E-01 | 2.18E-02 | 7.04E-16 | [NEBL] |
| 10 | 60338753 | rs4141671 | T | C | -2.73E-02 | 5.16E-03 | 8.71E-11 | [BICC1] |
| 10 | 62074139 | rs1471246 | G | A | 2.28E-02 | 5.25E-03 | 1.49E-09 | [ANK3] |
| 10 | 63641670 | rs2588924 | A | G | -2.32E-02 | 5.16E-03 | 2.26E-08 | CABCOCO1---[]--ARID5B |
| 10 | 69264839 | rs6480262 | C | T | -3.54E-02 | 8.02E-03 | 3.32E-08 | [CTNNA3] |
| 10 | 69838913 | rs117479359 | G | T | 6.45E-02 | 9.33E-03 | 5.02E-15 | HERC4-[]--MYPN |
| 10 | 69879366 | rs113337354 | A | G | 4.61E-02 | 7.75E-03 | 2.27E-09 | [MYPN] |
| 10 | 69902549 | rs3814180 | T | C | -2.96E-02 | 5.36E-03 | 5.79E-12 | [MYPN] |
| 10 | 69926319 | rs61854624 | C | A | 4.93E-02 | 7.01E-03 | 7.84E-16 | [MYPN] |
| 10 | 69938965 | rs10997980 | C | T | 8.73E-02 | 1.08E-02 | 5.11E-20 | [MYPN] |
| 10 | 69952035 | rs10998007 | C | A | -3.76E-02 | 8.98E-03 | 4.34E-08 | [MYPN] |
| 10 | 69974599 | 10:69974599_CG_C | CG | C | -4.84E-02 | 6.05E-03 | 9.90E-19 | MYPN-[]--ATOH7 |
| 10 | 69991853 | rs7916697 | A | G | -1.22E-01 | 6.01E-03 | 2.14E-131 | [ATOH7] |
| 10 | 70007008 | rs1900020 | A | G | 6.20E-02 | 1.62E-02 | 8.10E-11 | ATOH7--[]--PBLD |
| 10 | 70019201 | rs117897884 | C | T | 1.34E-01 | 1.77E-02 | 3.33E-19 | ATOH7--[]--PBLD |
| 10 | 70041695 | rs374085072 | G | A | 1.62E-01 | 2.87E-02 | 1.80E-09 | ATOH7--[]PBLD |
| 10 | 70171957 | rs10998137 | G | A | 6.94E-02 | 1.32E-02 | 1.43E-12 | RUFY2-[]-DNA2 |
| 10 | 70206657 | rs367873689 | C | CA | 1.10E-01 | 1.29E-02 | 1.30E-18 | [DNA2] |
| 10 | 70208719 | rs200774593 | T | C | 7.64E-02 | 1.50E-02 | 1.60E-08 | [DNA2] |
| 10 | 70399109 | rs7095472 | A | G | -4.40E-02 | 5.14E-03 | 6.87E-21 | [TET1] |
| 10 | 70774039 | rs7078237 | A | T | 3.47E-02 | 5.68E-03 | 7.93E-12 | [KIFBP] |
| 10 | 70794817 | rs561566466 | G | GAA | -5.57E-02 | 1.08E-02 | 1.50E-08 | [KIFBP] |
| 10 | 94950713 | rs17108260 | A | G | -3.32E-02 | 5.20E-03 | 5.30E-15 | CYP26A1---[]---MYOF |
| 10 | 94951081 | rs61861121 | G | A | 4.74E-02 | 9.24E-03 | 7.44E-09 | CYP26A1---[]---MYOF |
| 10 | 96012950 | rs7080472 | G | T | -3.30E-02 | 5.22E-03 | 5.86E-17 | [PLCE1] |
| 10 | 96071561 | rs2077218 | G | A | 3.78E-02 | 6.04E-03 | 2.79E-13 | [PLCE1] |
| 10 | 98967596 | rs4919084 | G | A | -2.73E-02 | 5.21E-03 | 2.73E-09 | [ARHGAP19-SLIT1] |
| 10 | 112028766 | rs7077557 | T | C | 3.13E-02 | 6.27E-03 | 2.56E-10 | [MXI1] |
| 10 | 118562545 | rs1637553 | T | G | -5.69E-02 | 1.03E-02 | 3.62E-10 | [HSPA12A] |
| 10 | 118563329 | rs1681739 | C | T | -4.57E-02 | 5.24E-03 | 2.03E-23 | [HSPA12A] |
| 10 | 118918956 | rs72840231 | A | T | -3.49E-02 | 5.24E-03 | 3.00E-11 | [MIR3663HG] |
| 11 | 19960147 | rs12807015 | G | T | -3.00E-02 | 5.25E-03 | 3.20E-10 | [NAV2] |
| 11 | 31570861 | rs34618943 | T | A | -5.27E-02 | 6.07E-03 | 1.30E-21 | [ELP4] |
| 11 | 31807524 | rs3026401 | C | T | -2.95E-02 | 6.31E-03 | 9.33E-09 | [ELP4,PAX6] |
| 11 | 33406776 | rs3898926 | T | C | -2.80E-02 | 5.13E-03 | 1.60E-09 | [KIAA1549L] |
| 11 | 57544484 | rs17455626 | T | C | -3.13E-02 | 5.17E-03 | 2.81E-11 | [AP001931.2,AP001931.1,CTNND1] |
| 11 | 58413910 | rs1938598 | T | C | 3.14E-02 | 5.99E-03 | 4.94E-08 | [GLYAT] |
| 11 | 63678128 | rs199826712 | T | TA | 5.11E-02 | 1.01E-02 | 9.90E-09 | [MARK2] |

Supplementary Information - Table 6. ML-based meta VCDR (hits)

| CHR | POS | SNP | EA | NEA | BETA | SE | P | GENE_CONTEXT |
|---|---|---|---|---|---|---|---|---|
| 11 | 65091708 | 11:65091708_AGTGT_A | AGTGT | A | 3.12E-02 | 5.70E-03 | 1.60E-08 | [AP000944.5] |
| 11 | 65146496 | rs585210 | G | C | -2.48E-02 | 5.15E-03 | 4.30E-09 | [SLC25A45] |
| 11 | 65240979 | 11:65240979_TAA_T | TAA | T | -7.38E-02 | 1.40E-02 | 1.30E-08 | NEAT1--[]--SCYL1 |
| 11 | 65326154 | rs12789028 | G | A | 6.75E-02 | 6.45E-03 | 9.77E-40 | [LTBP3] |
| 11 | 86670842 | rs11234891 | A | G | 3.01E-02 | 5.54E-03 | 1.78E-09 | FZD4-[]--TMEM135 |
| 11 | 86740573 | rs4944662 | C | T | -3.69E-02 | 6.70E-03 | 8.11E-12 | FZD4--[]-TMEM135 |
| 11 | 94533444 | rs138059525 | G | A | 2.34E-01 | 3.07E-02 | 2.60E-14 | [AMOTL1] |
| 11 | 95292922 | rs11021217 | G | A | 4.06E-02 | 5.40E-03 | 1.49E-16 | SESN3---[]---FAM76B |
| 11 | 100645211 | rs7123718 | G | C | 4.18E-02 | 7.95E-03 | 3.66E-09 | [ARHGAP42] |
| 11 | 100650559 | rs588615 | C | A | 2.44E-02 | 5.19E-03 | 5.82E-09 | [ARHGAP42] |
| 11 | 130262144 | rs12806983 | G | A | -6.10E-02 | 1.08E-02 | 3.83E-11 | ZBTB44--[]--ADAMTS8 |
| 11 | 130280725 | rs4936099 | C | A | -5.33E-02 | 5.25E-03 | 7.80E-32 | [ADAMTS8] |
| 12 | 3364640 | rs147867843 | A | ACTTTCT | -6.95E-02 | 1.28E-02 | 3.70E-09 | [TSPAN9] |
| 12 | 26392080 | rs16930371 | A | G | 3.74E-02 | 6.63E-03 | 2.20E-11 | [SSPN] |
| 12 | 31065843 | rs200103122 | A | AAAAT | 3.73E-02 | 7.09E-03 | 1.80E-08 | CAPRIN2---[]--TSPAN11 |
| 12 | 43548638 | rs1399377 | G | A | 1.84E-02 | 5.13E-03 | 2.34E-08 | PRICKLE1---[]---ADAMTS20 |
| 12 | 48011244 | rs679789 | C | T | -2.68E-02 | 5.29E-03 | 9.07E-10 | PCED1B---[]--RPAP3 |
| 12 | 48153944 | rs12426774 | T | C | 4.69E-02 | 7.02E-03 | 3.81E-16 | [RAPGEF3,SLC48A1] |
| 12 | 76114872 | rs6582298 | G | A | 4.27E-02 | 5.44E-03 | 4.21E-14 | [AC078923.1] |
| 12 | 83589688 | rs117090733 | C | A | 4.95E-02 | 9.86E-03 | 2.41E-08 | TMTC2--[] |
| 12 | 83604120 | rs61931863 | A | G | 6.67E-02 | 1.28E-02 | 1.70E-08 | TMTC2--[] |
| 12 | 83673711 | rs11115686 | G | A | 3.49E-02 | 5.15E-03 | 9.08E-15 | TMTC2---[] |
| 12 | 83856540 | rs12826083 | G | A | -5.66E-02 | 7.55E-03 | 1.26E-15 | TMTC2---[] |
| 12 | 83858003 | rs76005250 | G | A | 8.02E-02 | 1.12E-02 | 4.23E-15 | TMTC2---[] |
| 12 | 83861261 | rs77725841 | A | G | 6.41E-02 | 9.37E-03 | 4.33E-14 | TMTC2---[] |
| 12 | 83928366 | rs17653396 | C | T | 6.21E-02 | 1.01E-02 | 1.38E-10 | TMTC2---[] |
| 12 | 84011749 | rs11115933 | C | G | -6.53E-02 | 9.46E-03 | 1.94E-13 | TMTC2---[] |
| 12 | 84049853 | rs10506895 | G | A | 1.18E-01 | 5.16E-03 | 4.41E-133 | TMTC2---[] |
| 12 | 84132842 | rs1380758 | A | G | 5.58E-02 | 9.38E-03 | 3.49E-11 | TMTC2---[] |
| 12 | 84135457 | rs142121892 | C | CA | 5.65E-02 | 5.57E-03 | 6.30E-27 | TMTC2---[] |
| 12 | 84154996 | rs71450946 | A | G | -6.57E-02 | 9.87E-03 | 1.49E-14 | TMTC2---[] |
| 12 | 84225749 | rs200654546 | C | CT | 9.65E-02 | 1.83E-02 | 2.68E-15 | TMTC2---[] |
| 12 | 84225935 | rs7137822 | A | T | 5.46E-02 | 5.82E-03 | 8.50E-23 | TMTC2---[] |
| 12 | 84237124 | 12:84237124_CTAA_C | CTAA | C | 6.89E-02 | 1.18E-02 | 4.50E-11 | TMTC2---[] |
| 12 | 84359892 | rs12423401 | C | A | -7.10E-02 | 8.93E-03 | 4.02E-16 | TMTC2---[]---SLC6A15 |
| 12 | 84404719 | rs4772012 | T | C | 3.50E-02 | 6.11E-03 | 2.90E-11 | TMTC2---[]---SLC6A15 |
| 12 | 84890667 | rs71445008 | C | CT | -3.97E-02 | 5.34E-03 | 4.10E-16 | []---SLC6A15 |
| 12 | 91816926 | rs147377344 | C | CTTTTAC | 3.04E-02 | 5.30E-03 | 2.10E-08 | DCN---[]---LINC01619 |
| 12 | 106716441 | rs17218455 | C | T | 3.44E-02 | 6.80E-03 | 2.46E-08 | [TCP11L2] |
| 12 | 107250252 | rs17038814 | A | G | 5.54E-02 | 7.59E-03 | 5.13E-18 | [RIC8B] |
| 12 | 108134273 | rs4964616 | T | A | -4.00E-02 | 5.32E-03 | 2.26E-13 | [PRDM4] |
| 12 | 108988757 | rs17040818 | G | T | -2.97E-02 | 6.52E-03 | 3.21E-08 | [TMEM119] |
| 12 | 109874230 | rs2075432 | A | G | 2.45E-02 | 5.15E-03 | 1.82E-09 | [MYO1H] |
| 12 | 111800258 | rs3809272 | G | A | -2.82E-02 | 5.59E-03 | 9.62E-09 | [PHETA1] |
| 12 | 124666527 | rs7134138 | A | G | -3.64E-02 | 5.16E-03 | 3.64E-18 | [RFLNA] |
| 13 | 25766614 | rs9507473 | G | C | -4.22E-02 | 7.88E-03 | 6.88E-09 | AMER2--[]LINC01076 |
| 13 | 36683268 | rs9546383 | T | C | -4.49E-02 | 5.99E-03 | 1.50E-19 | [DCLK1] |
| 13 | 51913708 | rs9535646 | C | T | 4.00E-02 | 6.98E-03 | 4.01E-11 | [SERPINE3] |
| 13 | 109264870 | rs139237435 | A | ACATTTA | 5.18E-02 | 5.88E-03 | 1.20E-21 | [MYO16] |
| 13 | 110718555 | rs12875868 | A | G | -2.80E-02 | 6.03E-03 | 4.37E-08 | IRS2---[]--COL4A1 |
| 13 | 110778747 | 13:110778747_CCTTTT_C | CCTTTT | C | -4.57E-02 | 5.51E-03 | 9.90E-18 | IRS2---[]--COL4A1 |
| 14 | 23452128 | rs3811183 | C | G | -3.35E-02 | 5.28E-03 | 7.26E-13 | AJUBA[]-C14orf93 |

Supplementary Information - Table 6. ML-based meta VCDR (hits)

| CHR | POS | SNP | EA | NEA | BETA | SE | P | GENE_CONTEXT |
|---|---|---|---|---|---|---|---|---|
| 14 | 53989952 | rs11623384 | C | T | 4.10E-02 | 5.45E-03 | 4.10E-17 | DDHD1---[]--AL163953.1 |
| 14 | 59583906 | rs61985994 | C | G | 3.79E-02 | 7.39E-03 | 1.88E-08 | DACT1---[]--DAAM1 |
| 14 | 60806759 | rs7493429 | A | C | -5.70E-02 | 5.62E-03 | 2.59E-35 | PPM1A--[]--C14orf39 |
| 14 | 60914325 | rs139811951 | G | A | 4.77E-02 | 7.81E-03 | 1.43E-11 | [C14orf39] |
| 14 | 61238781 | rs12147818 | G | A | 2.17E-01 | 3.21E-02 | 8.10E-13 | [MNAT1] |
| 14 | 65074869 | rs8006017 | A | G | -5.53E-02 | 6.90E-03 | 1.23E-19 | PPP1R36--[]--PLEKHG3 |
| 14 | 85668732 | rs11626115 | G | A | -7.93E-02 | 1.19E-02 | 8.52E-10 | []---FLRT2 |
| 14 | 85862281 | rs34662716 | T | TAC | 3.25E-02 | 5.43E-03 | 6.30E-10 | []---FLRT2 |
| 14 | 85922578 | rs1289426 | A | G | -5.21E-02 | 6.14E-03 | 2.69E-18 | []--FLRT2 |
| 14 | 86021748 | rs2018653 | A | G | -4.40E-02 | 6.03E-03 | 3.06E-17 | [FLRT2] |
| 14 | 95957694 | rs11160251 | T | G | 3.13E-02 | 5.63E-03 | 2.53E-08 | SYNE3--[]--GLRX5 |
| 15 | 71840327 | rs35194812 | T | C | -3.73E-02 | 7.07E-03 | 2.81E-11 | [THSD4] |
| 15 | 71886234 | rs11632300 | T | C | 2.92E-02 | 5.55E-03 | 1.37E-09 | [THSD4] |
| 15 | 71945128 | rs17797245 | T | C | 2.65E-02 | 5.52E-03 | 5.32E-09 | [THSD4] |
| 15 | 74228391 | rs4077284 | A | G | 3.07E-02 | 5.35E-03 | 1.41E-11 | [LOXL1] |
| 15 | 84484384 | rs59199978 | A | G | -4.63E-02 | 6.71E-03 | 3.11E-11 | [ADAMTSL3] |
| 15 | 99458902 | rs28612945 | C | T | 4.67E-02 | 6.38E-03 | 1.54E-15 | [IGF1R] |
| 15 | 101200962 | rs11452536 | T | TA | -2.92E-02 | 5.36E-03 | 1.50E-09 | ASB7-[]---ALDH1A3 |
| 15 | 101201604 | rs4299136 | G | C | -7.17E-02 | 7.55E-03 | 7.59E-29 | ASB7-[]---ALDH1A3 |
| 15 | 101753394 | rs28623369 | T | G | 3.27E-02 | 5.99E-03 | 5.90E-10 | [CHSY1] |
| 16 | 51188433 | rs2052284 | T | A | -3.46E-02 | 5.27E-03 | 1.20E-11 | SALL1-[]---HNRNPA1P48 |
| 16 | 51341412 | rs117537696 | T | G | -1.30E-01 | 1.86E-02 | 1.38E-15 | SALL1---[]---HNRNPA1P48 |
| 16 | 51401342 | rs58577768 | C | T | -1.19E-01 | 1.55E-02 | 1.05E-16 | SALL1---[]---HNRNPA1P48 |
| 16 | 51416004 | rs1111196 | A | T | -7.16E-02 | 5.88E-03 | 5.35E-42 | SALL1---[]---HNRNPA1P48 |
| 16 | 51469726 | rs8053277 | T | C | 8.35E-02 | 5.61E-03 | 6.28E-65 | SALL1---[]---HNRNPA1P48 |
| 16 | 51470928 | rs71386559 | C | T | 1.02E-01 | 1.34E-02 | 1.33E-19 | SALL1---[]---HNRNPA1P48 |
| 16 | 51500120 | rs1080791 | A | G | 5.60E-02 | 6.00E-03 | 9.00E-22 | SALL1---[]--HNRNPA1P48 |
| 16 | 51562293 | rs537414151 | C | G | 1.48E-01 | 2.50E-02 | 2.20E-10 | SALL1---[]--HNRNPA1P48 |
| 16 | 51568425 | rs62039775 | G | T | -1.04E-01 | 8.62E-03 | 3.41E-40 | SALL1---[]--HNRNPA1P48 |
| 16 | 51649796 | 16:51649796_CTCTT_C | CTCTT | C | -3.55E-02 | 5.62E-03 | 8.90E-10 | [HNRNPA1P48] |
| 16 | 51667131 | rs8057507 | T | C | -5.22E-02 | 5.16E-03 | 1.44E-26 | [HNRNPA1P48] |
| 16 | 51854068 | rs11861489 | T | C | -3.47E-02 | 5.42E-03 | 1.60E-09 | HNRNPA1P48---[]---TOX3 |
| 16 | 74279778 | rs4889487 | G | C | 3.22E-02 | 5.72E-03 | 1.05E-11 | ZFHX3---[]--PSMD7 |
| 16 | 74456717 | rs199895842 | T | G | -8.15E-02 | 1.50E-02 | 3.60E-08 | [AC009053.4] |
| 16 | 74465514 | rs11648326 | C | A | 3.76E-02 | 7.46E-03 | 4.21E-09 | [AC009053.4] |
| 16 | 86382120 | rs1687626 | A | G | 2.95E-02 | 5.30E-03 | 2.42E-09 | IRF8---[]---FOXF1 |
| 16 | 86386675 | rs1728368 | C | T | 9.05E-02 | 8.98E-03 | 4.66E-28 | IRF8---[]---FOXF1 |
| 16 | 86439374 | rs12935509 | G | A | -6.15E-02 | 9.42E-03 | 1.42E-11 | IRF8---[]---FOXF1 |
| 16 | 86465590 | rs13332095 | G | A | -4.76E-02 | 8.54E-03 | 2.23E-10 | IRF8---[]--FOXF1 |
| 16 | 86511858 | rs7187191 | T | G | 4.58E-02 | 6.67E-03 | 1.24E-16 | IRF8---[]--FOXF1 |
| 17 | 10026855 | rs12936070 | C | T | 2.69E-02 | 5.93E-03 | 9.97E-10 | [GAS7] |
| 17 | 40867365 | rs115818584 | C | G | 1.22E-01 | 2.06E-02 | 4.01E-11 | [EZH1] |
| 17 | 45703433 | rs7220935 | C | T | 2.81E-02 | 5.15E-03 | 1.84E-11 | NPEPPS-[]--KPNB1 |
| 17 | 48225686 | rs4794104 | C | G | -4.13E-02 | 7.00E-03 | 6.55E-13 | [PPP1R9B] |
| 17 | 55419687 | rs792401 | G | A | 2.71E-02 | 5.60E-03 | 2.91E-08 | [MSI2] |
| 17 | 55564211 | rs277065 | G | A | 3.32E-02 | 7.30E-03 | 3.95E-08 | [MSI2] |
| 17 | 61865670 | 17:61865670_CT_C | CT | C | -3.20E-02 | 5.48E-03 | 2.40E-10 | [DDX42] |
| 17 | 65264966 | rs12939113 | C | T | -3.25E-02 | 5.70E-03 | 6.93E-11 | HELZ--[]--PSMD12 |
| 17 | 79602063 | rs9905786 | G | T | -2.52E-02 | 5.36E-03 | 2.56E-09 | [NPLOC4] |
| 17 | 80169426 | rs796355894 | A | AT | 2.99E-02 | 5.34E-03 | 1.60E-08 | [CCDC57] |
| 18 | 8799828 | rs568267 | C | T | 3.34E-02 | 5.92E-03 | 3.69E-10 | [MTCL1] |
| 18 | 23063159 | rs766791666 | T | TATC | -3.03E-02 | 5.31E-03 | 4.00E-10 | ZNF521---[]---SS18 |

| CHR | POS | SNP | EA | NEA | BETA | SE | P | GENE_CONTEXT |
|---|---|---|---|---|---|---|---|---|
| 18 | 34289285 | rs61735998 | G | T | 1.03E-01 | 1.64E-02 | 7.55E-12 | [FHOD3] |
| 18 | 56943484 | rs77759734 | C | T | -6.58E-02 | 1.20E-02 | 2.92E-08 | [CPLX4] |
| 19 | 817708 | rs7250902 | A | G | 3.93E-02 | 5.61E-03 | 3.65E-15 | [PLPPR3] |
| 19 | 14639064 | rs112614575 | C | CT | 3.96E-02 | 7.30E-03 | 6.10E-09 | [DNAJB1,TECR] |
| 19 | 32027330 | rs8102936 | G | A | 5.12E-02 | 5.47E-03 | 1.85E-21 | TSHZ3---[]---ZNF507 |
| 19 | 33477716 | 19:33477716_AAT_A | AAT | A | -4.52E-02 | 8.08E-03 | 1.40E-09 | [RHPN2] |
| 19 | 39195302 | rs757940594 | A | AGGAG | -3.15E-02 | 5.16E-03 | 7.40E-10 | [ACTN4] |
| 19 | 46356548 | rs7258364 | T | C | -2.64E-02 | 5.51E-03 | 2.14E-08 | [SYMPK] |
| 19 | 47455315 | rs311384 | A | G | 2.51E-02 | 5.66E-03 | 2.42E-08 | [ARHGAP35] |
| 20 | 1029686 | rs4816177 | A | G | -4.10E-02 | 6.79E-03 | 1.83E-08 | RSPO4--[]--PSMF1 |
| 20 | 6137310 | rs6076954 | A | G | 4.62E-02 | 5.71E-03 | 1.01E-19 | FERMT1--[]---LINC01713 |
| 20 | 6230570 | rs6133302 | C | T | -4.58E-02 | 7.33E-03 | 1.29E-13 | FERMT1---[]---LINC01713 |
| 20 | 6292315 | rs6076968 | C | T | 2.78E-02 | 5.18E-03 | 7.65E-10 | FERMT1---[]---LINC01713 |
| 20 | 6411069 | rs4815897 | A | G | -4.63E-02 | 6.57E-03 | 2.31E-12 | FERMT1---[]---LINC01713 |
| 20 | 6470094 | rs2326788 | G | A | 8.85E-02 | 5.33E-03 | 1.10E-67 | FERMT1---[]---LINC01713 |
| 20 | 6474916 | rs11483156 | C | CT | -5.47E-02 | 8.27E-03 | 4.80E-11 | FERMT1---[]---LINC01713 |
| 20 | 6514692 | rs6038531 | G | A | -1.45E-01 | 2.05E-02 | 4.81E-15 | FERMT1---[]---LINC01713 |
| 20 | 6535065 | rs78004679 | G | A | 8.75E-02 | 1.11E-02 | 1.81E-16 | FERMT1---[]---LINC01713 |
| 20 | 6544738 | rs73077173 | C | T | 7.42E-02 | 1.30E-02 | 3.44E-08 | FERMT1---[]---LINC01713 |
| 20 | 6631055 | rs1358805 | A | G | -4.30E-02 | 6.79E-03 | 1.42E-12 | FERMT1---[]--LINC01713 |
| 20 | 6650790 | rs6054446 | G | C | -7.35E-02 | 1.38E-02 | 9.81E-10 | FERMT1---[]--LINC01713 |
| 20 | 6759115 | rs235768 | A | T | -3.83E-02 | 5.28E-03 | 1.82E-15 | [BMP2] |
| 20 | 31157394 | rs4911242 | A | T | 3.17E-02 | 5.54E-03 | 2.02E-10 | [NOL4L] |
| 20 | 31438954 | rs4911268 | A | G | 5.00E-02 | 6.67E-03 | 5.16E-17 | MAPRE1[]-EFCAB8 |
| 20 | 45796660 | rs2903940 | A | G | -3.24E-02 | 5.16E-03 | 3.31E-13 | [EYA2] |
| 21 | 29506261 | rs6516818 | T | A | 2.43E-02 | 5.22E-03 | 4.48E-08 | LINC01673---[]---N6AMT1 |
| 22 | 28208528 | rs11704137 | C | G | -5.16E-02 | 5.91E-03 | 6.91E-22 | MN1--[]--PITPNB |
| 22 | 28324924 | rs62235636 | C | T | 9.94E-02 | 9.97E-03 | 3.47E-27 | [TTC28-AS1] |
| 22 | 28629713 | rs16986177 | C | T | -6.80E-02 | 7.02E-03 | 2.75E-31 | [TTC28] |
| 22 | 28653727 | 22:28653727_CATAT_C | CATAT | C | 3.55E-02 | 5.55E-03 | 7.50E-11 | [TTC28] |
| 22 | 28990300 | rs117456789 | C | T | 7.28E-02 | 1.20E-02 | 1.03E-10 | [TTC28] |
| 22 | 29032115 | rs71316851 | A | ATT | -3.45E-02 | 5.19E-03 | 7.60E-11 | [TTC28] |
| 22 | 29051261 | rs73170612 | A | C | 9.70E-02 | 1.18E-02 | 1.16E-20 | [TTC28] |
| 22 | 29063037 | rs542574575 | C | T | 3.46E-02 | 5.43E-03 | 2.00E-11 | [TTC28] |
| 22 | 29098417 | rs112564028 | A | G | -7.69E-02 | 1.85E-02 | 6.73E-09 | [CHEK2] |
| 22 | 29100301 | rs527965516 | T | A | 2.26E-01 | 1.79E-02 | 5.40E-41 | [CHEK2] |
| 22 | 29115066 | rs4822983 | C | T | 1.12E-01 | 5.49E-03 | 1.37E-110 | [CHEK2] |
| 22 | 29127402 | rs8184952 | T | C | 6.06E-02 | 7.06E-03 | 2.43E-19 | [CHEK2] |
| 22 | 29162191 | rs145346186 | A | C | 9.01E-02 | 1.26E-02 | 8.20E-12 | HSCB-[]-CCDC117 |
| 22 | 29276230 | 22:29276230_AT_A | AT | A | 4.93E-02 | 7.76E-03 | 6.00E-12 | Z93930.2--[]-ZNRF3 |
| 22 | 29441350 | 22:29441350_CA_C | CA | C | -3.90E-02 | 5.63E-03 | 1.60E-13 | [ZNRF3] |
| 22 | 30442880 | rs79955051 | G | A | -7.92E-02 | 1.60E-02 | 9.95E-09 | [HORMAD2-AS1] |
| 22 | 30503827 | rs17648370 | A | G | 9.23E-02 | 1.91E-02 | 1.51E-08 | [HORMAD2] |
| 22 | 30570022 | rs1003342 | A | G | 4.80E-02 | 5.16E-03 | 3.50E-27 | [HORMAD2] |
| 22 | 30606986 | rs9614164 | C | T | 3.75E-02 | 6.25E-03 | 2.87E-10 | AC002378.1-[]--LIF |
| 22 | 30620627 | rs6006405 | A | G | -5.65E-02 | 7.09E-03 | 1.75E-16 | AC002378.1--[]--LIF |
| 22 | 30649229 | rs73166584 | C | T | -3.85E-02 | 6.82E-03 | 1.32E-09 | LIF-[]-OSM |
| 22 | 30653991 | rs9620961 | C | T | 2.83E-02 | 5.35E-03 | 1.12E-08 | LIF--[]-OSM |
| 22 | 37819985 | rs133726 | T | G | -3.82E-02 | 7.08E-03 | 8.78E-10 | [ELFN2] |
| 22 | 37872262 | rs75110580 | G | A | -9.21E-02 | 1.40E-02 | 8.60E-12 | [MFNG] |
| 22 | 37904251 | rs2235334 | T | C | -2.56E-02 | 5.16E-03 | 2.40E-08 | [CARD10] |
| 22 | 37907069 | rs2092172 | G | A | -7.49E-02 | 6.21E-03 | 5.81E-40 | [CARD10] |

Supplementary Information - Table 6. ML-based meta VCDR (hits)

| CHR | POS | SNP | EA | NEA | BETA | SE | P | GENE_CONTEXT |
|---|---|---|---|---|---|---|---|---|
| 22 | 37925332 | rs549756240 | A | T | -5.60E-02 | 9.78E-03 | 1.90E-08 | CARD10-[]--CDC42EP1 |
| 22 | 37939510 | rs9610795 | G | T | -4.87E-02 | 5.42E-03 | 4.03E-21 | CARD10--[]--CDC42EP1 |
| 22 | 38057338 | rs9622678 | A | G | -3.75E-02 | 7.00E-03 | 1.90E-09 | [Z83844.3,PDXP] |
| 22 | 38076063 | rs62236673 | G | A | -2.60E-02 | 5.48E-03 | 1.91E-08 | LGALS1[]-NOL12 |
| 22 | 38177004 | rs12166106 | T | C | 3.71E-02 | 7.11E-03 | 6.65E-09 | TRIOBP-[]--H1-0 |
| 22 | 38180407 | 22:38180407_CAA_C | CAA | C | -6.60E-02 | 5.51E-03 | 2.70E-36 | TRIOBP-[]--H1-0 |
| 22 | 38594126 | rs147906180 | C | CAAAAA | -4.98E-02 | 5.51E-03 | 1.20E-20 | [PLA2G6] |
| 22 | 38674541 | rs8184979 | G | A | -3.93E-02 | 7.24E-03 | 2.81E-08 | TMEM184B-[]--CSNK1E |
| 22 | 39322264 | rs9306330 | C | T | -3.43E-02 | 6.67E-03 | 2.19E-09 | AL022318.1-[]--APOBEC3A |
| 22 | 46376985 | rs77164166 | G | A | 4.73E-02 | 5.55E-03 | 3.90E-17 | WNT7B-[]--LINC00899 |
| 22 | 46381414 | rs62228064 | G | A | -4.09E-02 | 7.25E-03 | 1.79E-09 | WNT7B-[]--LINC00899 |

| CHR | POS | SNP | EA | NEA | BETA | SE | P | GENE_CONTEXT | CRAIG | CRAIG_META |
|---|---|---|---|---|---|---|---|---|---|---|
| 1 | 3056222 | rs35271327 | A | AT | -1.18E-01 | 1.08E-02 | 2.10E-29 | [PRDM16] | TRUE | TRUE |
| 1 | 8486131 | rs302714 | A | C | -3.46E-02 | 5.37E-03 | 1.61E-14 | [RERE,RERE-AS1] | FALSE | TRUE |
| 1 | 12614029 | rs6541032 | T | C | -3.85E-02 | 5.14E-03 | 2.30E-18 | VPS13D--[]--DHRS3 | TRUE | TRUE |
| 1 | 47923058 | rs767682581 | C | CT | -3.22E-02 | 5.33E-03 | 1.40E-10 | FOXD2--[]---TRABD2B | FALSE | FALSE |
| 1 | 53565054 | rs58924052 | T | C | -6.34E-02 | 1.22E-02 | 3.70E-08 | [SLC1A7] | FALSE | FALSE |
| 1 | 56955386 | rs4638151 | C | T | -2.57E-02 | 5.51E-03 | 2.76E-08 | [AC119674.2] | FALSE | FALSE |
| 1 | 68848681 | rs1925953 | A | T | -5.63E-02 | 5.20E-03 | 4.49E-34 | WLS---[]--RPE65 | TRUE | TRUE |
| 1 | 89295765 | rs786914 | C | A | -3.23E-02 | 5.23E-03 | 1.67E-11 | [PKN2] | FALSE | TRUE |
| 1 | 92077097 | rs1192415 | G | A | 1.33E-01 | 6.50E-03 | 8.55E-110 | CDC7--[]--TGFBR3 | TRUE | TRUE |
| 1 | 103441814 | rs2061705 | A | G | 2.31E-02 | 5.13E-03 | 2.81E-08 | [COL11A1] | FALSE | FALSE |
| 1 | 110632536 | rs11102052 | A | T | 3.49E-02 | 5.32E-03 | 3.52E-12 | STRIP1--[]--UBL4B | FALSE | FALSE |
| 1 | 113046395 | rs351365 | T | C | -2.96E-02 | 5.89E-03 | 4.81E-09 | [WNT2B] | FALSE | FALSE |
| 1 | 155033308 | rs11589479 | G | A | 3.67E-02 | 6.91E-03 | 2.11E-08 | [ADAM15] | FALSE | FALSE |
| 1 | 165715300 | rs6426939 | C | T | 4.10E-02 | 7.68E-03 | 4.96E-09 | [TMCO1] | FALSE | FALSE |
| 1 | 167691909 | rs7548746 | A | G | 5.22E-02 | 1.08E-02 | 1.76E-08 | [MPZL1] | FALSE | FALSE |
| 1 | 169551682 | rs6028 | T | C | -3.04E-02 | 5.61E-03 | 1.64E-12 | [F5] | FALSE | TRUE |
| 1 | 183849739 | rs41263652 | G | C | 5.42E-02 | 8.45E-03 | 1.54E-11 | [RGL1] | FALSE | FALSE |
| 1 | 218520995 | rs6658835 | A | G | -4.50E-02 | 5.76E-03 | 5.42E-15 | [TGFB2] | TRUE | TRUE |
| 1 | 219451442 | rs7536147 | A | G | 2.89E-02 | 5.58E-03 | 3.26E-11 | AL360093.1--[]---ZC3H11B | FALSE | FALSE |
| 1 | 227585983 | rs6670351 | G | A | -5.02E-02 | 6.35E-03 | 6.82E-16 | CDC42BPA--[]---ZNF678 | TRUE | TRUE |
| 2 | 5680539 | rs7575439 | C | A | 3.08E-02 | 5.48E-03 | 5.18E-10 | LINC01249---[]--AC108025.1 | FALSE | FALSE |
| 2 | 12891476 | rs730126 | A | C | 3.00E-02 | 5.24E-03 | 1.58E-09 | TRIB2-[] | FALSE | FALSE |
| 2 | 19420060 | rs851321 | C | T | -1.98E-01 | 2.21E-02 | 6.99E-22 | NT5C1B---[]---OSR1 | FALSE | TRUE |
| 2 | 28383841 | rs10165930 | G | A | -2.90E-02 | 6.03E-03 | 1.80E-08 | [BABAM2] | FALSE | FALSE |
| 2 | 42509829 | rs6723361 | T | C | -3.00E-02 | 5.50E-03 | 8.98E-10 | [EML4] | FALSE | FALSE |
| 2 | 56013061 | rs17047234 | A | C | 5.97E-02 | 6.17E-03 | 1.34E-24 | PNPT1--[]--EFEMP1 | TRUE | TRUE |
| 2 | 111680818 | rs2880192 | A | G | 3.84E-02 | 5.37E-03 | 1.65E-16 | [ACOXL] | TRUE | TRUE |
| 2 | 180196027 | rs12620141 | C | A | -3.06E-02 | 5.41E-03 | 2.50E-09 | AC093911.1--[]---ZNF385B | FALSE | FALSE |
| 2 | 190269957 | 2:190269957_CTTTT_C | CTTTT | C | 3.13E-02 | 5.57E-03 | 3.00E-09 | COL5A2---[]--WDR75 | FALSE | FALSE |
| 2 | 239273949 | rs56330821 | C | T | 4.50E-02 | 9.14E-03 | 3.34E-09 | [TRAF3IP1] | FALSE | TRUE |
| 2 | 241931723 | rs12694992 | G | A | 2.87E-02 | 5.36E-03 | 9.14E-11 | [CROCC2] | FALSE | FALSE |
| 3 | 20059749 | rs2948098 | G | A | 2.99E-02 | 5.47E-03 | 2.00E-08 | PP2D1-[]--KAT2B | FALSE | FALSE |
| 3 | 25159627 | rs1604012 | A | T | -4.18E-02 | 5.14E-03 | 3.01E-20 | THRB-AS1---[]--RARB | TRUE | TRUE |
| 3 | 29493443 | rs1946825 | A | G | -2.78E-02 | 5.29E-03 | 2.71E-09 | [AC098650.1,RBMS3] | FALSE | FALSE |
| 3 | 32879823 | rs56131903 | A | T | 4.66E-02 | 5.53E-03 | 2.12E-21 | [TRIM71] | TRUE | TRUE |
| 3 | 48743342 | rs551116669 | T | TA | 3.29E-02 | 5.38E-03 | 2.80E-10 | [IP6K2] | FALSE | FALSE |
| 3 | 58035497 | rs12494328 | G | A | -4.28E-02 | 6.14E-03 | 5.89E-16 | [FLNB] | FALSE | TRUE |
| 3 | 71182447 | rs77877421 | A | T | -8.71E-02 | 1.13E-02 | 1.06E-13 | [FOXP1,AC097634.4] | FALSE | FALSE |
| 3 | 88387796 | rs9879264 | A | G | 4.65E-02 | 5.15E-03 | 7.27E-24 | [CSNKA2IP] | TRUE | TRUE |
| 3 | 99086375 | rs34814291 | G | A | -6.42E-02 | 5.89E-03 | 2.37E-34 | DCBLD2---[]---AC107029.1 | TRUE | TRUE |
| 3 | 106118371 | rs11424801 | C | CT | 3.08E-02 | 5.71E-03 | 5.30E-09 | CBLB---[]--LINC00882 | FALSE | FALSE |
| 3 | 126717964 | rs7644947 | A | G | -2.63E-02 | 5.59E-03 | 2.79E-08 | [PLXNA1] | FALSE | FALSE |
| 3 | 134089758 | rs143351962 | C | T | -1.62E-01 | 2.56E-02 | 2.01E-10 | [AMOTL2] | TRUE | FALSE |
| 4 | 7919903 | rs4696780 | A | G | -3.09E-02 | 5.32E-03 | 3.61E-11 | [AFAP1] | FALSE | FALSE |
| 4 | 53732586 | rs8287 | T | C | 3.21E-02 | 5.97E-03 | 2.29E-10 | [RASL11B] | FALSE | FALSE |
| 4 | 54979046 | rs1158401 | C | T | 4.43E-02 | 5.30E-03 | 3.68E-20 | [AC058822.1] | TRUE | TRUE |
| 4 | 79122800 | rs17003043 | G | A | 2.89E-02 | 5.41E-03 | 4.66E-09 | [FRAS1] | FALSE | FALSE |
| 4 | 106911742 | rs13112725 | G | C | -3.17E-02 | 5.99E-03 | 2.96E-08 | [NPNT] | FALSE | FALSE |
| 4 | 112399511 | rs2661764 | A | T | 3.04E-02 | 5.34E-03 | 2.30E-09 | PITX2---[]---FAM241A | FALSE | FALSE |
| 4 | 126399998 | rs77531977 | A | G | 4.48E-02 | 5.63E-03 | 2.19E-18 | [FAT4] | FALSE | TRUE |
| 4 | 128053375 | 4:128053375_AACAC_A | AACAC | A | 2.64E-02 | 5.25E-03 | 1.60E-08 | []---INTU | FALSE | FALSE |
| 4 | 166579647 | rs2611206 | G | A | -3.61E-02 | 7.32E-03 | 4.47E-09 | CPE---[]---TLL1 | FALSE | TRUE |
| 5 | 3645864 | rs13184559 | A | G | -2.82E-02 | 5.46E-03 | 4.57E-10 | IRX1--[]---LINC02063 | FALSE | TRUE |
| 5 | 15274048 | rs7709148 | C | T | 2.64E-02 | 5.20E-03 | 8.61E-09 | LINC02149-[]---FBXL7 | FALSE | FALSE |
| 5 | 31952051 | rs72759609 | T | C | 1.02E-01 | 8.52E-03 | 1.68E-40 | [PDZD2] | TRUE | TRUE |
| 5 | 55578661 | rs158653 | G | A | 3.81E-02 | 5.16E-03 | 1.26E-17 | ANKRD55--[]---LINC01948 | TRUE | TRUE |
| 5 | 82742118 | rs12188947 | A | C | 3.53E-02 | 5.23E-03 | 3.15E-17 | XRCC4--[]--VCAN | FALSE | TRUE |
| 5 | 87826536 | rs56755309 | T | C | -6.74E-02 | 8.98E-03 | 1.51E-19 | TMEM161B---[]---MEF2C | FALSE | TRUE |
| 5 | 121768585 | rs304380 | G | A | 3.19E-02 | 5.22E-03 | 5.47E-10 | [SNCAIP] | FALSE | FALSE |
| 5 | 125345974 | rs10075656 | A | C | -3.26E-02 | 6.03E-03 | 6.52E-10 | []---GRAMD2B | FALSE | FALSE |

Supplementary Information - Table 7. ML-based meta VCDR (loci)

| CHR | POS | SNP | EA | NEA | BETA | SE | P | GENE_CONTEXT | CRAIG | CRAIG_META |
|---|---|---|---|---|---|---|---|---|---|---|
| 5 | 129054770 | rs32819 | A | G | 8.94E-02 | 1.14E-02 | 2.79E-20 | [ADAMTS19] | TRUE | TRUE |
| 5 | 133411871 | rs187380 | C | T | 5.94E-02 | 8.63E-03 | 1.73E-15 | VDAC1--[]--TCF7 | TRUE | TRUE |
| 5 | 146925367 | rs7715946 | A | G | 3.91E-02 | 6.94E-03 | 3.11E-12 | DPYSL3--[]--JAKMIP2 | FALSE | FALSE |
| 5 | 172197790 | rs34013988 | C | T | 1.10E-01 | 1.32E-02 | 1.27E-23 | [AC022217.4,DUSP1] | TRUE | TRUE |
| 6 | 619600 | rs1150856 | A | C | -4.61E-02 | 6.81E-03 | 7.94E-17 | [EXOC2] | FALSE | TRUE |
| 6 | 1548369 | rs2745572 | A | G | 3.28E-02 | 5.47E-03 | 1.55E-13 | FOXF2---[]--FOXC1 | FALSE | FALSE |
| 6 | 1983440 | rs6914444 | T | C | 7.01E-02 | 7.59E-03 | 1.77E-22 | [GMDS] | TRUE | TRUE |
| 6 | 7205796 | rs4960295 | G | A | -4.37E-02 | 5.21E-03 | 3.53E-24 | [RREB1] | TRUE | TRUE |
| 6 | 11411838 | rs7742703 | C | T | 4.85E-02 | 8.73E-03 | 1.64E-11 | NEDD9--[]---TMEM170B | FALSE | FALSE |
| 6 | 31133577 | rs145919884 | A | AAAGCCC | 3.35E-02 | 5.41E-03 | 3.40E-10 | [TCF19,POU5F1] | FALSE | FALSE |
| 6 | 36552592 | rs200252984 | G | A | -5.46E-02 | 6.38E-03 | 9.90E-21 | STK38--[]-SRSF3 | TRUE | TRUE |
| 6 | 39531474 | rs9369127 | T | A | -4.80E-02 | 5.42E-03 | 3.46E-20 | [KIF6] | FALSE | TRUE |
| 6 | 75348855 | rs2485070 | A | T | -3.17E-02 | 7.11E-03 | 4.65E-09 | CD109---[]---COL12A1 | FALSE | FALSE |
| 6 | 122392511 | rs2684249 | T | C | 4.81E-02 | 5.24E-03 | 7.95E-25 | GJA1---[]---HSF2 | TRUE | TRUE |
| 6 | 126730543 | rs576049 | T | G | -4.48E-02 | 5.19E-03 | 9.04E-19 | CENPW--[]---RSPO3 | FALSE | FALSE |
| 6 | 148832343 | rs139973521 | A | ATGAG | -5.54E-02 | 8.25E-03 | 3.80E-13 | [SASH1] | FALSE | FALSE |
| 6 | 149979416 | rs1125 | G | A | 3.76E-02 | 5.47E-03 | 7.56E-14 | [LATS1] | FALSE | FALSE |
| 6 | 151295133 | rs6900628 | A | G | 3.17E-02 | 5.69E-03 | 4.82E-09 | [MTHFD1L] | FALSE | FALSE |
| 7 | 4780514 | rs3087749 | G | T | -2.74E-02 | 5.16E-03 | 1.13E-08 | [FOXK1] | FALSE | FALSE |
| 7 | 14237240 | rs10260511 | C | A | -5.86E-02 | 7.06E-03 | 1.52E-23 | [DGKB] | TRUE | TRUE |
| 7 | 19624489 | rs574793622 | A | AT | -3.24E-02 | 5.37E-03 | 2.00E-11 | FERD3L---[]---TWISTNB | TRUE | TRUE |
| 7 | 28393403 | rs7805378 | A | C | 3.47E-02 | 5.17E-03 | 7.25E-14 | [CREB5] | TRUE | TRUE |
| 7 | 28854950 | rs6964597 | T | A | 2.97E-02 | 5.77E-03 | 7.04E-09 | [CREB5] | FALSE | FALSE |
| 7 | 42117040 | rs2072201 | A | T | 2.82E-02 | 5.28E-03 | 9.74E-11 | [GLI3] | FALSE | FALSE |
| 7 | 101777382 | rs201530 | A | G | 3.11E-02 | 5.14E-03 | 1.90E-14 | [CUX1] | FALSE | TRUE |
| 7 | 116140931 | rs28503222 | G | C | -3.03E-02 | 6.77E-03 | 1.64E-08 | [CAV2] | FALSE | FALSE |
| 7 | 117635382 | rs2188836 | C | T | -2.42E-02 | 5.24E-03 | 4.02E-08 | CTTNBP2---[]-AC003084.1 | FALSE | FALSE |
| 8 | 8254590 | rs2945880 | A | G | -7.03E-02 | 8.14E-03 | 1.81E-21 | PRAG1--[]---AC114550.3 | TRUE | TRUE |
| 8 | 17526359 | rs11203888 | C | T | -3.15E-02 | 5.45E-03 | 4.36E-10 | [MTUS1] | FALSE | FALSE |
| 8 | 30445960 | rs79527387 | T | C | 3.95E-02 | 7.71E-03 | 9.11E-10 | [GTF2E2] | FALSE | FALSE |
| 8 | 61911070 | rs10957177 | A | G | 3.45E-02 | 5.98E-03 | 2.83E-12 | CHD7---[]--CLVS1 | FALSE | FALSE |
| 8 | 72579250 | rs10453110 | C | T | -7.49E-02 | 7.81E-03 | 7.27E-24 | EYA1--[]---AC104012.2 | TRUE | TRUE |
| 8 | 75519048 | 8:75519048_TTAAAA_T | TTAAAA | T | 3.73E-02 | 5.30E-03 | 1.18E-13 | [MIR2052HG] | FALSE | FALSE |
| 8 | 78945804 | rs10646223 | A | AAC | 2.92E-02 | 5.36E-03 | 2.50E-09 | []---PKIA | FALSE | FALSE |
| 8 | 88761223 | rs12547416 | C | T | 2.97E-02 | 5.16E-03 | 2.53E-13 | CNBD1---[]---DCAF4L2 | FALSE | FALSE |
| 8 | 131636781 | rs4565471 | C | T | 2.58E-02 | 5.16E-03 | 1.30E-10 | ASAP1---[]---ADCY8 | FALSE | FALSE |
| 8 | 143765414 | rs2920293 | C | G | 2.24E-02 | 5.16E-03 | 1.67E-09 | PSCA-[]--LY6K | FALSE | FALSE |
| 9 | 16619529 | rs13290470 | A | G | 4.16E-02 | 5.25E-03 | 3.34E-16 | [BNC2] | FALSE | FALSE |
| 9 | 18089832 | rs78542921 | T | A | -8.83E-02 | 1.32E-02 | 2.43E-13 | [ADAMTSL1] | TRUE | TRUE |
| 9 | 22051670 | rs944801 | G | C | -1.18E-01 | 5.20E-03 | 5.11E-144 | [CDKN2B-AS1] | TRUE | TRUE |
| 9 | 76622068 | rs11143754 | C | A | 3.23E-02 | 5.20E-03 | 1.18E-09 | AL451127.1---[]---AL355674.1 | FALSE | FALSE |
| 9 | 89252706 | rs10512176 | T | C | -4.89E-02 | 5.85E-03 | 3.32E-20 | TUT7---[]---GAS1 | FALSE | TRUE |
| 9 | 134572638 | rs35424590 | A | G | 3.88E-02 | 5.69E-03 | 5.82E-13 | [RAPGEF1] | FALSE | FALSE |
| 9 | 136145414 | rs587611953 | C | A | -5.06E-02 | 7.69E-03 | 4.80E-13 | [ABO] | FALSE | TRUE |
| 10 | 21437861 | rs190927291 | C | G | -1.60E-01 | 2.18E-02 | 7.04E-16 | [NEBL] | TRUE | TRUE |
| 10 | 60338753 | rs4141671 | T | C | -2.73E-02 | 5.16E-03 | 8.71E-11 | [BICC1] | FALSE | FALSE |
| 10 | 62074139 | rs1471246 | G | A | 2.28E-02 | 5.25E-03 | 1.49E-09 | [ANK3] | FALSE | FALSE |
| 10 | 63641670 | rs2588924 | A | G | -2.32E-02 | 5.16E-03 | 2.26E-08 | CABCOCO1---[]--ARID5B | FALSE | FALSE |
| 10 | 69991853 | rs7916697 | A | G | -1.22E-01 | 6.01E-03 | 2.14E-131 | [ATOH7] | TRUE | TRUE |
| 10 | 94950713 | rs17108260 | A | G | -3.32E-02 | 5.20E-03 | 5.30E-15 | CYP26A1---[]---MYOF | TRUE | TRUE |
| 10 | 96012950 | rs7080472 | G | T | -3.30E-02 | 5.22E-03 | 5.86E-17 | [PLCE1] | FALSE | TRUE |
| 10 | 98967596 | rs4919084 | G | A | -2.73E-02 | 5.21E-03 | 2.73E-09 | [ARHGAP19-SLIT1] | FALSE | FALSE |
| 10 | 112028766 | rs7077557 | T | C | 3.13E-02 | 6.27E-03 | 2.56E-10 | [MXI1] | FALSE | FALSE |
| 10 | 118563329 | rs1681739 | C | T | -4.57E-02 | 5.24E-03 | 2.03E-23 | [HSPA12A] | TRUE | TRUE |
| 11 | 19960147 | rs12807015 | G | T | -3.00E-02 | 5.25E-03 | 3.20E-10 | [NAV2] | FALSE | FALSE |
| 11 | 31570861 | rs34618943 | T | A | -5.27E-02 | 6.07E-03 | 1.30E-21 | [ELP4] | TRUE | TRUE |
| 11 | 33406776 | rs3898926 | T | C | -2.80E-02 | 5.13E-03 | 1.60E-09 | [KIAA1549L] | TRUE | TRUE |
| 11 | 57544484 | rs17455626 | T | C | -3.13E-02 | 5.17E-03 | 2.81E-11 | [AP001931.2,AP001931.1,CTNND1] | FALSE | FALSE |
| 11 | 63678128 | rs199826712 | T | TA | 5.11E-02 | 1.01E-02 | 9.90E-09 | [MARK2] | FALSE | FALSE |
| 11 | 65326154 | rs12789028 | G | A | 6.75E-02 | 6.45E-03 | 9.77E-40 | [LTBP3] | TRUE | TRUE |

| CHR | POS | SNP | EA | NEA | BETA | SE | P | GENE_CONTEXT | CRAIG | CRAIG_META |
|---|---|---|---|---|---|---|---|---|---|---|
| 11 | 86740573 | rs4944662 | C | T | -3.69E-02 | 6.70E-03 | 8.11E-12 | FZD4--[]-TMEM135 | TRUE | TRUE |
| 11 | 94533444 | rs138059525 | G | A | 2.34E-01 | 3.07E-02 | 2.60E-14 | [AMOTL1] | FALSE | FALSE |
| 11 | 95292922 | rs11021217 | G | A | 4.06E-02 | 5.40E-03 | 1.49E-16 | SESN3---[]---FAM76B | TRUE | TRUE |
| 11 | 100645211 | rs7123718 | G | C | 4.18E-02 | 7.95E-03 | 3.66E-09 | [ARHGAP42] | FALSE | FALSE |
| 11 | 130280725 | rs4936099 | C | A | -5.33E-02 | 5.25E-03 | 7.80E-32 | [ADAMTS8] | TRUE | TRUE |
| 12 | 3364640 | rs147867843 | A | ACTTTCT( | -6.95E-02 | 1.28E-02 | 3.70E-09 | [TSPAN9] | FALSE | FALSE |
| 12 | 26392080 | rs16930371 | A | G | 3.74E-02 | 6.63E-03 | 2.20E-11 | [SSPN] | FALSE | TRUE |
| 12 | 31065843 | rs200103122 | A | AAAAT | 3.73E-02 | 7.09E-03 | 1.80E-08 | CAPRIN2---[]--TSPAN11 | FALSE | FALSE |
| 12 | 43548638 | rs1399377 | G | A | 1.84E-02 | 5.13E-03 | 2.34E-08 | PRICKLE1---[]---ADAMTS20 | FALSE | FALSE |
| 12 | 48153944 | rs12426774 | T | C | 4.69E-02 | 7.02E-03 | 3.81E-16 | [RAPGEF3,SLC48A1] | FALSE | TRUE |
| 12 | 76114872 | rs6582298 | G | A | 4.27E-02 | 5.44E-03 | 4.21E-14 | [AC078923.1] | TRUE | FALSE |
| 12 | 84049853 | rs10506895 | G | A | 1.18E-01 | 5.16E-03 | 4.41E-133 | TMTC2---[] | TRUE | TRUE |
| 12 | 91816926 | rs147377344 | C | CTTTTAC( | 3.04E-02 | 5.30E-03 | 2.10E-08 | DCN---[]---LINC01619 | FALSE | FALSE |
| 12 | 107250252 | rs17038814 | A | G | 5.54E-02 | 7.59E-03 | 5.13E-18 | [RIC8B] | TRUE | TRUE |
| 12 | 108134273 | rs4964616 | T | A | -4.00E-02 | 5.32E-03 | 2.26E-13 | [PRDM4] | FALSE | FALSE |
| 12 | 108988757 | rs17040818 | G | T | -2.97E-02 | 6.52E-03 | 3.21E-08 | [TMEM119] | FALSE | FALSE |
| 12 | 109874230 | rs2075432 | A | G | 2.45E-02 | 5.15E-03 | 1.82E-09 | [MYO1H] | FALSE | FALSE |
| 12 | 111800258 | rs3809272 | G | A | -2.82E-02 | 5.59E-03 | 9.62E-09 | [PHETA1] | FALSE | FALSE |
| 12 | 124666527 | rs7134138 | A | G | -3.64E-02 | 5.16E-03 | 3.64E-18 | [RFLNA] | TRUE | TRUE |
| 13 | 25766614 | rs9507473 | G | C | -4.22E-02 | 7.88E-03 | 6.88E-09 | AMER2--[]LINC01076 | FALSE | FALSE |
| 13 | 36683268 | rs9546383 | T | C | -4.49E-02 | 5.99E-03 | 1.50E-19 | [DCLK1] | TRUE | TRUE |
| 13 | 51913708 | rs9535646 | C | T | 4.00E-02 | 6.98E-03 | 4.01E-11 | [SERPINE3] | FALSE | FALSE |
| 13 | 109264870 | rs139237435 | A | ACATTTA | 5.18E-02 | 5.88E-03 | 1.20E-21 | [MYO16] | TRUE | TRUE |
| 13 | 110778747 | 13:110778747_CCTTTT_C | CCTTTT | C | -4.57E-02 | 5.51E-03 | 9.90E-18 | IRS2---[]--COL4A1 | TRUE | TRUE |
| 14 | 23452128 | rs3811183 | C | G | -3.35E-02 | 5.28E-03 | 7.26E-13 | AJUBA[]-C14orf93 | TRUE | TRUE |
| 14 | 53989952 | rs11623384 | C | T | 4.10E-02 | 5.45E-03 | 4.10E-17 | DDHD1---[]--AL163953.1 | TRUE | TRUE |
| 14 | 59583906 | rs61985994 | C | G | 3.79E-02 | 7.39E-03 | 1.88E-08 | DACT1---[]--DAAM1 | FALSE | FALSE |
| 14 | 60806759 | rs7493429 | A | C | -5.70E-02 | 5.62E-03 | 2.59E-35 | PPM1A--[]--C14orf39 | TRUE | TRUE |
| 14 | 65074869 | rs8006017 | A | G | -5.53E-02 | 6.90E-03 | 1.23E-19 | PPP1R36--[]--PLEKHG3 | FALSE | TRUE |
| 14 | 85922578 | rs1289426 | A | G | -5.21E-02 | 6.14E-03 | 2.69E-18 | []--FLRT2 | TRUE | TRUE |
| 14 | 95957694 | rs11160251 | T | G | 3.13E-02 | 5.63E-03 | 2.53E-08 | SYNE3--[]--GLRX5 | FALSE | FALSE |
| 15 | 71840327 | rs35194812 | T | C | -3.73E-02 | 7.07E-03 | 2.81E-11 | [THSD4] | FALSE | TRUE |
| 15 | 74228391 | rs4077284 | A | G | 3.07E-02 | 5.35E-03 | 1.41E-11 | [LOXL1] | TRUE | TRUE |
| 15 | 84484384 | rs59199978 | A | G | -4.63E-02 | 6.71E-03 | 3.11E-11 | [ADAMTSL3] | FALSE | FALSE |
| 15 | 99458902 | rs28612945 | C | T | 4.67E-02 | 6.38E-03 | 1.54E-15 | [IGF1R] | FALSE | TRUE |
| 15 | 101201604 | rs4299136 | G | C | -7.17E-02 | 7.55E-03 | 7.59E-29 | ASB7-[]---ALDH1A3 | TRUE | TRUE |
| 15 | 101753394 | rs28623369 | T | G | 3.27E-02 | 5.99E-03 | 5.90E-10 | [CHSY1] | FALSE | FALSE |
| 16 | 51469726 | rs8053277 | T | C | 8.35E-02 | 5.61E-03 | 6.28E-65 | SALL1---[]---HNRNPA1P48 | TRUE | TRUE |
| 16 | 74279778 | rs4889487 | G | C | 3.22E-02 | 5.72E-03 | 1.05E-11 | ZFHX3---[]--PSMD7 | FALSE | FALSE |
| 16 | 86386675 | rs1728368 | C | T | 9.05E-02 | 8.98E-03 | 4.66E-28 | IRF8---[]---FOXF1 | TRUE | TRUE |
| 17 | 10026855 | rs12936070 | C | T | 2.69E-02 | 5.93E-03 | 9.97E-10 | [GAS7] | FALSE | FALSE |
| 17 | 40867365 | rs115818584 | C | G | 1.22E-01 | 2.06E-02 | 4.01E-11 | [EZH1] | FALSE | FALSE |
| 17 | 45703433 | rs7220935 | C | T | 2.81E-02 | 5.15E-03 | 1.84E-11 | NPEPPS-[]--KPNB1 | FALSE | FALSE |
| 17 | 48225686 | rs4794104 | C | G | -4.13E-02 | 7.00E-03 | 6.55E-13 | [PPP1R9B] | TRUE | TRUE |
| 17 | 55419687 | rs792401 | G | A | 2.71E-02 | 5.60E-03 | 2.91E-08 | [MSI2] | FALSE | FALSE |
| 17 | 61865670 | 17:61865670_CT_C | CT | C | -3.20E-02 | 5.48E-03 | 2.40E-10 | [DDX42] | FALSE | FALSE |
| 17 | 65264966 | rs12939113 | C | T | -3.25E-02 | 5.70E-03 | 6.93E-11 | HELZ--[]--PSMD12 | FALSE | TRUE |
| 17 | 79602063 | rs9905786 | G | T | -2.52E-02 | 5.36E-03 | 2.56E-09 | [NPLOC4] | FALSE | FALSE |
| 17 | 80169426 | rs796355894 | A | AT | 2.99E-02 | 5.34E-03 | 1.60E-08 | [CCDC57] | FALSE | FALSE |
| 18 | 8799828 | rs568267 | C | T | 3.34E-02 | 5.92E-03 | 3.69E-10 | [MTCL1] | FALSE | FALSE |
| 18 | 23063159 | rs766791666 | T | TATC | -3.03E-02 | 5.31E-03 | 4.00E-10 | ZNF521---[]---SS18 | FALSE | FALSE |
| 18 | 34289285 | rs61735998 | G | T | 1.03E-01 | 1.64E-02 | 7.55E-12 | [FHOD3] | FALSE | FALSE |
| 18 | 56943484 | rs77759734 | C | T | -6.58E-02 | 1.20E-02 | 2.92E-08 | [CPLX4] | FALSE | FALSE |
| 19 | 817708 | rs7250902 | A | G | 3.93E-02 | 5.61E-03 | 3.65E-15 | [PLPPR3] | TRUE | TRUE |
| 19 | 14639064 | rs112614575 | C | CT | 3.96E-02 | 7.30E-03 | 6.10E-09 | [DNAJB1,TECR] | FALSE | FALSE |
| 19 | 32027330 | rs8102936 | G | A | 5.12E-02 | 5.47E-03 | 1.85E-21 | TSHZ3---[]---ZNF507 | TRUE | TRUE |
| 19 | 33477716 | 19:33477716_AAT_A | AAT | A | -4.52E-02 | 8.08E-03 | 1.40E-09 | [RHPN2] | FALSE | FALSE |
| 19 | 39195302 | rs757940594 | A | AGGAG | -3.15E-02 | 5.16E-03 | 7.40E-10 | [ACTN4] | FALSE | FALSE |
| 19 | 46356548 | rs7258364 | T | C | -2.64E-02 | 5.51E-03 | 2.14E-08 | [SYMPK] | FALSE | FALSE |
| 19 | 47455315 | rs311384 | A | G | 2.51E-02 | 5.66E-03 | 2.42E-08 | [ARHGAP35] | FALSE | FALSE |

Supplementary Information - Table 7. ML-based meta VCDR (loci)

| CHR | POS | SNP | EA | NEA | BETA | SE | P | GENE_CONTEXT | CRAIG | CRAIG_META |
|---|---|---|---|---|---|---|---|---|---|---|
| 20 | 1029686 | rs4816177 | A | G | -4.10E-02 | 6.79E-03 | 1.83E-08 | RSPO4--[]--PSMF1 | FALSE | FALSE |
| 20 | 6470094 | rs2326788 | G | A | 8.85E-02 | 5.33E-03 | 1.10E-67 | FERMT1---[]---LINC01713 | TRUE | TRUE |
| 20 | 31438954 | rs4911268 | A | G | 5.00E-02 | 6.67E-03 | 5.16E-17 | MAPRE1[]-EFCAB8 | TRUE | TRUE |
| 20 | 45796660 | rs2903940 | A | G | -3.24E-02 | 5.16E-03 | 3.31E-13 | [EYA2] | FALSE | FALSE |
| 21 | 29506261 | rs6516818 | T | A | 2.43E-02 | 5.22E-03 | 4.48E-08 | LINC01673---[]---N6AMT1 | FALSE | FALSE |
| 22 | 29115066 | rs4822983 | C | T | 1.12E-01 | 5.49E-03 | 1.37E-110 | [CHEK2] | TRUE | TRUE |
| 22 | 37907069 | rs2092172 | G | A | -7.49E-02 | 6.21E-03 | 5.81E-40 | [CARD10] | TRUE | TRUE |
| 22 | 39322264 | rs9306330 | C | T | -3.43E-02 | 6.67E-03 | 2.19E-09 | AL022318.1-[]--APOBEC3A | FALSE | FALSE |
| 22 | 46376985 | rs77164166 | G | A | 4.73E-02 | 5.55E-03 | 3.90E-17 | WNT7B-[]--LINC00899 | FALSE | FALSE |

| **Ontology** | **Term ID** | **Description** | **ML-based P-val** | **Craig *et al.* P-val** |
|---|---|---|---|---|
| GOBP | GO:0048598 | embryonic morphogenesis | 1.1x10-5 | 0.093 |
| GOBP | GO:0007423 | sensory organ development | 6.4x10-4 | 0.011 |
| GOBP | GO:0060021 | palate development | 7.2x10-4 | 1 |
| MP1KO | MP:0001297 | microphthalmia | 0.0025 | 1 |
| MP1KO | MP:0001286 | abnormal eye development | 0.0036 | 1 |
| MP1KO | MP:0002697 | abnormal eye size | 0.0047 | 1 |
| GOBP | GO:0001655 | urogenital system development | 0.006 | 1 |
| MP1KO | MP:0002081 | perinatal lethality | 0.0062 | 1 |
| GOBP | GO:0048566 | embryonic digestive tract development | 0.011 | 0.22 |
| MP1KO | MP:0004508 | abnormal pectoral girdle bone morphology | 0.012 | 1 |
| MP1KO | MP:0003257 | abnormal abdominal wall morphology | 0.012 | 0.062 |
| MP1KO | MP:0011087 | neonatal lethality, complete penetrance | 0.012 | 1 |
| GOBP | GO:0043010 | camera-type eye development | 0.015 | 0.048 |
| MP1KO | MP:0000455 | abnormal maxilla morphology | 0.016 | 1 |
| MP1KO | MP:0002058 | neonatal lethality | 0.028 | 1 |
| GOBP | GO:0048562 | embryonic organ morphogenesis | 0.03 | 1 |
| GOBP | GO:0060537 | muscle tissue development | 0.032 | 1 |
| GOBP | GO:0001654 | eye development | 0.033 | 0.18 |
| MP1KO | MP:0002925 | abnormal cardiovascular development | 0.037 | 1 |
| HP | HP:0008056 | Aplasia/Hypoplasia affecting the eye | 0.041 | 1 |
| GOBP | GO:0014706 | striated muscle tissue development | 0.042 | 1 |
| MP1KO | MP:0003942 | abnormal urinary system development | 0.049 | 0.54 |
| GOBP | GO:0055123 | digestive system development | 1 | 0.041 |

Supplementary Information - Table 9. ML-based glaucoma | VCDR (hits)

| CHR | POS | SNP | EA | NEA | EAF | BETA | SE | P | NUM_INDV | SRC | INFO | GENE_CONTEXT |
|---|---|---|---|---|---|---|---|---|---|---|---|---|
| 2 | 19486169 | rs1658243 | A | G | 0.455 | -3.57E-03 | 6.63E-04 | 4.60E-08 | 65741 | Genotyped | 1 | NT5C1B---[]--OSR1 |
| 6 | 151296166 | rs6906912 | A | G | 0.71 | 4.02E-03 | 7.31E-04 | 2.90E-08 | 65896 | Imputed | 0.996 | [MTHFD1L] |
| 7 | 99939050 | rs118119933 | C | T | 0.827 | 5.94E-03 | 8.91E-04 | 2.00E-11 | 65896 | Imputed | 0.963 | SPDYE3--[]--PILRB |
| 7 | 100457578 | rs80308281 | T | C | 0.994 | -2.71E-02 | 4.38E-03 | 7.50E-10 | 65774 | Genotyped | 1 | [SLC12A9] |
| 11 | 89017961 | rs1126809 | G | A | 0.699 | -5.32E-03 | 7.38E-04 | 5.80E-13 | 63346 | Genotyped | 1 | [TYR] |
| 13 | 36663302 | rs9315385 | T | G | 0.805 | 4.79E-03 | 8.39E-04 | 8.00E-09 | 65192 | Genotyped | 1 | [DCLK1] |
| 15 | 28337939 | rs4778239 | C | T | 0.01 | -2.24E-02 | 3.35E-03 | 1.80E-11 | 65896 | Imputed | 0.942 | [OCA2] |
| 15 | 28350407 | rs111155258 | G | A | 0.982 | 1.61E-02 | 2.53E-03 | 1.70E-10 | 65896 | Imputed | 0.97 | OCA2-[]-HERC2 |
| 15 | 28365618 | rs12913832 | A | G | 0.216 | -1.38E-02 | 8.07E-04 | 2.20E-66 | 65666 | Genotyped | 1 | [HERC2] |
| 15 | 28534777 | rs117325217 | C | T | 0.973 | 1.40E-02 | 2.19E-03 | 2.50E-10 | 65896 | Imputed | 0.867 | [HERC2] |
| 15 | 28566742 | rs2525964 | G | A | 0.045 | -1.44E-02 | 1.75E-03 | 1.40E-16 | 65896 | Imputed | 0.863 | [HERC2] |
| 17 | 79530993 | rs8070929 | G | T | 0.638 | -3.81E-03 | 6.90E-04 | 1.50E-08 | 65896 | Imputed | 0.987 | [NPLOC4] |
| 22 | 29094084 | rs5762750 | T | A | 0.372 | 3.93E-03 | 6.98E-04 | 9.70E-09 | 65896 | Imputed | 0.963 | [CHEK2] |

Supplementary Information - Table 10. ML-based glaucoma | VCDR (loci)

| CHR | POS | SNP | EA | NEA | EAF | BETA | SE | P | NUM_INDV | SRC | INFO | GENE_CONTEXT |
|---|---|---|---|---|---|---|---|---|---|---|---|---|
| 2 | 19486169 | rs1658243 | A | G | 0.455 | -3.57E-03 | 6.63E-04 | 4.60E-08 | 65741 | Genotyped | 1 | NT5C1B---[]--OSR1 |
| 6 | 151296166 | rs6906912 | A | G | 0.71 | 4.02E-03 | 7.31E-04 | 2.90E-08 | 65896 | Imputed | 0.996 | [MTHFD1L] |
| 7 | 99939050 | rs118119933 | C | T | 0.827 | 5.94E-03 | 8.91E-04 | 2.00E-11 | 65896 | Imputed | 0.963 | SPDYE3--[]--PILRB |
| 11 | 89017961 | rs1126809 | G | A | 0.699 | -5.32E-03 | 7.38E-04 | 5.80E-13 | 63346 | Genotyped | 1 | [TYR] |
| 13 | 36663302 | rs9315385 | T | G | 0.805 | 4.79E-03 | 8.39E-04 | 8.00E-09 | 65192 | Genotyped | 1 | [DCLK1] |
| 15 | 28365618 | rs12913832 | A | G | 0.216 | -1.38E-02 | 8.07E-04 | 2.20E-66 | 65666 | Genotyped | 1 | [HERC2] |
| 17 | 79530993 | rs8070929 | G | T | 0.638 | -3.81E-03 | 6.90E-04 | 1.50E-08 | 65896 | Imputed | 0.987 | [NPLOC4] |
| 22 | 29094084 | rs5762750 | T | A | 0.372 | 3.93E-03 | 6.98E-04 | 9.70E-09 | 65896 | Imputed | 0.963 | [CHEK2] |